\documentclass[12pt]{article}

\font\grande=cmr9.5 scaled \magstep4
\font\medio=cmr9.5 scaled \magstep2
\outer\def\beginsection#1\par{\medbreak\bigskip
      \message{#1}\leftline{\bf#1}\nobreak\medskip
\vskip-\parskip
      \noindent}
\usepackage{graphicx} 
\begin{document}
\bibliographystyle {unsrt}

\titlepage

\begin{flushright}
CERN-PH-TH/2009-039
\end{flushright}

\vspace{15mm}
\begin{center}
{\grande Parameter dependence  of magnetized CMB observables}\\
\vspace{1.5cm}
 Massimo Giovannini 
 \footnote{Electronic address: massimo.giovannini@cern.ch} \\
\vspace{1cm}
{{\sl Department of Physics, 
Theory Division, CERN, 1211 Geneva 23, Switzerland }}\\
\vspace{0.5cm}
{{\sl INFN, Section of Milan-Bicocca, 20126 Milan, Italy}}
\vspace*{2cm}

\end{center}

\vskip 1cm
\centerline{\medio  Abstract}
Pre-decoupling magnetic fields affect the scalar modes of the geometry and produce observable effects which can be constrained also through the use of current (as opposed to forthcoming) data stemming from the Cosmic Microwave Background observations. The dependence of the temperature and polarization angular power spectra upon the parameters of an ambient magnetic field is encoded in the scaling properties  of a set of basic integrals whose derivation is simplified  in the limit of small angular scales. The magnetically-induced distortions patterns of the relevant observables can be computed analytically by employing scaling considerations which are corroborated by numerical results. The parameter 
space of the magnetized Cosmic Microwave background anisotropies is also discussed in the light of the obtained 
analytical results. 
\noindent

\vspace{5mm}

\vfill
\newpage
\renewcommand{\theequation}{1.\arabic{equation}}
\setcounter{equation}{0}
\section{Formulation of the problem}
\label{sec1}
There are two complementary approaches to the analysis of the Cosmic Microwave Background (CMB in what follows) observables. The first one is direct and it consists in computing the  angular power spectra by faithfully including all the relevant physical effects.  The second approach is indirect, i.e. it amounts to deriving the dependence of the 
(measured) temperature and polarization anisotropies upon the parameters of the the underlying model which needs to be falsified.  The recent WMAP 5yr data \cite{WMAP5a, WMAP5b,WMAP5c} (see also \cite{WMAP3a,WMAP3b}) have been confronted with a number of theoretical scenarios that are logically organized around the $\Lambda$CDM paradigm where 
$\Lambda$ stands for the dark-energy component and CDM stands 
for Cold Dark Matter.  Similar statements can be made for other recent CMB data such as the ACBAR
observations  \cite{ACBAR1,ACBAR2} and the QUAD measurements \cite{QUAD1,QUAD2,QUAD3,QUAD4}.

A useful bridge between the direct and the indirect approach is represented 
by a number of scaling relations which serve as a diagnostic for the 
dependence of the (observed) angular power spectra upon the  parameters of a pivotal model.  
The temperature and polarization autocorrelations (i.e., respectively, TT and EE angular 
power spectra) and their mutual cross-correlations (i.e. the TE angular power spectra) can be written, with shorthand notation, as 
\begin{equation}
 G^{(\mathrm{TT})}_{\ell} =  \frac{\ell (\ell  +1)}{2 \pi} C_{\ell}^{(\mathrm{TT})}, \qquad 
 G^{(\mathrm{EE})}_{\ell} =  \frac{\ell (\ell  +1)}{2 \pi} C_{\ell}^{(\mathrm{EE})},\qquad  G^{(\mathrm{TE})}_{\ell} =  \frac{\ell (\ell  +1)}{2 \pi} C_{\ell}^{(\mathrm{TE})}.
\label{Eq1}
\end{equation}
In the $\Lambda$CDM scenario the angular power spectra of Eq. (\ref{Eq1})  are functions of, at least, six physical quantities
\begin{equation}
G_{\ell}^{(\mathrm{XY})}= G_{\ell}^{(\mathrm{XY})}(n_{\mathrm{s}}, \,\Omega_{\mathrm{b}0}, \, \Omega_{\mathrm{c}0}, \Omega_{\Lambda},\, H_{0}, \epsilon_{\mathrm{re}}),
\label{Eq1a}
\end{equation}
where X and Y stand, respectively, for T and E and where the parameters  denote, with standard notations, the spectral index of (adiabatic) curvature perturbations (i.e. $n_{\mathrm{s}}$), the critical fractions of baryons, CDM and dark energy (i.e., respectively, 
$\Omega_{\mathrm{b}0}$, $\Omega_{\mathrm{c}0}$ and $\Omega_{\Lambda}$), 
the Hubble constant $H_{0}$ and the optical depth at reionization (i.e. $\epsilon_{\mathrm{re}}$).
 
\begin{figure}[!ht]
\centering
\includegraphics[height=6cm]{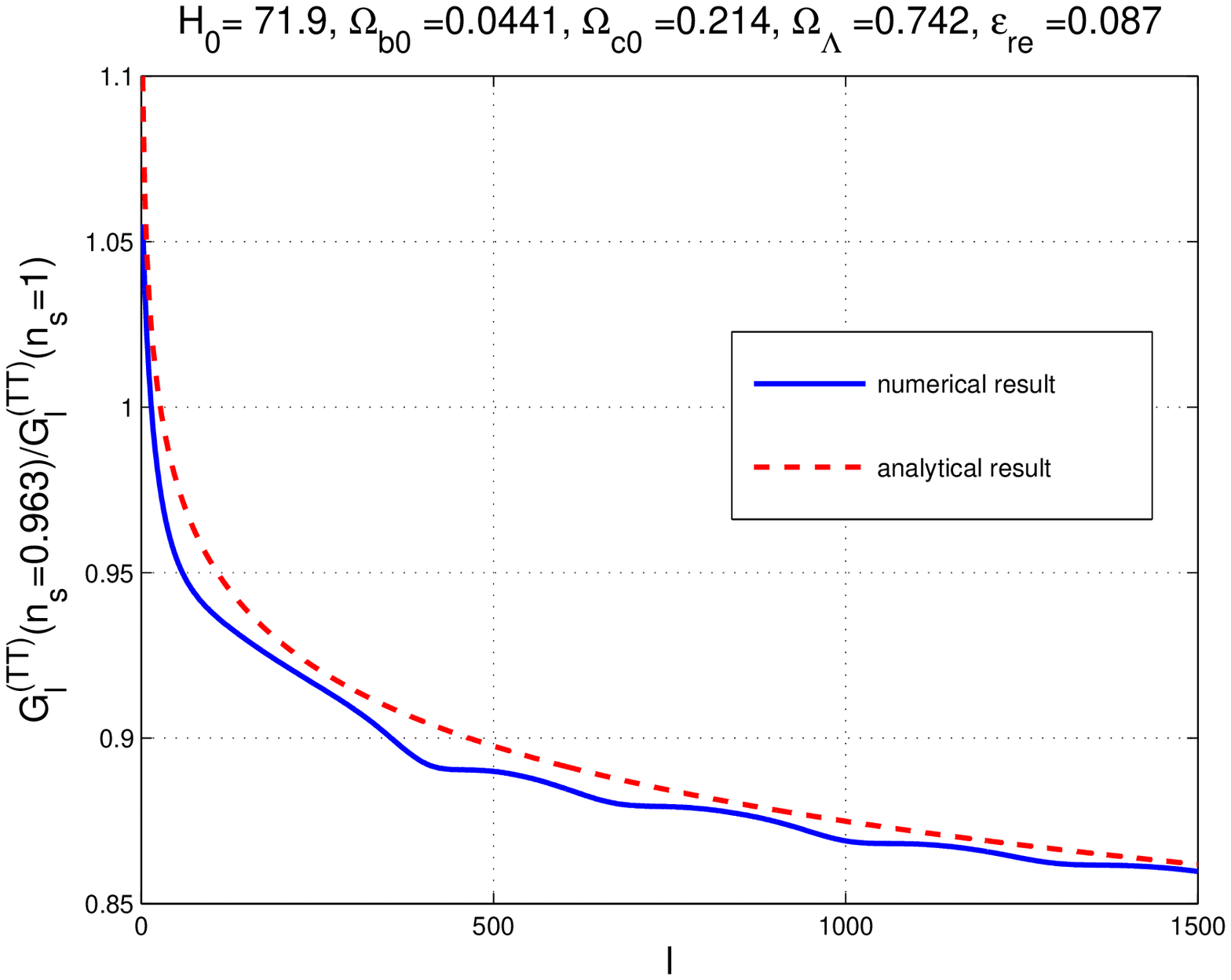}
\includegraphics[height=6cm]{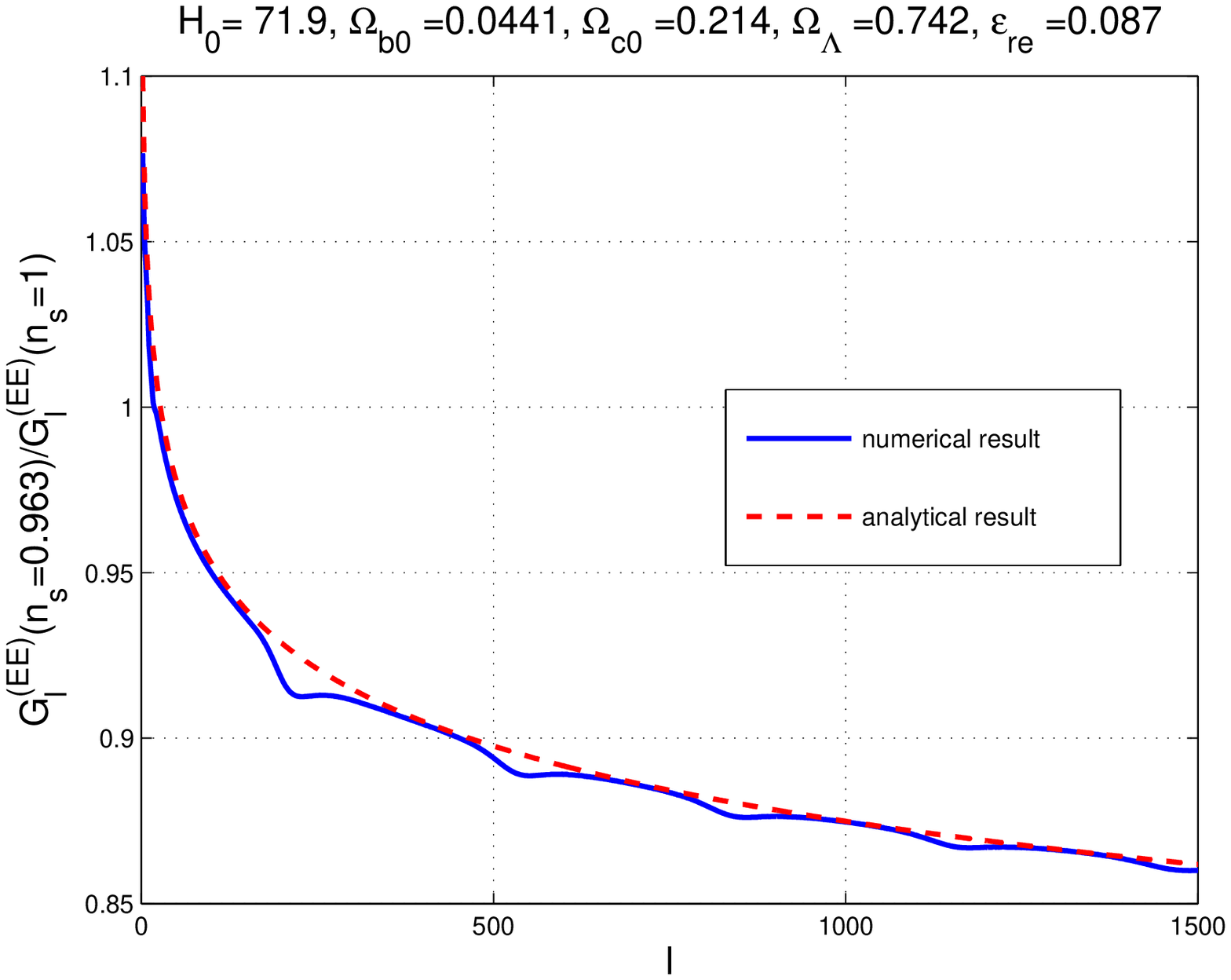}
\caption[a]{The usefulness of scaling relations is illustrated in a simplified situation involving the  variation of the scalar spectral index from the best-fit value of the WMAP 5yr data alone (i.e. $n_{\mathrm{s}} =0.963$) to an exact Harrison-Zeldovich spectrum (i.e. $n_{\mathrm{s}} = 1$).}
\label{figure1}      
\end{figure}

In the $\Lambda$CDM paradigm as well as in it extensions, the known scaling  
relations are often not the result of a numerical inference but are derived by means of analytical methods. 
Suppose, for sake of concreteness, 
that all the parameters of Eq. (\ref{Eq1a}) are fixed to the best fit of the WMAP 5yr data alone and just one (e.g. the spectral index) is allowed to scale.  
From semi-analytic considerations  it follows that 
\begin{equation}
G_{\ell}^{(\mathrm{TT})} \propto \biggl(\frac{\ell}{\ell_{\mathrm{p}}}\biggr)^{n_{\mathrm{s}} +1},\qquad G_{\ell}^{(\mathrm{EE})} \propto \biggl(\frac{\ell}{\ell_{\mathrm{p}}}\biggr)^{n_{\mathrm{s}} +1},\qquad G_{\ell}^{(\mathrm{TE})} \propto \biggl(\frac{\ell}{\ell_{\mathrm{p}}}\biggr)^{n_{\mathrm{s}}}
\label{Eq1b}
\end{equation}
where the notation $\propto$  signifies that the corresponding 
quantity scales with the multipole in a given manner\footnote{In Eq. (\ref{Eq1b}) $\ell_{\mathrm{p}}$ denotes 
the pivot multipole at which the initial conditions are customarily set. This scale is largely conventional 
and it will be hereby chosen to coincide with $\ell = 29$ which does correspond to the pivot 
wavenumber $k_{\mathrm{p}} = 0.002\, \mathrm{Mpc}^{-1}$.}. When the scalar spectral index changes 
from the best-fit value  (i.e. $n_{\mathrm{s}} =0.963$)  to a different value  $G_{\ell}^{(\mathrm{TT})}$ and $G_{\ell}^{(\mathrm{EE})}$ will be modified according to Eq. (\ref{Eq1b}).
On the vertical axis of the plots reported in  Fig. \ref{figure1} the ratios 
$G_{\ell}^{(\mathrm{TT})}(n_{\mathrm{s}} =0.963)/G_{\ell}^{(\mathrm{TT})}(n_{\mathrm{s}} =1)$ and 
 $G_{\ell}^{(\mathrm{EE})}(n_{\mathrm{s}} =0.963)/G_{\ell}^{(\mathrm{EE})}(n_{\mathrm{s}} =1)$ are computed numerically (full line) and analytically (as they emerge from Eq. (\ref{Eq1b})).  
 
There are  scaling relations involving, at once, different parameters. As it is  known from elementary 
considerations,  the height of the first peak in the acoustic oscillations of  $G_{\ell}^{(\mathrm{TT})}$ scales with first power the sound speed of the baryon photon fluid, which depends, in turn, upon the critical fraction of baryons; in formulae:
\begin{equation}
c_{\mathrm{sb}}(z_{*}) = \frac{1}{\sqrt{3[ 1 + R_{\mathrm{b}}(z_{*})]}}, \qquad  
R_{\mathrm{b}}(z) = \frac{3}{4} \frac{\rho_{\mathrm{b}}}{\rho_{\gamma}} = 
30.36 \,\omega_{\mathrm{b}} \,\biggl(\frac{10^{3}}{z_{*}}\biggr),
\label{Eq3}
\end{equation}
where $\omega_{\mathrm{b}} = h_{0}^2 \Omega_{\mathrm{b}0}$  and 
$z_{*}$ is the redshift of the last scattering. The examples can be multiplied 
by considering all the parameters of the $\Lambda$CDM scenario either 
alone or in some appropriate combinations. 
Instead of considering the dependence of the temperature and polarization 
anisotropies upon the various parameters listed at the right hand side of 
Eq. (\ref{Eq1a}),  it is often practical to consider a class of truly physical parameters emerging directly from the analysis of the various power spectra 
(see, e.g. \cite{norm1,norm2,norm3,norm4} and references therein). 
Concrete examples along this line are:
\begin{itemize}
\item{}  the relative heights of the first three peaks in $G_{\ell}^{(\mathrm{TT})}$;
\item{} the positions of the peaks in all the observed angular power spectra (i.e. 
 $G_{\ell}^{(\mathrm{TT})}$,  $G_{\ell}^{(\mathrm{EE})}$ and  $G_{\ell}^{(\mathrm{TE})}$)
  their heights, their depths, their mutual distances;
\item{} the numerical value of the acoustic multipole\footnote{The acoustic multipole is defined as  $\ell_{\mathrm{A}} = \pi D_{\mathrm{A}}(z_{*})/r_{\mathrm{s}}(z_{*})$ where $D_{\mathrm{A}}(z_{*})$ and $r_{\mathrm{s}}(z_{*})$ are, respectively,  the comoving angular diameter distance and  is the 
sound horizon at last scattering.};
\end{itemize}
and so on and so forth.
The height of the first acoustic peak does not have a simple dependence upon the parameters of Eq. (\ref{Eq1a}). Conversely, in terms of the quantities of 
 Eq. (\ref{Eq1a}), some power spectra exhibit rather contrived scaling properties  which become instead manifest as a function of appropriate sets of derived variables which are 
accessible to direct observations.  For instance, in the standard $\Lambda$CDM paradigm,  the numerical values of  the position of the anticorrelation peak in the TE power spectrum can be easily related to the position of the first Doppler peak in the TT power spectra; the height of the anticorrelation peak itself, however, does not have comparatively simple scaling with the parameters 
of Eq. (\ref{Eq1a}). 

\begin{figure}[!ht]
\centering
\includegraphics[height=6cm]{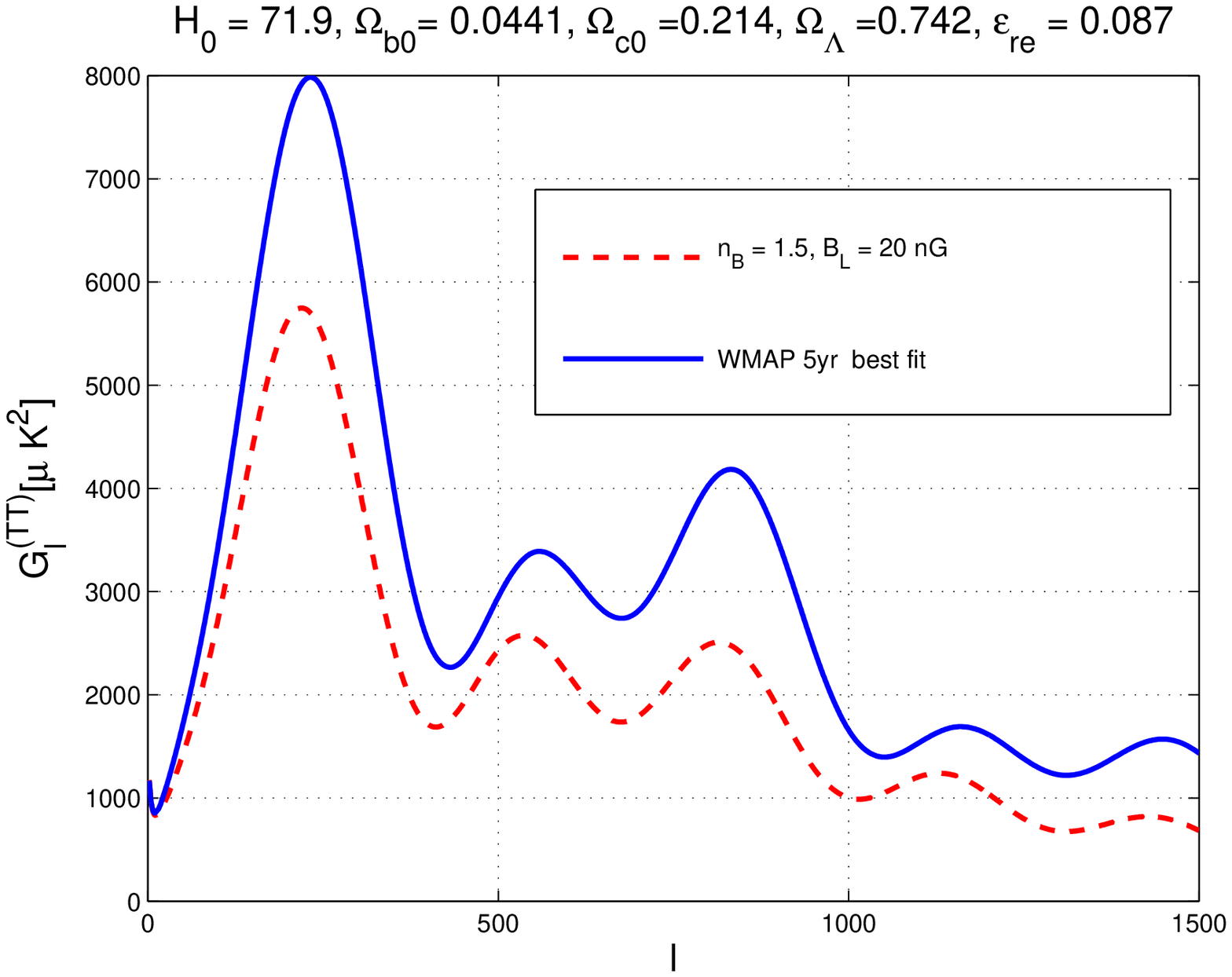}
\includegraphics[height=6cm]{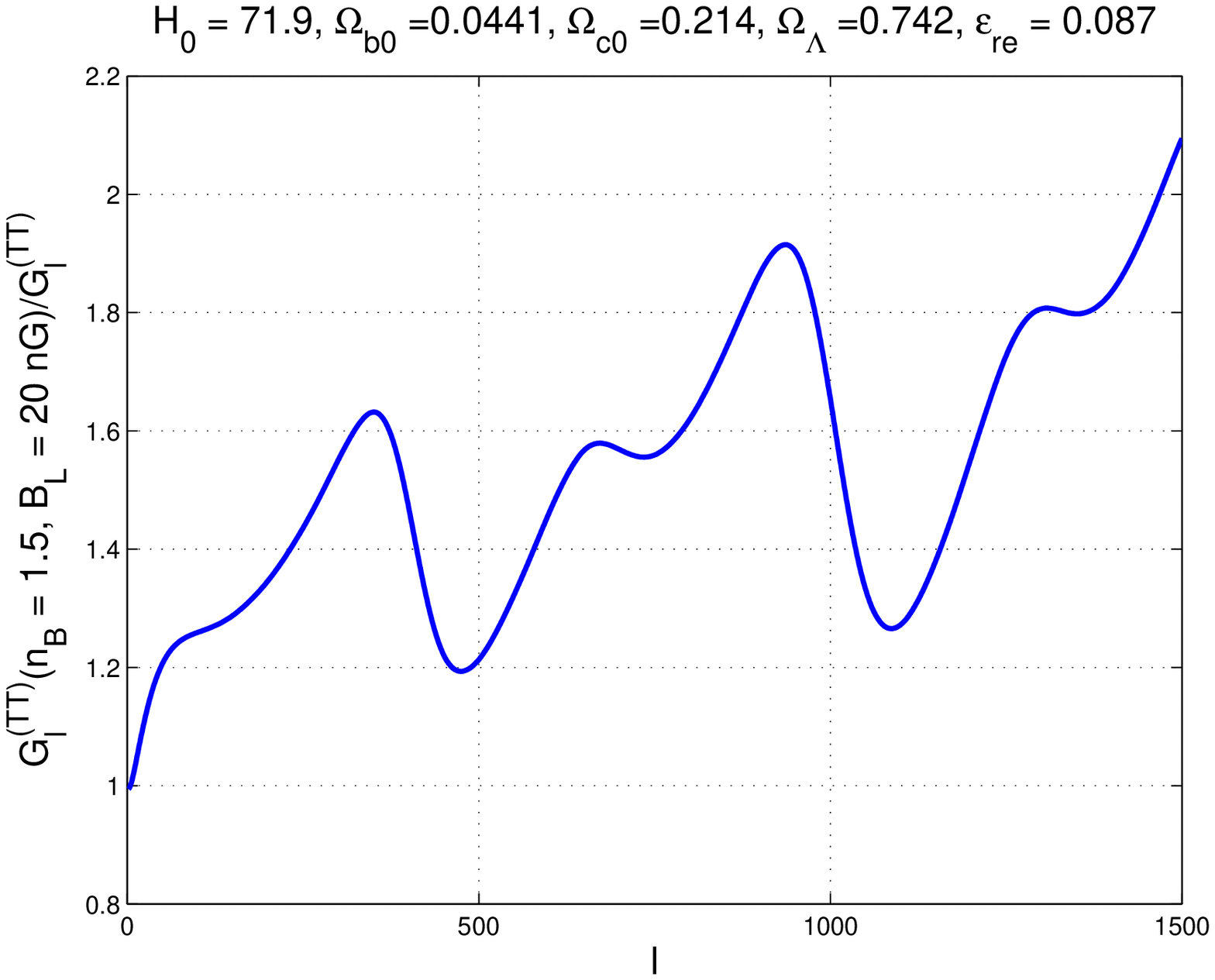}
\includegraphics[height=6cm]{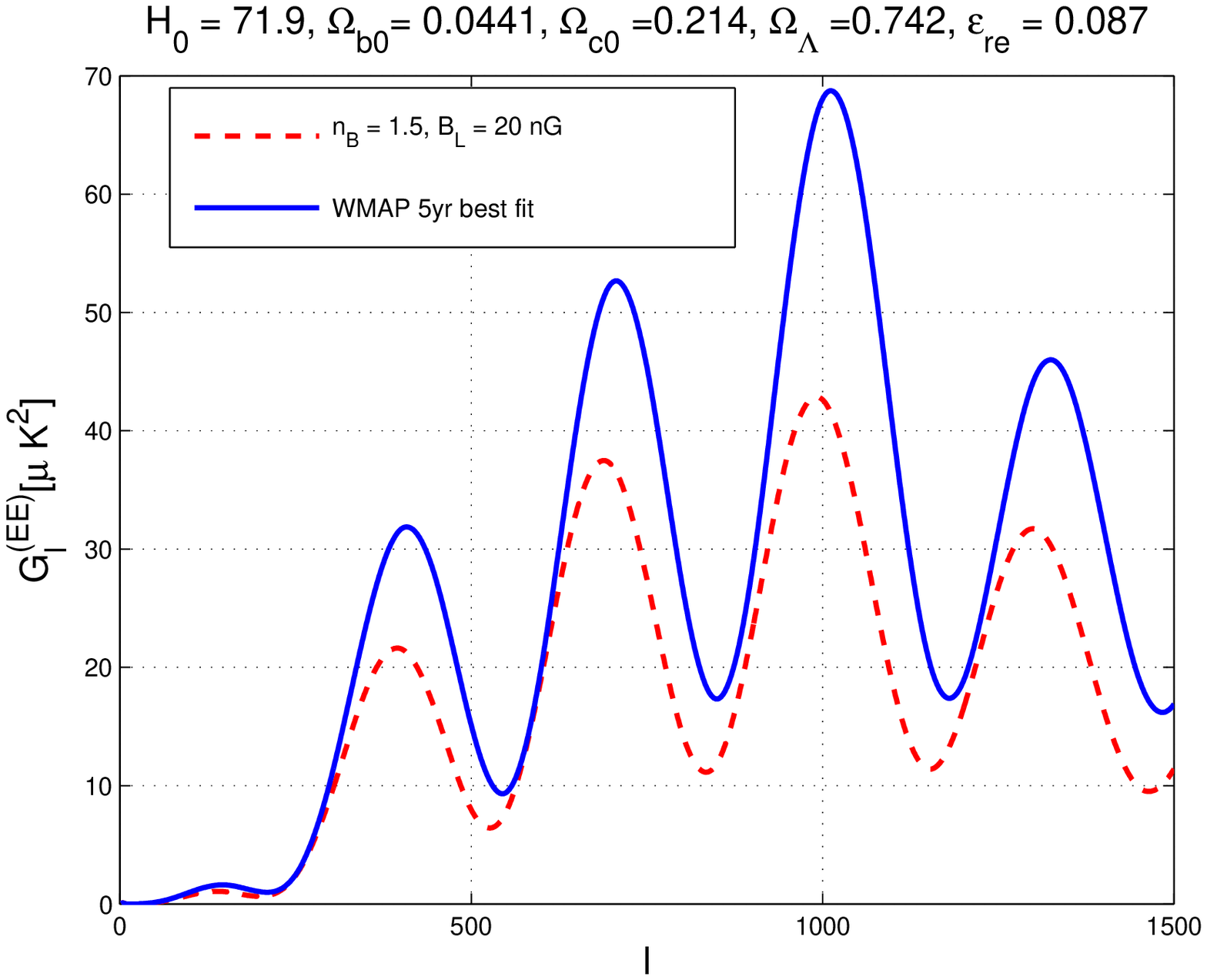}
\includegraphics[height=6cm]{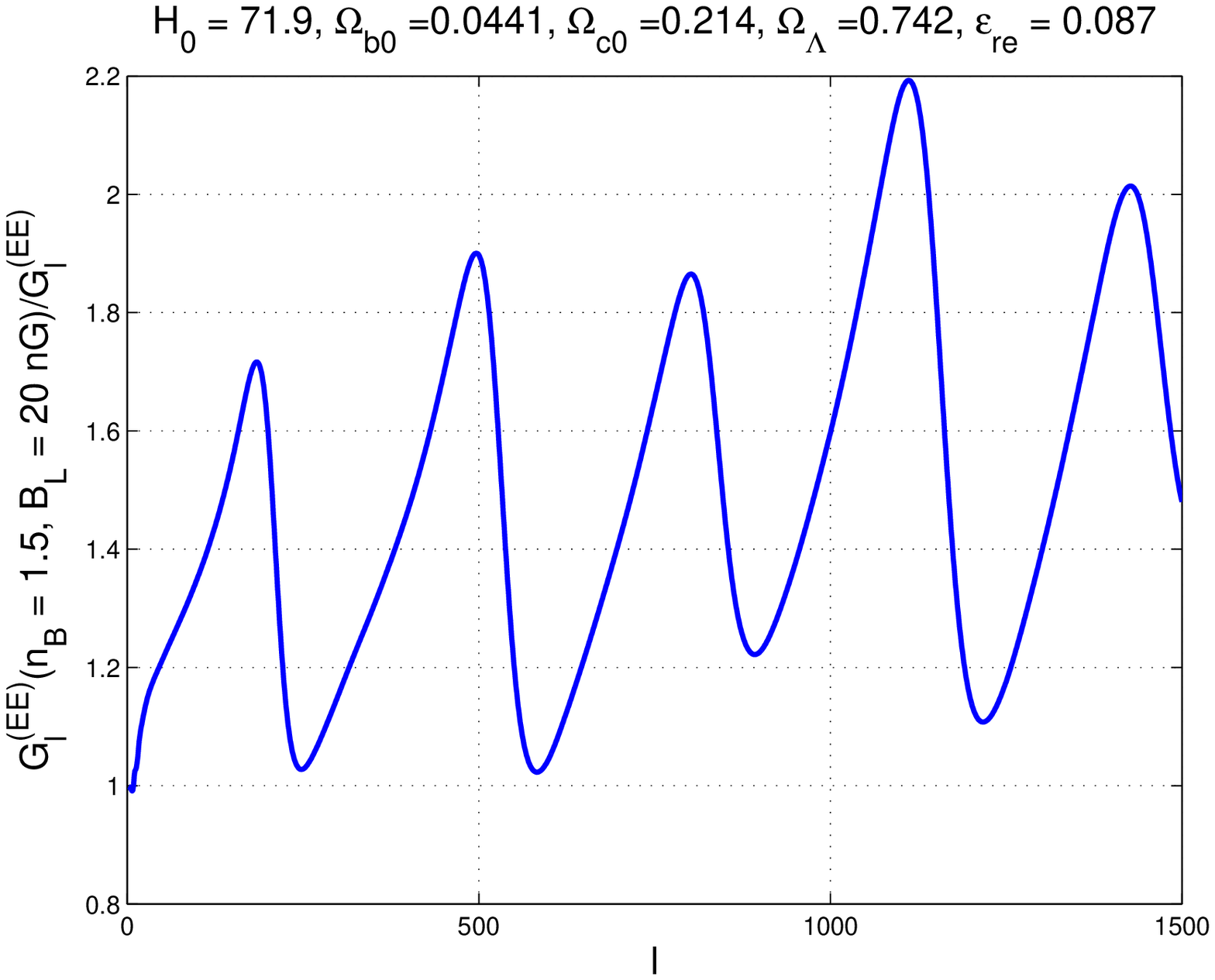}
\caption[a]{The temperature and polarization autocorrelation (plots at the left). In  the plots at the right we report 
the ratios $G_{\ell}^{(XX)}(n_{\mathrm{B}} \neq 0, \, B_{\mathrm{L}} \neq 0)/G_{\ell}^{(XX)}$ with $X = \mathrm{T}, \mathrm{E}$. 
By definition $G_{\ell}^{(XX)} = G_{\ell}^{(XX)}(n_{\mathrm{B}} = 0, \, B_{\mathrm{L}}= 0)$ i.e. $G_{\ell}^{XX}$ 
denotes either the temperature or the polarization autocorrelations in the absence of ambient magnetic field.}
\label{figure2}      
\end{figure}

The  CMB observables can be indeed studied in terms of a set of so-called normal parameters whose distinctive feature is that their mutual correlation is (or at least should be) very small.
One of the purposes of the  present paper is to look for similar types of scaling relations but in a qualitatively  different case, i.e. 
 when the model contains, on top of $\Lambda$CDM parameters, also an ambient magnetic field. 
To be even more specific we wish to consider the situation where 
the $\Lambda$CDM paradigm includes also a magnetized 
background whose presence necessarily entails supplementary parameters. 
The minimal situation, in this respect, contemplates two new parameters, i.e. the magnetic spectral index $n_{\mathrm{B}}$ and the magnetic field amplitude $B_{\mathrm{L}}$. In this case Eq. (\ref{Eq1a}) becomes
\begin{equation}
G_{\ell}^{(\mathrm{XY})}= G_{\ell}^{(\mathrm{XY})}(n_{\mathrm{B}},\, B_{\mathrm{L}},\, n_{\mathrm{s}}, \,\Omega_{\mathrm{b}0}, \, \Omega_{\mathrm{c}0}, \Omega_{\Lambda},\, H_{0}, \epsilon_{\mathrm{re}}).
\label{Eq1c}
\end{equation}
To formulate in visual terms the main problem addressed in the present investigation,  it is useful to look at Fig. \ref{figure2} which, in some way,  is the analog of Fig. \ref{figure1} but in the case when large-scale magnetic fields are consistently included in the pre-decoupling physics.
In the two plots at the left of Fig. \ref{figure2} 
the TT and EE angular power spectra are presented in two cases, i.e. 
in the absence of an ambient magnetic field (dashed line in both plots) and in 
the case when a magnetic field modifies the initial conditions and the evolution 
of CMB anisotropies (full lines in both plots at the left). 
Just for illustrative purposes the magnetic spectral index and the comoving magnetic field amplitude have been chosen to be, respectively, $n_{\mathrm{B}} = 1.5$ and 
$B_{\mathrm{L}} = 20$ nG.  Always in Fig. \ref{figure2} (but in the two plots at the right)
the TT and the EE angular power spectra have been divided by the corresponding 
power spectra but computed in the absence of magnetic fields. In other words, 
in the plots at the right the two curves stem from the ratio of the angular power 
spectra illustrated in the right plots of the same Fig. \ref{figure2}.
Already at a qualitative level, the right plot of Fig. \ref{figure2} shows interesting features like, for instance,  different periodicities but similar growth rates with the multipole 
number. Is it possible to understand these as well as other features in analytic terms?
This is one of the questions we ought to address. Another question 
could be: as in Fig. \ref{figure1} the scaling properties can be used 
to infer the variation of the power spectra with the scalar spectral index, can we do 
the same when large-scale magnetic fields are present?
To achieve such a purpose it will prove useful to employ different approximations schemes which served as basic ingredients for developing the numerical techniques that led to the results of Fig. \ref{figure2}.

Before going through  the plan of the investigation it is appropriate to swiftly remind the 
main motivations related to the study of the magnetized Universe (see, e.g. \cite{magnetized,rees1} for dedicated reviews on the subject) which is basically the research program pursued here. Magnetic fields 
in gravitationally bound systems are a rather mundane feature of our Universe 
but one of the least understood especially  when the typical correlation scale 
of the field is large as it happens, for instance, in the case of galaxies, clusters or even superclusters. Since the pioneering works of Alfv\'en and Fermi \cite{alv1,fermi1,fermi2} (see also \cite{alv2,alv3}) large-scale magnetic fields have been the subject of numerous speculations whose detailed nature cannot be summarized here. The interested reader is refereed to review articles such as \cite{magnetized} and to the references of more recent publications 
\cite{spectator,bamba1,bamba2,camp1,camp2}. 
 It would be difficult to account for the detailed arguments leading to 
the different ideas which can eventually originate large-scale magnetism. Furthermore, as it will be argued below, the 
overall spirit of the present approach is more pragmatic; instead of dwelling on the rigor of the speculations 
leading to large-scale magnetism, it seems more urgent, on a physical ground, to decide 
which are the measurable effects of pre-decoupling magnetism. The latter quest is also 
experimentally better founded  since the degree of rigor of a speculation is always rather 
complicated to assess and might strongly depend upon the theoretical prejudice 
of the authors.

The characteristics of the approaches discussed in this paper  do not exclude 
the possibility of magnetizing the early history of the Universe \cite{magnetized} which is a 
rather intriguing subject of speculations dating back to the pioneering works of Zeldovich \cite{zeldovich} and, independently, 
Harrison \cite{harrison1,harrison2} (see also \cite{magnetized} for further details). 
The early phases of the evolution 
of the Universe are often connected with an inflationary epoch \footnote{ Large-scale magnetic fields produced 
inside the Hubble radius after inflation will have a correlation scale 
bounded (from above) by the Hubble radius at the moment 
when some charge separation is produced (be it, for instance, 
the electroweak time).  Since the Hubble radius, during radiation, evolves much faster than the correlation scale of the produced field, the typical scale over which the magnetic field is coherent today is 
much shorter than the Mpc, obliterating, in this way, the 
possibility of successfully reproducing the galactic magnetic field \cite{magnetized}.}  where, however, it seems to be rather 
difficult to produce large-scale magnetic fields in four-dimensional Friedmann-Robertson-Walker Universes. 
This impasse is related to the peculiar form of the evolution equations of Maxwell fields in curved backgrounds which are, technically, invariant under the Weyl rescaling of the geometry.  To amplify gauge fields one might want to extend his 
model to higher-dimensional frameworks \cite{intdim} or couple the kinetic term of the gauge fields to a 
spectator field \cite{spectator}. Magnetogenesis models based on the dynamics of an appropriate spectator 
field delicately improve on the structure of conventional inflationary models and can be directly 
constrained by CMB data \cite{spectator}.  In spite of the specific model it seems relevant that 
the amplified gauge fields are Abelian. The only non-screened 
vector modes that are present at finite conductivity 
are the ones associated with the hypercharge field.
The non-Abelian fields develop actually a mass and they are screened as the Universe thermalizes.
After the electroweak phase transition the photon field remains unscreened 
with amplitude $\cos{\theta_{\mathrm{w}}} \vec{{\mathcal Y}}$ where $\vec{{\mathcal Y}}$ 
is the hypercharge field and $\theta_{\mathrm{w}}$ is the Weinberg's angle.
While it is certainly interesting to speculate on the origin of large-scale magnetism 
prior to matter-radiation equality (i.e. for redshifts larger than, approximately, $3200$)
it is also rather urgent, as it will be argued in section 2,  
to scrutinize the CMB observables with the purpose of falsifying the statement that large-scale 
magnetic fields were indeed present around matter-radiation equality and, later, at the epoch 
of photon decoupling (i.e. for a typical redshift of the order of $1100$).  

Large-scale magnetic fields are a well defined object of experimental study 
since the pioneering contributions of  Hiltner and Hall correctly interpreted in terms 
of a large-scale (interstellar) magnetic field by Davis and Greenstein \cite{gal1}.
For extended reviews on galactic magnetism see \cite{gal2} and \cite{gal3}. 
Large-scale magnetism is also a well established phenomenon in rich (i.e. 
Abell) clusters of galaxies \cite{cl1} (see also \cite{cl2,cl3})  as well as, still with 
large uncertainties, in superclusters \cite{scl1}.  Magnetic fields in normal galaxies 
at high redshifts \cite{highz} could be already strong and this would be theoretically suggestive, in spite 
of the necessary caveats stemming from the large observational uncertainties (similar, in nature, to the ones 
experience while observing supercluster magnetism). 
Since we do observe large-scale for progressively larger 
redshifts it is natural to ask what happens at the photon decoupling especially 
because of the wealth of CMB data. In investigating 
such a class of phenomena the idea is to frame the least number of hypotheses 
on the subsequent evolution of large-scale magnetic fields so that we will take 
as starting point a faithful plasma description which is exactly the one employed 
in terrestrial laboratories \cite{spitzer,krall,biskamp,boyd}. 

Having spelled out the general perspective of the present paper,  its layout is, in short, the following.  
In section 2 the physics of the strongly and weakly interacting species at the epoch of photon decoupling
 will be briefly summarized with emphasis on the role of large-scale magnetic fields. Such a discussion will provide, 
 in a reasonably self-contained perspective, all the equations which will be employed in the subsequent analysis. 
 Section 3 treats the analytical methods employed in the line of sight solutions of the Boltzmann hierarchy, while, 
 in section 4 the (magnetized) temperature and polarization anisotropies are computed in terms of a set of 8 
 basic integrals. Section 5 illustrates the comparison of analytical and numerical 
results while Section 6 deals with the analysis of the parameter space of magnetized CMB anisotropies.
 Section 7 contains the concluding considerations.

\renewcommand{\theequation}{2.\arabic{equation}}
\setcounter{equation}{0}
\section{Strongly and weakly interacting species}
\label{sec2}
Prior to decoupling the evolution of the plasma can be described in terms of charged and neutral species. 
Charged species interact directly with the ambient magnetic field whose effect, on the neutral species, 
is mediated by the relativistic fluctuations of the geometry.
The separate role of the electrons and ions is often overlooked when the relativistic fluctuations of the 
geometry are consistently taken into account. The approach 
initially formulated in \cite{max1,max2} and developed in \cite{max3} is based on 
a rather conservative perspective: instead to doing a lot of effort 
to compute exotic phenomena triggered by large-scale magnetic fields, it is logically  
more urgent to compute in detail how large-scale magnetic fields affect CMB observables (see \cite{maxrv} for a more detailed formulation of such a research program). If large-scale magnetic fields gravitate and interact, simultaneously, with electrons and protons the most relevant effects on the temperature and polarization observables will be the one coming from the scalar modes of the geometry \cite{max1,max2,max3}. 

The scalar modes of the geometry admit two kinds of initial data 
which allow for the inclusion of large-scale 
magnetic fields, i.e. adiabatic initial conditions leading to the magnetized
adiabatic mode (see \cite{max1,max2,max4}) and entropic initial conditions leading to various 
magnetized isocurvature modes (see \cite{max5}). 
In what follows the main focus will be on the magnetized adiabatic mode. However, the same considerations developed here can be easily extended to the case of entropic initial conditions. In the case of adiabatic initial 
conditions the fluctuations of the spatial curvature are due to the fluctuations of the energy density while  the fluctuations of the specific entropy are strictly vanishing at large scales. In the case of entropic initial conditions 
the curvature inhomogeneities are due to the fluctuations of the sound speed which are related, in turn, 
to the fluctuations of the specific entropy (see last part of section \ref{sec2}).  Non-adiabatic initial conditions 
can be observationally constrained in different ways \cite{h1,h2,h3,h4,h5} and may lead, in the magnetized case, to interesting shape effects on the CMB observables. There are, of course, also different themes which 
involve the physics of large-scale magnetic fields  in connection with CMB physics (see \cite{maxrv} for a dedicated review). For instance,  large-scale (tangled) magnetic fields 
might have also specific effects related to the vector and tensor modes of the geometry (which are minute at large scales). These effects have been analyzed, at various levels of concreteness,  in \cite{bs1,bs2,bs3} (see also \cite{cs1,cs2} and compare them to 
\cite{max1,max2,max3,max5,faraday1}). 

Depending upon their interaction rates, the constituents of the plasma can be classified into two groups: the strongly interacting species (such as the electrons, the protons and the photons) and the weakly interacting constituents (such as the CDM particles and the neutrinos).  The difference between the two aforementioned categories resides in electromagnetic interaction which strongly affects the evolution of the electron-photon-ion system while it affects only indirectly the evolution of the weakly interacting species. 
The indirect effect of large-scale magnetic fields on the weakly interacting species comes from gravitational interactions: since large-scale magnetic fields gravitate, the relativistic fluctuations of the geometry are modified by their presence via the enforcement of the various constraints stemming from Einstein equations.  There is a whole class 
of effects which are related to the high-frequncy branch of the spectrum of plasma excitations \cite{faraday1,faraday2} 
which can be treated within the same framework described here (i.e. the magnetized adiabatic mode). 
In what follows, however, the focus will be on the scaling properties of the TT, EE and TE correlations since the angular 
power spectra of the B-mode polarization have been the subject of a separate study (see \cite{faraday1}, second and third papers).
It should be stressed that the values of the magnetic fields 
used in this paper are, sometimes, extreme, e.g. intensities of $10$ nG are 
by far excluded both by direct limits stemming from the polarization observables 
\cite{faraday1,faraday2} and from the analysis of the peak structure of the TT correlations \cite{ESTIMATE}. In \cite{ESTIMATE} the WMAP 5yr data 
have been analyzed by including the effects of large-scale magnetic fields.
In this perspective, for instance, the parameters reported in Fig. \ref{figure2} 
are excluded. More specifically, the values\footnote{The magnetic field intensity and the magnetic spectral index are assigned as in \cite{max3,max4} (see also second and third paper of \cite{faraday1}). In the present context,  $B_{\mathrm{L}}$ is the comoving amplitude of the field regularized over a typical scale $k_{\mathrm{L}} = \mathrm{Mpc}^{-1}$. } $(n_{\mathrm{B}}, \, B_{\mathrm{L}} ) = 
(2,\, 10 \mathrm{nG})$ are excluded, by the analysis of the TT and TE, to $95$ \%  
confidence level. At the same time, it is useful to illustrate the results 
in terms of these extreme values since, in this way, the visual impact is 
more pronounced and the scaling of the results with the parameters of the ambient magnetic field more evident.
\subsection{Generalities}
The simplest description of the pre-decoupling plasma in the presence of large-scale magnetic fields 
can be derived from the general pair of equations:
\begin{eqnarray}
&& R_{\mu}^{\nu} - \frac{1}{2} \delta_{\mu}^{\nu} R = 8 \pi G T_{\mu}^{\nu}, 
\label{F1}\\
&& \nabla_{\mu} F^{\mu\nu} = 4 \pi j^{\nu}.
\label{F2}
\end{eqnarray}
In Eq. (\ref{F1}) $R_{\mu\nu}$ is the Ricci tensor, $R$ is the Ricci scalar and 
$T_{\mu}^{\nu}$ is the total energy-momentum tensor of the system. In Eq. (\ref{F2}) 
$F^{\mu\nu}$ is the Maxwell field strength and $j^{\nu}$ is the total current 
of the system. Both the total energy momentum tensor and the total current 
must be covariantly conserved, i.e. 
\begin{equation}
\nabla_{\mu} j^{\mu} =0, \qquad \nabla_{\mu} T^{\mu\nu} =0.
\label{F3}
\end{equation}
In Eq. (\ref{F3}) $\nabla_{\mu}$ denotes the covariant derivative. The total 
energy-momentum tensor is given by:
\begin{equation}
T^{\mu\nu} = T^{\mu\nu}_{(\mathrm{e})} + T^{\mu\nu}_{(\mathrm{i})} + 
T^{\mu\nu}_{(\nu)} + T^{\mu\nu}_{(\gamma)} + T^{\mu\nu}_{(\mathrm{c})} 
+ T^{\mu\nu}_{(\Lambda)} + T^{\mu\nu}_{(\mathrm{EM})},
\label{F4}
\end{equation}
where the subscripts denote, respectively, the contributions of electrons, ions, neutrinos, photons and CDM particles. More 
quantitatively the energy-momentum tensors of the different species are:
\begin{eqnarray}
&& T^{\alpha\beta}_{(\mathrm{e})} = \rho_{\mathrm{e}} \,u_{(\mathrm{e})}^{\alpha} u_{(\mathrm{e})}^{\beta}, \qquad T^{\alpha\beta}_{(\mathrm{i})} = \rho_{\mathrm{i}}\, u_{(\mathrm{i})}^{\alpha} u_{(\mathrm{i})}^{\beta},
\qquad T^{\alpha\beta}_{(\mathrm{c})} = \rho_{\mathrm{c}} \,u_{(\mathrm{c})}^{\alpha} u_{(\mathrm{c})}^{\beta}
\label{F4a}\\
&& T^{\alpha\beta}_{(\nu)}= \frac{4}{3} \rho_{\nu}\, u^{\alpha}_{(\nu)} u^{(\beta)}_{(\nu)} - \frac{\rho_{\nu}}{3} g^{\alpha\beta},\qquad 
T^{\alpha\beta}_{(\gamma)}= \frac{4}{3} \rho_{\gamma} \,u^{\alpha}_{(\gamma)} u^{(\beta)}_{(\gamma)} - \frac{\rho_{\gamma}}{3} g^{\alpha\beta},
\label{F4b}\\
&& T_{(\mathrm{EM})}^{\alpha\beta} = 
\frac{1}{4\pi} \biggl[ - F^{\alpha\mu} F^{\beta}_{\mu} + \frac{1}{4} 
g^{\alpha\beta} F_{\mu\nu} F^{\mu\nu}\biggr], \qquad T^{\alpha\beta}_{\Lambda} = \rho_{\Lambda} g^{\alpha\beta},
\label{F4c}
\end{eqnarray}
 where $F_{0 i} = - a^2 \, {\mathcal E}_{i}$ and $F_{ij} = - a^2
  \epsilon_{ijk} {\mathcal B}^{k}$ are the components of the electromagnetic field strengths expressed, respectively, in terms of the electric and magnetic fields.  In  Eq. (\ref{F4b}) the energy-momentum 
  tensor of the neutrinos should also contain a contribution from the anisotropic stress which 
 is, however, fully inhomogeneous and affects the evolution 
 of the curvature perturbations rather than the evolution of the background metric.
  The total current of the system is due to electrons and ions, i.e. 
\begin{eqnarray}
&& j^{\mu} = e \,\tilde{n}_{\mathrm{i}} u^{\mu}_{(\mathrm{i})} - 
e\, \tilde{n}_{\mathrm{e}} u^{\mu}_{(\mathrm{e})}, 
\nonumber\\
&& g_{\mu\nu}\,u^{\mu}_{(\mathrm{i})}\,u^{\nu}_{(\mathrm{i})} = 1, \qquad 
g_{\mu\nu}\,u^{\mu}_{(\mathrm{e})}\,u^{\nu}_{(\mathrm{e})} = 1,
\label{F5}
\end{eqnarray}
where $e$ denotes the electric charge\footnote{In this paper the units will be such that $e^{2}/(\hbar c) = 1/137$. Furthermore, as it is apparent from Eq. (\ref{F2}), in front of 
$F_{\alpha\beta} F^{\alpha\beta}$, in the action, there is a factor $1/(16\, \pi)$ which 
is reflected in the $4\pi$ of Eq. (\ref{F2}). Within these conventions and imposing 
the natural system of units $\hbar= c= 1$, $1\, \mathrm{Gauss} = 6.9241 \times 10^{-20}\, \mathrm{GeV}^2$. } ; $\tilde{n}_{\mathrm{e}}$ and 
$\tilde{n}_{\mathrm{i}}$ are the physical (as opposed to comoving) concentrations of the electrons and of the ions. 

The evolution equations of the background geometry follow directly 
from Eq. (\ref{F1}) by recalling that, in the $\Lambda$CDM paradigm, the 
geometry is conformally flat (i.e. $g_{\mu\nu} = a^2 \eta_{\mu\nu}$ where 
$\eta_{\mu\nu}$ is the Minkowski metric):
\begin{eqnarray}
&& {\mathcal H}^2 = \frac{8\pi G}{3} a^2 \rho_{\mathrm{t}}, 
\label{F6}\\
&& {\mathcal H}^2 - {\mathcal H}' = 4\pi G a^2 (p_{\mathrm{t}} + \rho_{\mathrm{t}}),
\label{F7}\\
&& \rho_{\mathrm{t}}' + 3 {\mathcal H} (\rho_{\mathrm{t}} + p_{\mathrm{t}}) =0.
\label{F8}
\end{eqnarray}
In Eqs. (\ref{F6}), (\ref{F7}) and (\ref{F8}) 
\begin{itemize}
\item{} the prime denotes a derivation with respect to the conformal time 
coordinate $\tau$;
\item{} ${\mathcal H} = a'/a$  which also implies ${\mathcal H}= a H$ where $H= \dot{a}/a$ (where the overdot 
denotes a derivation with respect to the cosmic time coordinate $t$; recall that $dt = a(\tau)\, d\tau$);
\item{} finally the total energy density and the total pressure are:
\begin{eqnarray}
\rho_{\mathrm{t}} &=& \rho_{\mathrm{e}} + \rho_{\mathrm{i}} + \rho_{\gamma} + \rho_{\nu} + \rho_{\mathrm{c}}+  \rho_{\Lambda},
\label{F9}\\
p_{\mathrm{t}} &=& \frac{\rho_{\gamma}}{3} + \frac{\rho_{\nu}}{3} -  \rho_{\Lambda}. 
\label{F10}
\end{eqnarray}
\end{itemize}
For purposes of presentation we started directly from the covariantly conserved evolution 
of the energy-momentum tensor. It can be shown that this description is fully 
equivalent to a truncated Vlasov-Landau description \cite{maxrv,max4}.

\subsection{Strongly interacting species}
The photons, the electrons and the ions interact electromagnetically and their velocities 
 are tied together by the presence of scattering terms. At the same time 
 photons, electrons and ions affect the evolution of the background geoemetry 
 (i.e. Eqs. (\ref{F6})--(\ref{F8}) and (\ref{F9})--(\ref{F10})) and of 
 its relativistic inhomogeneities. 
 From Eq. (\ref{F2}) the evolution of the Maxwell fields obeys 
\begin{eqnarray}
&& \vec{\nabla} \cdot \vec{E} = 4 \pi e (n_{\mathrm{i}} - n_{\mathrm{e}}),
\label{S1}\\
&&\vec{\nabla} \cdot \vec{B} =0,
\label{S2}\\
&& \vec{\nabla} \times \vec{E} = - \vec{B}^{\,'},
\label{S3}\\
&& \vec{\nabla}\times \vec{B} = 4 \pi e (n_{\mathrm{i}}\, \vec{v}_{\mathrm{i}} - 
n_{\mathrm{e}}\, \vec{v}_{\mathrm{e}} ) + \vec{E}^{\,'},
\label{S4}
\end{eqnarray}
where the comoving concentrations and the comoving electromagnetic fields are:
\begin{equation}
n_{\mathrm{i}} = a^3 \tilde{n}_{\mathrm{i}},\qquad 
n_{\mathrm{e}} = a^3 \tilde{n}_{\mathrm{e}},\qquad \vec{E} = a^2 \vec{{\mathcal E}}, \qquad  
\vec{B} = a^2 \vec{{\mathcal B}}.
\label{S4a}
\end{equation}
The evolution equations 
of the electrons, of the ions and of the photons must include the relevant scattering terms
governing their mutual momentum exchanges:
\begin{eqnarray}
&& \vec{v}_{\mathrm{e}}^{\,\prime} + {\mathcal H}\,\vec{v}_{\mathrm{e}} = - \frac{e}{m_{\mathrm{e}} \, a} [ \vec{E} + \vec{v}_{\mathrm{e}} \times \vec{B}] - \vec{\nabla} \phi 
+ 
\frac{4}{3} \frac{\rho_{\gamma}}{\rho_{\mathrm{e}}} a 
\Gamma_{\gamma \, \mathrm{e}} (\vec{v}_{\gamma} - \vec{v}_{\mathrm{e}}) + a \Gamma_{\mathrm{e\,i}} ( \vec{v}_{\mathrm{i}} - \vec{v}_{\mathrm{e}}),
\label{S5}\\
&&  \vec{v}_{\mathrm{i}}^{\, \prime} + {\mathcal H}\,\vec{v}_{\mathrm{i}} =  \frac{e}{m_{\mathrm{p}} \, a} [ \vec{E} + \vec{v}_{\mathrm{i}} \times \vec{B}] - \vec{\nabla} \phi 
+ 
\frac{4}{3} \frac{\rho_{\gamma}}{\rho_{\mathrm{i}}} a 
\Gamma_{\gamma \, \mathrm{i}} (\vec{v}_{\gamma}-\vec{v}_{\mathrm{i}} ) + a \Gamma_{\mathrm{e\,i}} \frac{\rho_{\mathrm{e}}}{\rho_{\mathrm{i}}}( \vec{v}_{\mathrm{e}} - \vec{v}_{\mathrm{i}}),
\label{S6}\\
&& \vec{v}_{\gamma}^{\,\prime} = - \frac{1}{4} \vec{\nabla} \delta_{\gamma} - \vec{\nabla} \phi 
+ a \Gamma_{\gamma\mathrm{i}} (\vec{v}_{\mathrm{i}} - \vec{v}_{\gamma}) + 
a \Gamma_{\gamma\mathrm{e}}  ( \vec{v}_{\mathrm{e}} - \vec{v}_{\gamma}).
\label{S7}
\end{eqnarray}
The relevant interaction rates between the different species appearing in Eqs. (\ref{S5}), (\ref{S6}) and (\ref{S7}) are given by
\begin{eqnarray}
&&\Gamma_{\gamma\mathrm{e}} = \tilde{n}_{\mathrm{e}} 
\sigma_{\mathrm{e}\gamma},\qquad 
\Gamma_{\gamma\mathrm{i}} = \tilde{n}_{\mathrm{i}} 
\sigma_{\mathrm{i}\gamma},\qquad \sigma_{\mathrm{e}\gamma} 
= \frac{8}{3}\pi \biggl(\frac{e^2}{m_{\mathrm{e}}}\biggr)^2, \qquad 
\sigma_{\mathrm{i}\gamma} 
= \frac{8}{3}\pi \biggl(\frac{e^2}{m_{\mathrm{i}}}\biggr)^2,
\label{S8}\\
&& \Gamma_{\mathrm{e\,i}} = \tilde{n}_{\mathrm{e}} \sqrt{\frac{T}{m_{\mathrm{e}}}} \, \sigma_{\mathrm{e\,i}} = \Gamma_{\mathrm{i\, e}},\qquad \sigma_{\mathrm{e\,i}} = 
\frac{e^4}{T^2} \ln{\Lambda_{\mathrm{C}}},\qquad \Lambda_{\mathrm{C}} = \frac{3}{2 e^3} \sqrt{\frac{T^3}{\tilde{n}_{\mathrm{e}}\pi}},
\label{S10}
\end{eqnarray}
where $T$ is the temperature and $\Lambda_{\mathrm{C}}$ is the Coulomb log \cite{spitzer,krall}. 
In Eqs. (\ref{S5}), (\ref{S6}) and (\ref{S7}) $\phi$ denotes one of the two longitudinal 
fluctuations of the geometry \cite{bardeen} whose explicit form is given by \footnote{In Eq. (\ref{S11})
and in what follows, $\delta_{\mathrm{s}}(...)$ denotes the scalar fluctuation of the corresponding quantity.}
\begin{equation}
\delta_{\mathrm{s}}\, g_{00} = 2 a^2 \,\phi,\qquad \delta_{\mathrm{s}} g_{ij} = 2 a^2 \psi \delta_{ij}.
\label{S11}
\end{equation}
The density contrasts of the strongly interacting species evolve, respectively, as 
\begin{eqnarray}
&& \delta_{\mathrm{e}}' = - \vec{\nabla} \cdot \vec{v}_{\mathrm{e}} + 3 \psi' 
- \frac{e}{m_{\mathrm{e}} a}\vec{E} \cdot \vec{v}_{\mathrm{e}},
\label{S12}\\
&& \delta_{\mathrm{i}}' = - \vec{\nabla} \cdot \vec{v}_{\mathrm{i}} + 3 \psi' 
+ \frac{e}{m_{\mathrm{p}} a}\vec{E} \cdot \vec{v}_{\mathrm{i}},
\label{S13}\\
&& \delta_{\gamma}' = 4\psi' - \frac{4}{3} \vec{\nabla} \cdot \vec{v}_{\gamma}.
\label{S14}
\end{eqnarray}
Equations (\ref{S5}), (\ref{S6}) and (\ref{S7}) describe, together with Eqs. (\ref{S12})--(\ref{S14}), a three-fluid system formed by photons, electrons and ions. At early times, i.e. well before photon decoupling, $\Gamma_{\mathrm{e\,i}}  \gg \Gamma_{\gamma \mathrm{e}} \gg H$ and $\Gamma_{\mathrm{\gamma\, e}} \gg \Gamma_{\mathrm{\gamma\, i}}$ (since $m_{\mathrm{p}} \gg m_{\mathrm{e}}$). The 
three fluid system can therefore be described in terms of two effective fluids. The main 
equations of the system, in this regime, are therefore the appropriate 
generalization of the familiar magnetohydrodynamical reduction 
\cite{spitzer,krall,biskamp,boyd} but in the case 
where the relativistic fluctuations of the geometry are consistently included 
in the original equations of the multicomponent plasma. 
By summing and subtracting Eqs. (\ref{S5}) and (\ref{S6}) the following pair 
of equations can be easily obtained, i.e. 
\begin{eqnarray}
&& \vec{v}_{\mathrm{b}}^{\,\prime} + {\mathcal H}  \vec{v}_{\mathrm{b}} = \frac{\vec{J}\times 
\vec{B}}{a^4 \rho_{\mathrm{b}}} - \vec{\nabla}\phi + \frac{4}{3}
 \frac{\rho_{\gamma}}{\rho_{\mathrm{b}}} a \Gamma_{\gamma\, \mathrm{e}} (\vec{v}_{\gamma} - \vec{v}_{\mathrm{b}}),
\label{S15}\\
&& \vec{J} = \sigma (\vec{E} + \vec{v}_{\mathrm{b}} \times \vec{B}),
\label{S16}
\end{eqnarray}
where $\rho_{\mathrm{b}} = (m_{\mathrm{e}} + m_{\mathrm{p}}) \tilde{n}_{0}$.
To derive Eqs. (\ref{S15}) and (\ref{S16}) it should be borne in mind 
that the plasma is globally neutral, i.e. that 
$\tilde{n}_{\mathrm{i}} = \tilde{n}_{\mathrm{e}}$. From the evolution equations of the density contrasts (i.e. Eqs. (\ref{S12}) and (\ref{S13}))
 it  follows that 
\begin{equation}
\delta_{\mathrm{b}}' = - \vec{\nabla}\cdot\vec{v}_{\mathrm{b}} + 3 \psi' + \frac{\vec{J} \cdot \vec{E}}{a^4 \rho_{\mathrm{b}}},
\label{S17}
\end{equation}
where, by definition, 
\begin{equation}
\delta_{\mathrm{b}} = \frac{m_{\mathrm{e}}}{m_{\mathrm{p}} + m_{\mathrm{e}}} \delta_{\mathrm{e}} + \frac{m_{\mathrm{p}}}{m_{\mathrm{p}} + m_{\mathrm{e}}} \delta_{\mathrm{i}}, \qquad \delta_{\mathrm{b}} = \frac{\delta\rho_{\mathrm{b}}}{\rho_{\mathrm{b}}}, \qquad \delta\rho_{\mathrm{b}} = 
\delta \rho_{\mathrm{e}} + \delta\rho_{\mathrm{i}}.
\label{S17def}
\end{equation}
Equations (\ref{S15}), (\ref{S16}) and (\ref{S17}) together with Eqs. (\ref{S7}) and (\ref{S14}) describe the baryon-photon fluid whose 
velocities (see Eqs. (\ref{S7}) and  (\ref{S15})) obey
\begin{eqnarray}
&& \vec{v}_{\mathrm{b}}^{\,\prime} + {\mathcal H}  \vec{v}_{\mathrm{b}} = \frac{\vec{J}\times 
\vec{B}}{a^4 \rho_{\mathrm{b}}} - \vec{\nabla}\phi + \frac{\epsilon'}{R_{\mathrm{b}}}(\vec{v}_{\gamma} - \vec{v}_{\mathrm{b}}),
\label{S17a}\\
&&\vec{v}_{\gamma}^{\,\prime} = - \frac{1}{4} \vec{\nabla} \delta_{\gamma} - \vec{\nabla} \phi 
+  \epsilon'  ( \vec{v}_{\mathrm{e}} - \vec{v}_{\mathrm{b}}),
\label{S17b}
\end{eqnarray}
where $R_{\mathrm{b}}$ is the baryon-to-photon ratio already introduced in Eq. (\ref{Eq3}) and 
where the differential optical depth $\epsilon' =a \Gamma_{\gamma\mathrm{e}} = a \tilde{n}_{\mathrm{0}} x_{\mathrm{e}} \sigma_{\mathrm{e}\gamma}$ has been introduced.

The evolution equations for the photon-baryon system are the basis for the magnetohydrodynamical 
description of the problem and for the analysis of the initial conditions of the Einstein-Boltzmann 
hierarchy \cite{max1,max2,max3,max4}.  The differential optical depth enters directly the visibility function which 
gives the probability that a photon is emitted between $\tau$ and $\tau + d\tau$:
\begin{equation}
{\mathcal K}(\tau) = \epsilon' e^{- \epsilon(\tau, \tau_{0})}, \qquad 
\epsilon(\tau,\tau_{0}) = \int_{\tau}^{\tau_{0}} a(\tau') \tilde{n}_{0}(\tau') \sigma_{\gamma\mathrm{e}} d\tau'.
\label{S17c}
\end{equation}
The visibility function which will be adopted for the analytic 
estimates can be approximated with a double Gaussian whose first peak arises around last scattering (i.e. for $\tau \simeq \tau_{*}$) 
\begin{equation}
{\mathcal K}(\tau) = {\mathcal N}(\sigma_{*}) \,e^{- \frac{(\tau - \tau_{*})^2}{2 \sigma_{*}^2}}, \qquad 0 < \tau < \tau_{x},
\label{LS5}
\end{equation}
where $\tau_{x}$ is an intermediate conformal time such that $\tau_{x} < \tau_{\mathrm{re}}$ where $\tau_{\mathrm{re}}$ is the reionization time.
In Eq. (\ref{LS5}), ${\mathcal N}(\sigma_{*})$ is determined by requiring that
the integral of ${\mathcal K}(\tau)$ over $\tau$ is normalized to $1$. The WMAP data suggest a thickness 
(in redshift space) $\Delta z_{*} \simeq 195 \pm 2$  which would imply 
that $\sigma_{*}$, in units of the (comoving) angular diameter distance to recombination, 
can be estimated as  $\sigma_{*}/\tau_{0} \simeq 1.43 \times 10^{-3}$. 
When $\tau_{0} \gg \tau_{*}$ and 
 $ \tau_{0} \gg \sigma_{*}$ the normalization appearing in Eq. (\ref{LS5}) can be estimated as ${\mathcal N}(\sigma_{*}) \to\sigma_{*}^{-1}\, \sqrt{2/\pi}$. The second  peak 
 occurs for the reionization epoch. Also in this case 
 the visibility function can be approximated with a Gaussian profile centered, 
 this time, around $\tau_{\mathrm{re}}$. The specific form of the 
 profile can be obtained from  Eq. (\ref{LS5}) by replacing 
 ($\tau_{*}$, $\sigma_{*}$) with ($\tau_{\mathrm{rec}}$, $\sigma_{\mathrm{rec}}$) and by taking into account that $z_{\mathrm{re}} = 11\pm 1.4$. The Gaussian (or double Gaussian) parametrization of the visibility has been used in several 
 works (see, e.g. \cite{zeld1,pav1} and also \cite{wyse,zalrec}). 

Prior to decoupling the system can be further simplified. The photon 
and the baryon velocities are quickly  synchronized because of the 
hierarchy between the scattering rate and the Hubble rate. Thus, the evolution 
equations of the photon-baryon system effectively reduce to:
\begin{eqnarray}
&& \delta_{\gamma}' = 4\psi' - \frac{4}{3} \vec{\nabla}\cdot \vec{v}_{\gamma\mathrm{b}},
\label{S18}\\
&& \delta_{\mathrm{b}}' = 3 \psi' - \vec{\nabla}\cdot \vec{v}_{\gamma\mathrm{b}} 
+ \frac{\vec{J} \cdot \vec{E}}{a^4 \rho_{\mathrm{b}}},
\label{S19}\\
&& \vec{v}_{\gamma\mathrm{b}}^{\,\prime} + \frac{{\mathcal H} R_{\mathrm{b}}}{R_{\mathrm{b}} +1} \vec{v}_{\gamma\mathrm{b}}  - \frac{\eta}{\rho_{\gamma} ( 1 + R_{\mathrm{b}})} \nabla^2 \vec{v}_{\gamma
\mathrm{b}} = - \frac{\vec{\nabla} \delta_{\gamma}}{4 ( 1 + R_{\mathrm{b}})} -\vec{\nabla} \phi + \frac{3}{4 a^4 \rho_{\gamma}}\vec{J} \times \vec{B}.
\label{S20}
\end{eqnarray}
In Eq. (\ref{S20}), the shear viscosity term  $\eta = (4/15) \rho_{\gamma} \lambda_{\mathrm{TH}}$ depends upon the photon mean free path $\lambda_{\mathrm{TH}}$ 
which is, in turn, inversely proportional to the differential optical depth. 

\subsection{Weakly interacting constituents}
The effect of the ambient magnetic field on the weakly interacting species 
is mediated by the relativistic fluctuations of the geometry which are 
affected by the scalar modes of the electromagnetic background. 
The evolution of the CDM is given by 
\begin{equation}
\delta_{\mathrm{c}}' = 3 \psi' - \vec{\nabla}\cdot \vec{v}_{\mathrm{c}},
\qquad \vec{v}_{\mathrm{c}}^{\,\prime} + {\mathcal H} \vec{v}_{\mathrm{c}} = - \vec{\nabla} \phi.
\label{Sc15}
\end{equation}
The evolution equation of the neutrinos can be written instead as 
\begin{eqnarray}
&& \delta_{\nu}' = - \frac{4}{3} \vec{\nabla} \cdot \vec{v}_{\nu} + 4 \psi',
\label{Sc16}\\
&& \vec{v}_{\nu}' = \vec{\nabla}\sigma_{\nu} - \frac{1}{4} \vec{\nabla} \delta_{\nu} 
- \vec{\nabla} \phi,
\label{Sc17}\\
&& \sigma_{\nu}' = \frac{4}{15} \vec{\nabla}\cdot\vec{v}_{\nu}.
\label{Sc18}
\end{eqnarray}
In Eqs. (\ref{Sc17}) and (\ref{Sc18}) $\sigma_{\nu}$ is related 
to the neutrino anisotropic stress as $\partial_{i}\partial^{j} \tilde{\Pi}_{j}^{i}= (p_{\nu} + \rho_{\nu})
\nabla^2 \sigma_{\nu}$.  The weakly interacting species are affected by
the action of large-scale magnetic fields through the evolution of the 
fluctuations of the geometry which obey the (perturbed) Einstein 
equations. The $(00)$ and $(0i)$ perturbed components 
of Eq. (\ref{F1}) are, in the gauge (\ref{S11}), 
\begin{eqnarray}
&& \nabla^2 \psi - 3 {\mathcal H} ( {\mathcal H} \phi + \psi') = 4 \pi G a^2 ( \delta_{\mathrm{s}} \rho_{\mathrm{t}} + 
\delta_{\mathrm{s}} \rho_{\mathrm{B}} + \delta_{\mathrm{s}} \rho_{{\mathrm{E}}}) ,
\label{HAM}\\
&& \vec{\nabla} ( {\mathcal H} \phi + \psi') + 4\pi G a^2\biggl[ (p_{\mathrm{t}} + \rho_{\mathrm{t}}) \vec{v}_{\mathrm{t}} + \frac{\vec{E} \times \vec{B}}{4 \pi a^4}\biggr]=0.
\label{MOM}
\end{eqnarray}
The $(ij)$ component of the perturbed Einstein equations can be 
broken, respectively,  into a trace full and a trace less part:
\begin{eqnarray}
&& \psi'' + {\mathcal H} (\phi' + 2 \psi') + ({\mathcal H}^2 + 2 {\mathcal H}') \phi  + \frac{1}{3} \nabla^2 (\phi - \psi) = 4\pi Ga^2 [ 
\delta_{\mathrm{s}} p_{\mathrm{t}} + \delta_{\mathrm{s}} p_{\mathrm{B}}
+ \delta_{\mathrm{s}} p_{\mathrm{E}}],
\label{tr1}\\
&& \partial^{i} \partial^{j}(\phi - \psi) - \frac{\delta^{ij}}{3} \nabla^2(\phi- \psi)= 
8 \pi G a^2 [ \tilde{\Pi}^{ij} + \Pi_{\mathrm{E}}^{ij}  + 
\Pi_{\mathrm{B}}^{ij}],
\label{tr2}
\end{eqnarray}
where  $\delta_{\mathrm{s}} \rho_{\mathrm{t}}$ and $\delta_{\mathrm{s}}
p_{\mathrm{t}}$ are the total fluctuations of the energy density and of the 
pressure while 
\begin{eqnarray}
&& \delta_{\mathrm{s}} \rho_{\mathrm{B}} = \frac{B^2}{8\pi a^4}, \qquad 
 \delta_{\mathrm{s}} \rho_{\mathrm{E}} = \frac{E^2}{8\pi a^4},\qquad  \delta_{\mathrm{s}} p_{\mathrm{B}} = \frac{\delta_{\mathrm{s}} \rho_{\mathrm{B}}}{3}, \qquad  \delta_{\mathrm{s}} p_{\mathrm{E}} = \frac{\delta_{\mathrm{s}} \rho_{\mathrm{E}}}{3},
 \label{tr4}\\
&& \Pi_{\mathrm{E}\,j}^{j} = \frac{1}{4\pi a^4} \biggl[ E_{i} E^{j} - \frac{\delta_{i}^{j}}{3} E^2 \biggr],\qquad 
 \Pi_{\mathrm{B}\,j}^{j} = \frac{1}{4\pi a^4} \biggl[
 B_{i} B^{j} - \frac{\delta_{i}^{j}}{3} B^2 \biggr],
\label{tr6}
\end{eqnarray}
where, as in Eq. (\ref{S4a}), $\vec{E}$ and $\vec{B}$ denote the electromagnetic fields and, by definition,  
$\vec{B} = a^2  \vec{{\mathcal B}}$ and $B^2= B_{i} B^{i}$ while $E^2 = E_{i} E^{i}$. Furthermore, following 
the same notation employed for the neutrino anisotropic stress we shall denote 
\begin{equation}
\partial_{i} \partial_{j} \Pi^{ij}_{\mathrm{E}}= (p_{\gamma} + \rho_{\gamma}) \nabla^2 \sigma_{\mathrm{E}},\qquad 
\partial_{i} \partial_{j} \Pi^{ij}_{\mathrm{B}}= (p_{\gamma} + \rho_{\gamma}) \nabla^2 \sigma_{\mathrm{B}}
\label{tr7}
\end{equation}
\subsection{The magnetized adiabatic mode}
In what follows the attention will be focussed on the 
magnetized adiabatic mode for which all the possible entropic fluctuations vanish. 
This requirement implies that ${\mathcal S}_{\mathrm{ij}} = - 3 (\zeta_{\mathrm{i}} - \zeta_{\mathrm{j}})$ 
where the indices run over all the constituents of the plasma and where $\zeta_{\mathrm{i}} = - \psi + \delta_{\mathrm{i}}/[3 (w_{\mathrm{i}} +  1)]$. The vanishing of the entropy fluctuations vanish at large-scale (i. e. for $k\tau \ll 1$) implies 
that $\zeta_{\mathrm{i}} = \zeta_{\mathrm{j}}$, where, again the indices run over all the species of the plasma.
The latter (gauge-invariant) condition reads off, in the gauge defined by Eq. (\ref{S11}) as 
\begin{equation}
\delta_{\mathrm{i}} = \frac{w_{\mathrm{i}} +1}{w_{\mathrm{j}} + 1} \delta_{\mathrm{j}},
\label{tr8}
\end{equation}
for any pair of constituents of the plasma. 

To set the initial conditions of the Einstein-Boltzmann hierarchy the consistent solution of Eqs. (\ref{S18})--(\ref{S20}), (\ref{Sc15}), (\ref{Sc16})--(\ref{Sc18}), (\ref{HAM})--(\ref{MOM})  and (\ref{tr1})--(\ref{tr2}) should be found at the initial integration 
time $\tau_{\mathrm{initial}}$ when the wavenumbers satisfy $k\tau_{\mathrm{initial}} \ll 1$. As it is well known 
the latter condition implies that the corresponding wavelengths are larger than the Hubble radius:
\begin{equation}
k \tau = \frac{k}{{\mathcal H}} = \frac{k}{a H} = \frac{k_{\mathrm{phys}}}{H} \ll 1;
\label{approx1}
\end{equation}
in Eq. (\ref{approx1}), $k$ is the comoving wavenumber and $k_{\mathrm{phys}}(\tau) = k/a(\tau)$ is the 
physical wavenumber; furthermore ${\mathcal H} = a H$ where $H = \dot{a}/a$ and ${\mathcal H} = a'/a$
(see the comments after Eqs. (\ref{F6}), (\ref{F7}) and (\ref{F8})). Equation (\ref{approx1}) stipulates 
that the physical wavenumbers, at a given time, are always smaller than the Hubble rate implying, by definition,
that the corresponding (physical) wavelengths are larger than the Hubble radius $H^{-1}$. Note that the 
first equality in Eq. (\ref{approx1}) is exact in a  pure radiation-dominated phase when ${\mathcal H} = \tau^{-1}$.
In the realistic situation, however, the scale factor interpolates between the radiation-dominated and the matter-dominated 
epochs and ${\mathcal H} = (2/\tau_{1}) \sqrt{\alpha + 1}/\alpha$ where $\alpha = a/a_{\mathrm{eq}} = [(\tau/\tau_{1})^2 + 
2  (\tau/\tau_{1})]$ (see also the discussion around Eqs. (\ref{TP7g}) and (\ref{TP7d})).

If the condition (\ref{approx1}) holds, then it is also true that $k/\sigma \ll k/{\mathcal H}$ where 
$\sigma = \sigma_{\mathrm{c}} a$ is the conductivity. Indeed recall that 
\begin{equation}
\sigma = \frac{T}{\alpha_{\mathrm{em}}} \biggl(\frac{T}{m_{\mathrm{e}} a}\biggr)^{1/2}, \qquad 
H= \sqrt{\frac{4 \pi^3 g_{\rho}}{45}}\, \frac{T_{\mathrm{phys}}^2}{M_{\mathrm{P}}},
\label{approx2}
\end{equation}
where $T = a T_{\mathrm{phys}}$, $g_{\rho}$ is the effective number of relativistic degrees 
of freedom and $M_{\mathrm{P}} =1/\sqrt{G} \simeq 1.22\times 10^{19}$ GeV. Then we can write, in the case 
of a cold plasma, 
\begin{equation}
\frac{k}{\sigma} = \alpha_{\mathrm{em}} \frac{k_{\mathrm{phys}}}{T_{\mathrm{phys}}} \sqrt{\frac{m_{\mathrm{e}}}{T_{\mathrm{phys}}}} \equiv  \alpha_{\mathrm{em}} \biggl(\frac{k_{\mathrm{phys}}}{H}\biggr) \biggl(\frac{4 \pi^3 g_{\rho}}{45}\biggr)^{1/2} \biggl(\frac{m_{\mathrm{e}}}{T_{\mathrm{phys}}}\biggr)^{1/2} \, \biggl(\frac{T_{\mathrm{phys}}}{M_{\mathrm{P}}}\biggr).
\label{approx3}
\end{equation}
According to Eq. (\ref{approx3}), 
for $T_{\mathrm{phys}} < \mathrm{MeV}$, $k/\sigma \ll 1$ provided  $k_{\mathrm{phys}}/H \ll 1$.
Equation (\ref{approx3})  shows also that that the condition $k/\sigma <1$ holds also for wavelengths which are shorter 
than the Hubble (i.e. $k_{\mathrm{phys}} > H$) since $T_{\mathrm{phys}}/M_{\mathrm{P}}$ is a really minute number 
(of the order of $10^{-28}$ for temperatures in the eV range).

The approximation scheme defined by Eqs. (\ref{approx1}), (\ref{approx2}) and (\ref{approx3}) allows for a 
consistent solution of Eqs. (\ref{S18})--(\ref{S20}), (\ref{Sc15}), (\ref{Sc16})--(\ref{Sc18}), (\ref{HAM})--(\ref{MOM})  and (\ref{tr1})--(\ref{tr2}); 
 the explicit form of the magnetized adiabatic mode can then be written as:
\begin{eqnarray} 
&& \phi(k)= -\frac{10 \,{\mathcal R}_{*}(k)}{4 R_{\nu} + 15} - 2  \frac{R_{\gamma} \{ 4 \sigma_{\mathrm{B}}(k) - R_{\nu} [\Omega_{\mathrm{B}}(k) + \Omega_{\mathrm{E}}(k)]\}}{ 4 R_{\nu} +15},
\nonumber\\
&& \psi(k) = \biggl( 1 + \frac{2}{5} R_{\nu}\biggr) \phi(k)  + \frac{R_{\gamma}}{5} \{ 4 [\sigma_{\mathrm{B}}(k)+  \sigma_{\mathrm{E}}(k)] - R_{\nu} [\Omega_{\mathrm{B}}(k) + \Omega_{\mathrm{E}}(k)]\},
\nonumber\\
&& \delta_{\gamma}(k,\tau) = -2 \phi(k) - R_{\gamma} [\Omega_{\mathrm{B}}(k)  + \Omega_{\mathrm{E}}(k)] ,
\nonumber\\
&& \delta_{\nu}(k) = -2 \phi(k) - R_{\gamma} [\Omega_{\mathrm{B}}(k) + \Omega_{\mathrm{E}}(k)],
\nonumber\\
&& \delta_{\mathrm{c}}(k) = - \frac{3}{2} \phi(k) - \frac{3}{4}R_{\gamma} [\Omega_{\mathrm{B}}(k) + \Omega_{\mathrm{E}}(k)],
\nonumber\\
&& \delta_{\mathrm{b}}(k) = - \frac{3}{2} \phi(k) - \frac{3}{4}R_{\gamma} [\Omega_{\mathrm{B}}(k) + \Omega_{\mathrm{E}}(k)],
\nonumber\\
&& \sigma_{\nu}(k,\tau) = - \frac{R_{\gamma}}{R_{\nu}} [ \sigma_{\mathrm{B}}(k) +\sigma_{\mathrm{E}}(k)]+ \frac{k^2 \tau^2}{6 R_{\nu}} [ \psi(k) - \phi(k)],
\nonumber\\
&& \theta_{\gamma\mathrm{b}}(k,\tau) = \frac{k^2 \tau}{2} \biggl[ \phi(k) + \frac{R_{\nu}}{2}  \Omega_{\mathrm{B}}(k)  - 
\frac{R_{\gamma}}{2} \Omega_{\mathrm{E}}(k) - 2 \sigma_{\mathrm{B}}(k) \biggr],
\nonumber\\
&& \theta_{\nu}(k,\tau)= \frac{k^2 \tau}{2}\biggl[ \phi(k) - \frac{R_{\gamma} \Omega_{\mathrm{B}}(k)}{2} + 2 \frac{R_{\gamma}}{R_{\nu}}( \sigma_{\mathrm{B}}(k) +\sigma_{\mathrm{E}}(k)) \biggr],
\nonumber\\
&& \theta_{\mathrm{c}}(k,\tau) = \frac{k^2 \tau}{2} \phi(k),
\label{L4}
\end{eqnarray}
where ${\mathcal R}_{*}(k)$ is the curvature perturbation on comoving orthogonal hypersurfaces, 
\begin{equation}
{\mathcal R}_{*}(k)=- \psi - \frac{{\mathcal H}({\mathcal H} \phi + \psi')}{{\mathcal H}^2 - {\mathcal H}'} \simeq  - \psi(k) - \frac{\phi(k)}{2}.
\label{L5}
\end{equation}
For notational convenience, in Eq. (\ref{L4}), $R_{\gamma} = 1 - R_{\nu}$ 
denotes the photon fraction in the radiation plasma and $R_{\nu}$ is given, by definition, as 
\begin{equation}
R_{\nu} = \frac{\rho_{\nu}}{\rho_{\gamma} + \rho_{\nu}} = \frac{3 \times (7/8)\times (4/11)^{4/3}}{ 1 + 3 \times (7/8)\times (4/11)^{4/3}} = 0.4052,
\label{L6}
\end{equation}
where $3$ counts the degrees of freedom associated with the 
massless neutrino families, $(7/8)$ arises because neutrinos follow 
the Fermi-Dirac statistics; the factor  $(4/11)^{4/3}$ stems 
from the relative reduction of the neutrino (kinetic) temperature (in comparison 
with the photon temperature) after weak interactions fall out of thermal 
equilibrium. 

In Eq. (\ref{L4}) the following dimensionless quantities have also been 
introduced:
\begin{equation}
\Omega_{\mathrm{E}} = \frac{\delta_{\mathrm{s}} \rho_{\mathrm{E}}}{\rho_{\gamma}}, 
\qquad \Omega_{\mathrm{B}} = \frac{\delta_{\mathrm{s}} \rho_{\mathrm{B}}}{\rho_{\gamma}};
\label{L7}
\end{equation}
see also Eqs. (\ref{tr4}) for a definition of $\delta\rho_{\mathrm{s}} \rho_{\mathrm{E}}$ 
and $\delta_{\mathrm{s}} \rho_{\mathrm{B}}$;  the quantities $\sigma_{\mathrm{E}}$ and $\sigma_{\mathrm{B}}$ have been already introduced in Eq. (\ref{tr7}).
It is finally useful to recall a pair of useful vector identities which connect $\sigma_{\mathrm{E}}$ and $\sigma_{\mathrm{B}}$ 
to $\Omega_{\mathrm{E}}$ and $\Omega_{\mathrm{B}}$, i.e. 
\begin{eqnarray}
&& \nabla^2 \sigma_{\mathrm{E}} = \frac{\nabla^2 \Omega_{\mathrm{E}}}{4} + \frac{3}{16 \pi \rho_{\gamma} a^4} \{ 
\vec{\nabla}\cdot[ (\vec{\nabla} \times \vec{E})\times \vec{E}] - 4\pi \vec{E} \cdot \vec{\nabla} \rho_{\mathrm{q}}\},
\label{L8}\\
&& \nabla^2 \sigma_{\mathrm{B}} = \frac{\nabla^2 \Omega_{\mathrm{B}}}{4} + \frac{R_{\mathrm{b}}}{a^4 \rho_{\mathrm{b}}} \vec{\nabla} \cdot[ \vec{J} \times \vec{B}], 
\label{L9}
\end{eqnarray}
where $\rho_{\mathrm{q}} = e ( n_{\mathrm{i}} - n_{\mathrm{e}})$; $\vec{J} = \vec{\nabla}\times \vec{B}/(4\pi)$ 
is the total current in the one-fluid limit; $R_{\mathrm{b}}$ (see Eq. (\ref{Eq3})) is the baryon-to-photon ratio.

Because of Eqs. (\ref{approx2})--(\ref{approx3}), the contribution of the electric field fluctuations to the initial conditions turns out to be almost always negligible. At the same time it is interesting to consider, in some detail, the transient regime where some putative electric field dies off thanks to the large values of the conductivity. In the latter case Eqs. (\ref{L4}) and (\ref{L5}) allow for the inclusion of the electric field spectra in  the initial conditions of the Einstein-Boltzmann hierarchy. This possibility will not be considered here but will be separately discussed. 

\subsection{Line of sight solution of the Boltzmann hierarchy}
 The temperature and polarization power spectra are, by definition, 
\begin{eqnarray}
&& C_{\ell}^{(\mathrm{TT})}= \frac{1}{2\ell + 1} \sum_{m} \langle 
 a^{(\mathrm{T})*}_{\ell\,m} a^{(\mathrm{T})}_{\ell\,m}\rangle ,
\label{E1}\\
&& C_{\ell}^{(\mathrm{EE})} = \frac{1}{2\ell + 1} \sum_{m} \langle 
a^{(\mathrm{E})*}_{\ell\,m} a^{(\mathrm{E})}_{\ell\,m}\rangle ,
\label{E2}\\
&& C_{\ell}^{(\mathrm{TE})} = \frac{1}{2\ell + 1} \sum_{m} \langle a^{(\mathrm{T})*}_{\ell\,m} a^{(\mathrm{E})}_{\ell\,m}\rangle.
\label{E3}
\end{eqnarray}
In terms of the intensity and polarization fluctuations in real space (i.e. 
$\Delta_{\mathrm{I}}(\hat{n},\tau)$ and $\Delta_{\mathrm{E}}(\hat{n}, \tau)$),
the coefficients $a_{\ell\,m}^{(\mathrm{T})}$ and $a_{\ell\,m}^{(\mathrm{E})}$ 
are: 
\begin{eqnarray}
&& a^{(\mathrm{T})}_{\ell m} 
=  \int d \hat{n}\, Y_{\ell m}^{*}(\hat{n}) \Delta_{\mathrm{I}}(\hat{n}, \tau),
\nonumber\\
&&  a^{(\mathrm{E})}_{\ell m} 
=N_{\ell} \int d \hat{n}\, Y_{\ell m}^{*}(\hat{n}) \Delta_{\mathrm{E}}(\hat{n}, \tau), 
\label{E5}
\end{eqnarray}
where $\hat{n}$ denotes the direction of propagation of the radiation and $N_{\ell} 
= \sqrt{(\ell -2)!/(\ell +2)!}$. 
The (real space) E-mode fluctuation is defined as \cite{Em1,Em2} 
(see also \cite{faraday1,Em3}):
\begin{equation}
\Delta_{\mathrm{E}}(\hat{n},\tau) = - \frac{1}{2} \{ K_{-}^{(1)}(\hat{n})[K_{-}^{(2)}(\hat{n})
\Delta_{+}(\hat{n},\tau)] +  K_{+}^{(-1)}(\hat{n})[K_{+}^{(-2)}(\hat{n}) \Delta_{-}(\hat{n},\tau)]\},
\label{E6}
\end{equation}
where $\Delta_{\pm}(\hat{n},\tau) = \Delta_{\mathrm{Q}}(\hat{n},\tau) \pm i \, \Delta_{\mathrm{U}}(\hat{n},\tau)$  and where $K_{\pm}^{\mathrm{s}}(\hat{n})$ are a pair of differential operators which can either raise or lower 
the spin-weight of a given function:
\begin{equation}
K_{\pm}^{\mathrm{s}}(\hat{n}) 
= - (\sin{\vartheta})^{\pm \mathrm{s}}\biggl[ \partial_{\vartheta} \pm
\frac{i}{\sin{\vartheta}} \partial_{\varphi}\biggr] 
(\sin{\vartheta})^{\mp \mathrm{s}}.
\label{E7}
\end{equation}
The known advantage of dealing directly with the E-mode polarization (rather than with the Stokes parameters) 
is that $\Delta_{\mathrm{E}}(\hat{n},\tau)$ is a scalar (i.e. a function 
of spin-weight $0$ \cite{Em2}) for rotations around the direction of propagation 
of the radiation field. In this sense $\Delta_{\mathrm{E}}(\hat{n},\tau)$ is fully 
analog to $\Delta_{\mathrm{I}}(\hat{n},\tau)$, i.e. the brightness perturbation 
of the intensity of the radiation field which is, of course, a function of 
spin-weight $0$. Defining the projection  of the Fourier 
mode in the direction of the photon momentum as $\mu = \cos{\vartheta}$, the 
evolution of the intensity of the radiation field reads (in Fourier space):
\begin{eqnarray}
&& \Delta_{\mathrm{I}}' + ( i k \mu + \epsilon') \Delta_{\mathrm{I}} = \tilde{S}_{\mathrm{I}}(k,\mu,\tau),
\label{E8}\\ 
&& v_{\mathrm{b}}' + {\mathcal H} v_{\mathrm{b}} = 
\tilde{S}_{v_{\mathrm{b}}}(k,\tau),
\label{E9}
\end{eqnarray}
where
\begin{eqnarray}
&& S_{\mathrm{P}}(k,\tau) 
= \Delta_{\mathrm{I}2} + \Delta_{\mathrm{P}0} + \Delta_{\mathrm{P}2} 
\label{E9a}\\
&& \tilde{S}_{\mathrm{I}}(k,\mu,\tau) = - i k \mu \phi + \psi' + \epsilon' \biggl[ \Delta_{\mathrm{I}0} + \mu v_{\mathrm{b}} - 
\frac{(3\mu^2-1)}{4}S_{\mathrm{P}}(k,\tau) \biggr],
\label{E10}\\
&& \tilde{S}_{v_{\mathrm{b}}}(k,\tau) = - \frac{\epsilon'}{R_{\mathrm{b}}} ( 3 i \Delta_{\mathrm{I}1} + v_{\mathrm{b}}) - 
i k \frac{\Omega_{\mathrm{B}} - 4 \sigma_{\mathrm{B}}}{4 R_{\mathrm{b}}}.
\label{E11}
\end{eqnarray}
Equation (\ref{E9}) is the Fourier space version of Eq. (\ref{S17a}) and $v_{\mathrm{b}}$ 
is the divergence-full part  $\vec{v}_{\mathrm{b}}$.
The Fourier transform of the intensity and of the E-mode polarization 
is defined, within the present conventions, as
\begin{equation}
\Delta_{\mathrm{I}}(\hat{n},\tau) = \frac{1}{(2\pi)^{3/2}} \int d^{3} k \,\Delta_{\mathrm{I}}(k,\mu,\tau), 
\qquad 
\Delta_{\mathrm{E}}(\hat{n},\tau) = \frac{1}{(2\pi)^{3/2}} \int d^{3} k\, \Delta_{\mathrm{E}}(k,\mu,\tau).
\label{E11a}
\end{equation}
From Eqs. (\ref{E6})--(\ref{E7})  the explicit form of the E-mode polarization 
can be written: 
\begin{eqnarray}
\Delta_{\mathrm{E}} (\hat{n},\tau) = - \biggl\{( 1 -\mu^2) \Delta_{\mathrm{Q}}'' - 4 \mu \Delta_{\mathrm{Q}}' - 2 \Delta_{\mathrm{Q}} - 
\frac{\partial_{\varphi}^2 \Delta_{\mathrm{Q}}}{1 - \mu^2 }  - 
2 \biggl[ \partial_{\varphi} \Delta_{\mathrm{U}}' - \frac{\mu}{1 - \mu^2} \partial_{\varphi} \Delta_{\mathrm{U}}\biggr] \biggr\}.
\label{E12}
\end{eqnarray}
In the case of the magnetized adiabatic mode $\Delta_{\mathrm{U}}$ and 
$\Delta_{\mathrm{Q}}$ do not have azimuthal dependence. Furthermore 
a B-mode polarization is only generated thanks to Faraday mixing which has 
been investigated analytically elsewhere \cite{faraday1,faraday2} and will not be repeated here.
The total polarization degree coincides with the contribution of $\Delta_{\mathrm{Q}}$, 
i.e. $\Delta_{\mathrm{Q}}(\hat{n},\tau) = \Delta_{\mathrm{P}}(\hat{n},\tau)$.
It follows from Eq. (\ref{E12}) that, in Fourier space, the E-mode polarization is 
\begin{equation}
\Delta_{\mathrm{E}}(k,\mu,\tau) = - \partial_{\mu}^2 [ ( 1 - \mu^2) \Delta_{\mathrm{P}}(k,\mu,\tau)],
\label{E13}
\end{equation}
where $\Delta_{\mathrm{P}}$ obeys
\begin{equation}
\Delta_{\mathrm{P}}' + ( ik \mu + \epsilon') \Delta_{\mathrm{P}} = \tilde{S}_{\mathrm{P}}(k,\mu,\tau),
\qquad \tilde{S}_{\mathrm{P}}(k,\mu,\tau) = 
\frac{3}{4} ( 1 - \mu^2) S_{\mathrm{P}}(k,\tau).
\label{E14}
\end{equation}
Using line of sight integration the formal solution of Eqs. (\ref{E8}) and (\ref{E14}) 
can be formally written as \cite{LS1,LS2}:
\begin{eqnarray}
&& \Delta_{\mathrm{I}}(k,\mu,\tau_{0}) = \int_{0}^{\tau_{0}} e^{i k \mu (\tau-\tau_{0})} e^{-\epsilon(\tau,\tau_{0})} \, \tilde{S}_{\mathrm{I}}(k,\mu,\tau)\, d\tau,
\label{E15}\\
&& \Delta_{\mathrm{P}}(k,\mu,\tau_{0}) = \int_{0}^{\tau_{0}} e^{i k \mu (\tau-\tau_{0})} e^{-\epsilon(\tau,\tau_{0})} \, \tilde{S}_{\mathrm{P}}(k,\mu,\tau)\, d\tau.
\label{E16}
\end{eqnarray}
The solution expressed by Eqs. (\ref{E15}) and (\ref{E16}) 
 assumes, implicitly, that the source terms can be 
independently evaluated either numerically or analytically. The approximation of tight Coulomb coupling will now be consistently used. The large-scale magnetic fields will then affect electrons and protons whose evolution  can be determined in the appropriate one-fluid limit.   It should be stressed that the approach discussed here is very similar, in spirit, to the various 
semi-analytic techniques which have been
 employed (in the absence of large-scale magnetic fields) by various authors 
\cite{sem2,sem3,sem4} starting with the pioneering work of Peebles and Yu \cite{sem5}.  

\renewcommand{\theequation}{3.\arabic{equation}}
\setcounter{equation}{0}
\section{Temperature and polarization anisotropies}
\label{sec3}
Assuming tight coupling between photons, electrons and baryons, the evolution of the monopole and of the dipole of the brightness perturbations determine the evolution of the source term in the temperature 
and polarization ansiotropies (i.e. Eqs. (\ref{E8})--(\ref{E9}) and (\ref{E14})).
The monopole and the dipole obey, in Fourier space, the following pair of equations
\begin{eqnarray}
&& (\psi - \Delta_{\mathrm{I}0})' = k \Delta_{\mathrm{I}1},
\label{TP1}\\
&&[(R_{\mathrm{b}} + 1) \Delta_{\mathrm{I}1}]'  + 2 \frac{k^2}{k_{\mathrm{D}}^2} (R_{\mathrm{b}} +1)
 \Delta_{\mathrm{I}1}= 
\frac{k}{3} \Delta_{\mathrm{I}0} + \frac{k(R_{\mathrm{b}} +1)}{3}\phi + \frac{k (\Omega_{\mathrm{B}} - 4 \sigma_{\mathrm{B}})}{12},
\label{TP2}
\end{eqnarray}
where $R_{\mathrm{b}}$ has been already introduced in Eq. (\ref{Eq3}) and 
where $k_{\mathrm{D}}$ is the wave-number corresponding to diffusive 
damping, i.e. the wave-number at which diffusive effects start being important. 
To lowest-order in the photon-baryon coupling the diffusive 
damping is simply proportional to the shear viscosity coefficient $\eta$ which has 
been already introduced in Eq. (\ref{S20}). More precisely, to lowest 
order in the tight-coupling, $k_{\mathrm{D}}^{-2} = \eta/[\rho_{\gamma} ( 1 + R_{\mathrm{b}})]$ where $\eta$ has been defined right after Eq. (\ref{S20}) and is 
proportional to the photon mean free path. The estimates based on shear 
viscosity can be improved by going to higher order in the tight-coupling expansion 
and by further refining the estimates depending upon the explicit values 
of the $\Lambda$CDM parameters. In particular, 
for typical values of the parameters close to the best-fit provided 
by the $\Lambda$CDM model the values of $k_{\mathrm{D}}$ and 
$\ell_{\mathrm{D}}$ (i.e. the diffusive multipole) can be estimated as
 \begin{equation}
\ell_{\mathrm{D}} = k_{\mathrm{D}}\, D_{\mathrm{A}}(z_{*})= 
\frac{2240 \, d_{\mathrm{A}}(z_{*})}{\sqrt{\sqrt{r_{\mathrm{R}*} +1} - \sqrt{r_{\mathrm{R}*}}}} 
\biggl(\frac{z_{*}}{10^{3}} \biggr)^{5/4} \, \omega_{\mathrm{b}}^{0.24} \omega_{\mathrm{M}}^{-0.11}.
\label{TP3}
\end{equation}
The (comoving) angular diameter distance at $z_{*}$ 
has been rescaled, in Eq. (\ref{TP3}) as
\begin{equation}
D_{\mathrm{A}}(z_{*}) = \frac{2}{\sqrt{\Omega_{\mathrm{M}0}} H_{0} } d_{\mathrm{A}}(z_{*}).
\label{TP4}
\end{equation}
Furthermore, always in Eq. (\ref{TP3}) $r_{\mathrm{R}*}$ is the ratio of the radiation and matter energy densities at $z_{*}$, i.e. 
\begin{equation}
r_{\mathrm{R}*} = \frac{\rho_{\mathrm{R}}(z_{*})}{\rho_{\mathrm{M}}(z_{*})} = \frac{a_{\mathrm{eq}}}{a_{*}}=
4.15 \times 10^{-2} \, \omega_{\mathrm{M}}^{-1}\, \biggl(\frac{z_{*}}{10^{3}}\biggr).
\label{TP5}
\end{equation}
where, following the customary notation, 
$\omega_{\mathrm{M}} = h_{0}^2\,\Omega_{\mathrm{M}0}$.  The numerical 
content of Eqs. (\ref{TP3})--(\ref{TP5}) is fully specified in terms of $z_{*}$ whose 
explicit form can be written as 
\begin{eqnarray}
z_{*} &=& 1048[ 1 + (1.24 \times 10^{-3})\, \omega_{\mathrm{b}}^{- 0.738}] [ 1 + g_{1} \omega_{\mathrm{M}}^{\,\,\,g_2}],
\label{TP6}\\
g_{1} &=& \frac{0.0783 \, \omega_{\mathrm{b}}^{-0.238}}{[1 + 39.5 \,\,
\omega_{\mathrm{b}}^{\,\,0.763}]},\qquad 
g_{2} = \frac{0.560}{1 + 21.1 \, \omega_{\mathrm{b}}^{\,\,1.81}}.
\label{TP7}
\end{eqnarray}
Equations (\ref{TP6})--(\ref{TP7}) imply 
$z_{*} =1090.5$ in excellent agreement with the estimate of  
\cite{WMAP5a,WMAP5b,WMAP5c}, i.e. $z_{*} = 1090.51 \pm 0.95$.
The evolution of the monopole and of the dipole can be determined from the 
WKB solution of Eqs. (\ref{TP1}) and (\ref{TP2}), i.e. 
\begin{eqnarray}
&& \Delta_{\mathrm{I}0}(k,\tau) + \phi(k,\tau) = {\mathcal L}(k,\tau) + \sqrt{c_{\mathrm{sb}}} {\mathcal M}(k,\tau) 
\cos{[k r_{\mathrm{s}}(\tau)]} \, e^{- \frac{k^2}{k_{\mathrm{D}}^2}},
\label{TP7a}\\
&& \Delta_{\mathrm{I}1}(k,\tau) = c_{\mathrm{sb}}^{3/2} {\mathcal M}(k,\tau) 
\sin{[k\, r_{\mathrm{s}}(\tau)]} e^{- \frac{k^2}{k_{\mathrm{D}}^2}},
\label{TP7b}
\end{eqnarray}
where ${\mathcal L}(k,\tau)$ and ${\mathcal M}(k,\tau)$ are fixed once 
the initial conditions of the Einstein-Boltzmann hierarchy are specified. In what 
follows, as already mentioned, the initial conditions shall correspond 
to the magnetized adiabatic mode. In Eqs. (\ref{TP7a}) and (\ref{TP7b}) 
$r_{\mathrm{s}}(\tau)$, is the sound horizon
 \begin{equation}
 r_{\mathrm{s}}(\tau_{*}) = \int_{0}^{\tau_{*}} d\tau \, c_{\mathrm{s}\mathrm{b}}(\tau) = 
 \int_{0}^{\tau_{*}} \,\frac{d\tau}{\sqrt{3 [ R_{\mathrm{b}}(\tau) +1]}},
\label{TP7c}
\end{equation}
whose explicit form will be determined as a function of $z_{*}$. 
The explicit solution of Eqs. (\ref{F6}), (\ref{F7}) and (\ref{F8}) for the 
matter radiation transition implies that $a(x) = a_{\mathrm{eq}}[x^2 + 2 x]$ 
with  $x= \tau/\tau_{1}$. This also means that:
\begin{equation}
 \biggl(\frac{\tau_{*}}{\tau_{1}} + 1\biggr) = \sqrt{1 + \frac{a_{*}}{a_{\mathrm{eq}}}}
=  \sqrt{\frac{ 1 + r_{\mathrm{R}*}}{r_{\mathrm{R}*} }}.
\label{TP7g}
\end{equation}
where $\tau_{1} = 2 \sqrt{(a_{\mathrm{eq}}/\Omega_{\mathrm{M}0})} /H_{0}$. By definition of baryon-to-photons ratio 
(see, e.g., Eq. (\ref{Eq3})) we have that 
$R_{\mathrm{b}}(x) = R_{\mathrm{b}*} r_{\mathrm{R}*} ( x^2 + 2 x)$. Thus, defining Eq.  $y = x+1$,  Eq. (\ref{TP7c}) becomes easily
\begin{eqnarray}
r_{\mathrm{s}}(\tau_{*}) &=& \frac{\tau_{1}}{\sqrt{3}} \int_{0}^{\tau_{*}/\tau_{1}}
\frac{d x}{\sqrt{R_{\mathrm{b}*} r_{\mathrm{R}*} (x^2 + 2 x) +1}} 
\nonumber\\
&=&\frac{\tau_{1}}{\sqrt{3}\,R_{\mathrm{b}*} r_{\mathrm{R}*}} \int_{1}^{(\tau_{*}/\tau_{1}) +1} \frac{d y}{\sqrt{y^2 + y_{0}^2}},\qquad y_{0} = 
\sqrt{\frac{1 - R_{\mathrm{b}*} r_{\mathrm{R}*}}{  R_{\mathrm{b}*} r_{\mathrm{R}*}}},
\label{TP7d}
\end{eqnarray}
which can be integrated via a further change of variables (i.e. 
 $y = y_{0} \sinh{w}$); the result is:
 \begin{equation}
r_{\mathrm{s}}(\tau_{*}) = \frac{\tau_{1}}{\sqrt{3}\,R_{\mathrm{b}*} r_{\mathrm{R}*}} \biggl\{ \mathrm{arcsinh}\biggl[ 
\frac{(\tau_{*}/\tau_{1}) +1}{y_{0}}\biggr] - \biggl[ \frac{1}{y_{0}}\biggr]\biggr\}.
\label{TP7e}
\end{equation}
Since, by definition, $ \mathrm{arcsinh}(\alpha) = \ln{[\alpha + \sqrt{\alpha^2 +1}]}$ and $\tau_{1}$ is given after Eq. (\ref{TP7g})
the sound horizon at $\tau_{*}$ is given by:
\begin{equation}
r_{\mathrm{s}}(\tau_{*}) = \frac{2}{H_{0}} \frac{1}{\sqrt{\Omega_{\mathrm{M}0}}} 
\frac{1}{\sqrt{3 R_{\mathrm{b}*} ( z_{*} +1)}} \ln{\biggl[ \frac{\sqrt{1 +  R_{\mathrm{b}*}} + 
\sqrt{R_{\mathrm{b}*}}\sqrt{1 + r_{\mathrm{R}*}}}{1 + \sqrt{r_{\mathrm{R}*}\,R_{\mathrm{b}*}}}\biggr]}.
\label{TP7h}
\end{equation}
Having determined the monopole and the dipole by solving Eqs. (\ref{TP1}) and 
(\ref{TP2}) the polarization observables depend chiefly upon the value of the dipole as it 
arises to lowest order in the tight-coupling expansion. However, as it was 
already observed long ago \cite{LS1}, the first-order tight-coupling estimate 
is not satisfactory from the numerical point of view and must be 
improved. Following this logic, Eqs. (\ref{E8}) and (\ref{E14}) imply 
the following relations between the multipoles of the intensity and polarization 
brightness perturbations:
\begin{eqnarray}
 && \Delta_{{\rm P}0} ' - \frac{\epsilon'}{2} [ \Delta_{{\rm P}2} + \Delta_{{\rm I}2} - \Delta_{{\rm P}0} ]= - k \Delta_{{\rm P}1},
 \label{TP8}\\
 && \Delta_{{\rm I} 2}'  + \epsilon'\biggl[ \frac{9}{10} \Delta_{{\rm I}2} - \frac{1}{10} (\Delta_{{\rm P}0} + 
\Delta_{{\rm P} 2} )\biggr] = - \frac{3}{5} k \Delta_{{\rm I}3} + \frac{2}{5} k \Delta_{{\rm I} 1},
\label{TP9}\\
&&  \Delta_{{\rm P} 2}' +  \epsilon'\biggl[ \frac{9}{10} \Delta_{{\rm P}2} - \frac{1}{10} (\Delta_{{\rm P}0} + 
\Delta_{{\rm I} 2} )\biggr] = - \frac{3}{5} k \Delta_{{\rm P}3} + \frac{2}{5} k \Delta_{{\rm P} 1}.
\label{TP10}
\end{eqnarray}
Summing up Eqs. (\ref{TP8}), (\ref{TP9}) and (\ref{TP10}) and recalling that, by definition,  $S_{\mathrm{P}}= 
(\Delta_{\mathrm{I}2} + \Delta_{\mathrm{P}0} + \Delta_{\mathrm{P}2})$, Eqs. (\ref{TP8})--(\ref{TP10}) imply
 \begin{equation}
 S_{\rm P}' + \frac{3}{10} \epsilon' S_{\rm P} = k \biggl[ \frac{2}{5} \Delta_{{\rm I}1} - 
 \frac{3}{5}\biggl( \Delta_{{\rm P}1}+ \Delta_{{\rm P}3} + \Delta_{{\rm I}3}\biggr)\biggr].
\label{TP11}
 \end{equation}
The result of the solution of Eq. (\ref{TP11}) turns out to be more accurate than the lowest order tight-coupling result. Indeed, neglecting $ \Delta_{{\rm P}1}$, 
$\Delta_{{\rm P}3}$ and  $\Delta_{{\rm I}3}$ (which are all smaller than 
$\Delta_{{\rm I}1}$) Eq. (\ref{TP11}) can be  formally integrated:
\begin{equation}
S_{\rm P}(k,\tau) =\frac{2}{5} k e^{ 3 \epsilon(\tau,\tau_0)/10} \int_{0}^{\tau}  d \tau' \overline{\Delta}_{{\rm I}1}(k,\tau') e^{-3 \epsilon(\tau',\tau_{0})/10},
\label{TP12}
\end{equation}
which also implies that 
\begin{equation}
\Delta_{\rm P}(k,\mu,\tau_{0}) = - 0.515\, (k \,\,\sigma_{*}) ( 1 - \mu^2) e^{i k\mu(\tau_{*} -\tau_{0})} \overline{\Delta}_{{\rm I}1}(k,\tau_{*}).
\label{TP13}
\end{equation}
The coefficients $a_{\ell m}^{(\mathrm{T})}$ and $a_{\ell m}^{(\mathrm{E})}$ can be determined in terms of Eqs. (\ref{TP7a})--(\ref{TP7b}) and (\ref{TP13}) following the standard techniques. More specifically the coefficient $a_{\ell m}^{(\mathrm{T})}$ can be expressed as:
\begin{equation}
a^{(\mathrm{T})}_{\ell m} = \frac{\sqrt{4 \pi}}{(2\pi)^{3/2}} \, (-i)^{\ell} \, \sqrt{2 \ell + 1} 
\int d^3 k e^{- \frac{k^2}{k_{\mathrm{t}}^2}} \biggl[ 
(\Delta_{\mathrm{I}0} + \phi) j_{\ell}(x) + 3 \Delta_{\mathrm{I}1} \biggl(\frac{d j_{\ell}}{dx}\biggr)\biggr],
\label{TP14}
\end{equation}
where $ x = k(\tau_{0} - \tau_{*})$ and  where $j_{\ell}(x)$ are the spherical Bessel 
functions \cite{abr1,abr2} of argument $x$. In Eq. (\ref{TP14})  $k_{\mathrm{t}} = \sqrt{3}/\sigma_{*}$ arises from the integration over $\tau$ of the Gaussian visibility function.
The  coefficient  $a_{\ell m}^{(\mathrm{E})}$ turns out to be:
\begin{equation}
a_{\ell m}^{\mathrm{E}}= \frac{3}{4} \frac{(-i)^{\ell}}{(2\pi)^{3/2}} 
\sqrt{\frac{(\ell -2)!}{(\ell + 2)!}} \sqrt{4\pi} \sqrt{2\ell + 1} 
\int d^{3} k\,\, x^2 \,\, [ ( 1 + \partial_{x}^2)^2] j_{\ell}(x) \int_{0}^{\tau_{0}}
{\mathcal K}(\tau) S_{\mathrm{P}}(k,\tau) d\tau.
\label{TP15}
\end{equation}
In Eqs. (\ref{TP14}) and (\ref{TP15}) the following two results have been 
repeatedly  used:
\begin{eqnarray}
&& \int d\hat{n} \,Y_{\ell\, m}^{*}(\hat{n}) \, e^{- i \mu\, x} = 
\sqrt{4\pi} \, (-i)^{\ell} \, \sqrt{2 \ell+1} j_{\ell}(x),
\label{TP16}\\
&& \partial_{\mu}^2[ ( 1 -\mu^2)^2\, e^{- i \mu x}] = \partial_{\mu}^2[( 1 + \partial_{x}^2)^2 
e^{-i \mu x}] = - ( 1 + \partial_{x}^2) \,x^2 \,e^{- i \mu x}.
\label{TP17}
\end{eqnarray}
Furthermore, in Eq. (\ref{TP15}), the equation of the spherical Bessel functions \cite{abr1,abr2} has been repeatedly used.
Notice that, in Eq. (\ref{TP15}) the integral over $\tau$ of the visibility function can be simplified by using, for $S_{\mathrm{P}}(k,\mu,\tau)$, the expression of Eq. (\ref{TP12}) and by performing exactly the same integral leading to Eq. (\ref{TP13}). 

In the present discussion we are interested in the scaling properties of the correlation functions over 
relatively small scales where simplifying expressions for the Bessel functions can be used. In this limit, the reionization effects can be parametrized as follows. In Eqs. (\ref{E15}) and (\ref{E16})  the integral over $\tau$ 
can be separated in two distinct contributions. For sake of concreteness consider the polarization integral 
which gives 
\begin{equation}
\Delta_{\mathrm{P}}(k,\mu,\tau_{0}) = \int_{0}^{\tau_{\mathrm{re}}} d\tau {\mathcal K}(\tau) e^{- i \mu x} \tilde{S}_{\mathrm{P}}(k,\mu,\tau) + \int_{\tau_{\mathrm{re}}}^{\tau_{0}} d\tau {\mathcal K}(\tau) e^{- i \mu x} \tilde{S}_{\mathrm{P}}(k,\mu,\tau).
\label{TP18}
\end{equation}
The first term at the right hand side  Eq. (\ref{TP18})  is the most relevant for $\ell \gg 20$ and it 
is given by $e^{- \epsilon_{\mathrm{re}}} \overline{\Delta}_{\mathrm{P}}(k,\mu,\tau_{0})$ where 
$\overline{\Delta}_{\mathrm{P}}(k,\mu,\tau_{0})$ is the value of the polarization in the 
absence of reionization. For small $\ell$ the second term in Eq. (\ref{TP18}) is the most relevant \cite{zalrec}
and it leads to supplementary peaks in the angular power spectra (i.e. the so-called reionization peaks). 
Within the approximations of this section, the integrand of the second term in Eq. (\ref{TP18}) is simply 
proportional to the quadrupole of the intensity which can be evaluated, for $k < k_{\mathrm{D}}$, as \cite{zalrec}  
\begin{equation}
\Delta_{\mathrm{I}2}(k,\mu,\tau_{\mathrm{re}}) = \{{\mathcal L}(k,\tau_{*}) + \sqrt{c_{\mathrm{sb}}} {\mathcal M}(k,\tau_{*}) \cos{[k r_{\mathrm{s}}(\tau_{*})]}\} j_{2}(x_{\mathrm{re}})
\label{TP19}
\end{equation}
where $j_{2}(x_{\mathrm{re}})$ is the spherical Bessel function for $\ell =2$ and where $x_{\mathrm{re}} = [k(\tau_{\mathrm{re}} - \tau_{*})]$. The reionization peaks arise, roughly, at the first peak of $j_{2}(x_{\mathrm{re}})$, i.e. 
for $x_{\mathrm{re}} \simeq 2$. 
\renewcommand{\theequation}{4.\arabic{equation}}
\setcounter{equation}{0}
\section{The basic integrals}
\label{sec4}
\subsection{Generalities}
The considerations of the previous section depend upon two sorts of scales i.e. 
\begin{itemize}
\item{} damping scales, (e. g. $\ell_{\mathrm{D}}$, $\ell_{\mathrm{t}}$, $\epsilon_{\mathrm{re}}$...)
which control the falloff of the temperature 
and polarization angular power spectra;
\item{} oscillatory scales (e.g. $\ell_{\mathrm{A}}$)  which control the structure of the  peak and depths in the TT, EE and TE correlations.  
\end{itemize}
The thermal diffusivity multipole $\ell_{\mathrm{D}}$, already introduced in Eq. (\ref{TP3}),  can be estimated 
using the best fit to the WMAP 5yr data alone \cite{WMAP5a,WMAP5b,WMAP5c}; 
Eq. (\ref{TP3}) leads to $\ell_{\mathrm{D}} =  1422.08$. The finite thickness of the visibility function leads to an effective multipole which can be estimated as $\ell_{\mathrm{t}} = \sqrt{3}/(k_{0} \sigma_{*})$. Again 
using the estimated thickness in the visibility function the WMAP 5yr data 
allow to estimate $\ell_{\mathrm{t}} = 1211.22$. The typical scales 
$\ell_{\mathrm{D}}$ and $\ell_{\mathrm{t}}$ can be combined 
in what is often called Silk damping scale, i.e. 
\begin{equation}
\frac{1}{\ell_{\mathrm{S}}^2} = \frac{1}{\ell_{\mathrm{t}}^2} + \frac{1}{\ell_{\mathrm{D}}^2},\qquad \ell_{\mathrm{S}} = \sqrt{\frac{\ell_{\mathrm{t}}^2 \, \ell_{\mathrm{D}}^2}{\ell_{\mathrm{t}}^2 + \ell_{\mathrm{D}}^2}}.
\label{B1}
\end{equation}
In the case of the numerical values listed above $\ell_{\mathrm{S}} = 922.09$.
The oscillatory patterns in the angular power spectra are determined by the acoustic 
multipole, i.e. 
\begin{eqnarray}
\ell_{\mathrm{A}} &=& \frac{\pi D_{\mathrm{A}}(z_{*})}{r_{\mathrm{s}}(z_{*})}= 
 \frac{2 \pi \, d_{\mathrm{A}}(z_{*})}{ H_{0} \sqrt{\Omega_{\mathrm{M}0}} \, r_{\mathrm{s}}(z_{*})} \equiv \frac{ \sqrt{3R_{\mathrm{b}*}}\, \pi \, \sqrt{z_{*} +1} \, d_{\mathrm{A}}(z_{*})}{ \ln{\biggl[
\frac{\sqrt{1 + R_{\mathrm{b}*}} + \sqrt{( 1 +  r_{\mathrm{R}*})R_{\mathrm{b}*}}}{1 + 
\sqrt{r_{\mathrm{R}*} R_{\mathrm{b}*}}}\biggr]}},
\nonumber\\
&\equiv& \biggl(\frac{z_{*}}{10^{3}}\biggr)^{1/2} \frac{\sqrt{R_{\mathrm{b}*}}\,d_{\mathrm{A}}(z_{*})}{\ln{\biggl[
\frac{\sqrt{1 + R_{\mathrm{b}*}} + \sqrt{( 1 +  r_{\mathrm{R}*})R_{\mathrm{b}*}}}{1 + 
\sqrt{r_{\mathrm{R}*} R_{\mathrm{b}*}}}\biggr]}}.
\label{B2}
\end{eqnarray}
The first equality of Eq. (\ref{B2}) is just the definition of the acoustic multipole 
while the second equality uses a more explicit form of the (comoving) 
angular diameter distance. Note that, in Eq. (\ref{B2}), the (reduced) angular diameter distance 
$d_{\mathrm{A}}(z_{*})$ goes asymptotically to $0.89$ for $z_{*}>500$ and for the standard 
values of the cosmological parameters.  In the explicit expressions of some integrand 
it will prove useful to have an explicit expression also for $\gamma_{\mathrm{A}} = \pi/\ell_{\mathrm{A}}$:
indeed, in various oscillating factors, the combination $\gamma_{\mathrm{A}}\ell$ arises naturally.
Thus Eq. (\ref{B2}) also implies  
\begin{equation}
\gamma_{\mathrm{A}} = \frac{\pi}{\ell_{\mathrm{A}}} = \frac{1}{d_{\mathrm{A}}(z_{*})\sqrt{ 3 R_{\mathrm{b}*} (z_{*} +1)}} \ln{\biggl[
\frac{\sqrt{1 + R_{\mathrm{b}*}} + \sqrt{( 1 +  r_{\mathrm{R}*})R_{\mathrm{b}*}}}{1 + 
\sqrt{r_{\mathrm{R}*} R_{\mathrm{b}*}}}\biggr]}.
\label{B3}
\end{equation}
 According to Eq. (\ref{B2}),
 $\ell_{\mathrm{A}} =  301.57$ while the WMAP 5yr data imply that 
 the acoustic scale at decoupling is given by $\ell_{\mathrm{A}} = 302.08^{0.83}_{-0.84}$.  
 The acoustic multipole can be compared with the equality multipole, i.e. 
 \begin{equation}
\ell_{\mathrm{eq}} = \sqrt{2 \Omega_{\mathrm{M}0}} H_{0} \sqrt{z_{\mathrm{eq}} +1} D_{\mathrm{A}}(z_{\mathrm{eq}}) = 2 \sqrt{2} \sqrt{z_{\mathrm{eq}}} \sqrt{\omega_{\mathrm{M}}} = 2\sqrt{2} \sqrt{\frac{z_{*}}{r_{\mathrm{R}*}}}
d_{\mathrm{A}}(z_{*}), 
\label{sc12}
\end{equation}
where we recalled that  ${\mathcal H}_{\mathrm{eq}} = a_{\mathrm{eq}}\, H_{\mathrm{eq}} =
\sqrt{2 \Omega_{\mathrm{M}0}}  (a_{0}/a_{\mathrm{eq}})^{1/2}$,
Combining the last equality of Eq. (\ref{sc12}) with Eq. (\ref{TP5}) we get 
$\ell_{\mathrm{eq}} = 439.057 \,\sqrt{\omega_{\mathrm{M}}}\, d_{\mathrm{A}}(z_{*})$ which also 
implies, in the case of the WMAP 5yr parameters, $\ell_{\mathrm{eq}} \simeq 136.95$.
In terms of $\ell_{\mathrm{A}}$ the position of the first three Doppler peaks can be obtained 
approximately obtained from \cite{norm4} $\ell_{m} = \ell_{\mathrm{A}} ( m - \varphi_{m})$ where
\begin{eqnarray}
&& \ell_{1} = \ell_{\mathrm{A}}( 1 - \varphi_{1}), \qquad 
\varphi_{1} = 0.267 \, \biggl(\frac{r_{\mathrm{R}*}}{0.3} \biggr)^{0.1},
\label{sc8}\\
&& \ell_{2} = \ell_{\mathrm{A}}(2  - \varphi_{2}), \qquad 
\varphi_{2} = 0.24 \, \biggl(\frac{r_{\mathrm{R}*}}{0.3} \biggr)^{0.1},
\label{sc9}\\
&&  \ell_{3} = \ell_{\mathrm{A}}(3 - \varphi_{3}), \qquad 
\varphi_{3} = 0.35 \, \biggl(\frac{r_{\mathrm{R}*}}{0.3} \biggr)^{0.1}.
\label{sc10}
\end{eqnarray}
The values of $\ell_{1}$, $\ell_{2}$ and $\ell_{3}$ are deduced in the case $n_{\mathrm{s}} =1$. When 
$n_{\mathrm{s}}\neq 1$ the positions are shifted as $\ell_{m} \to \ell_{m} + \Delta \ell_{m}$
\begin{equation}
\Delta \ell_{1} = 0.13 \, |n_{\mathrm{s}} -1| \ell_{1}, \qquad \Delta \ell_{2} = 0.33 \, |n_{\mathrm{s}} -1|\ell_{2},\qquad \Delta \ell_{3} = 0.61 \, |n_{\mathrm{s}} -1| \ell_{3}.
\label{sc10a}
\end{equation}
In the vanilla $\Lambda$CDM and for the WMAP 5yr 
best fit we have that 
\begin{equation}
\ell_{1} = 219,\qquad \ell_{2} =535, \qquad \ell_{3} = 814.
\label{sc10b}
\end{equation}
which is approximately what could be obtained from Eqs. (\ref{sc10})--(\ref{sc10a})

\subsection{The angular power spectra}

Using Eqs. (\ref{TP14}) and (\ref{TP15}) into Eqs. (\ref{E1}), (\ref{E2}) and (\ref{E3}) 
a more explicit expression of the angular power spectra can be obtained: 
\begin{eqnarray}
&& C_{\ell}^{(\mathrm{TT})} = 4\pi \int \frac{d k}{k} \frac{k^3}{2\pi^2} |\Delta^{(\mathrm{TT})}_{\ell}(k,\tau_{0})|^2,
\label{GEN1}\\
&& C_{\ell}^{(\mathrm{EE})} = 4\pi \int \frac{d k}{k} \frac{k^3}{2\pi^2} |\Delta^{(\mathrm{EE})}_{\ell}(k,\tau_{0})|^2,
\label{GEN2}\\
&& C_{\ell}^{(\mathrm{TE})} = 4\pi \int \frac{d k}{k} \frac{k^3}{2\pi^2} |\Delta^{(\mathrm{TE})}_{\ell}(k,\tau_{0})|^2,
\label{GEN3}
\end{eqnarray}
where the following quantities have been introduced:
\begin{eqnarray}
&&  |\Delta^{(\mathrm{TT})}_{\ell}(k,\tau_{0})|^2 = \biggl\{ |\Delta_{\mathrm{I}0} +\psi|^2 + 9 |\Delta_{\mathrm{I}1}|^2 \biggl[1 - \frac{\ell (\ell+1)}{x^2}\biggr]\biggr\} j^2_{\ell}(x)\,\,e^{- 2 \frac{k^2}{k_{\mathrm{t}}^2}},
\label{GEN4}\\
&& |\Delta^{(\mathrm{EE})}_{\ell}(k,\tau_{0})|^2= 0.265 \, (k \sigma_{*})^2 |\Delta_{\mathrm{I}1}|^2 
\ell (\ell -1) (\ell +1) (\ell +2) \frac{j^2_{\ell}(x)}{x^4},
\label{GEN5}\\
&& |\Delta^{(\mathrm{TE})}_{\ell}(k,\tau_{0})|^2= 0.515 \sqrt{\ell (\ell -1) (\ell +1) (\ell +2)} (k \sigma_{*}) \Delta_{\mathrm{I}1} 
(\Delta_{\mathrm{I}0} + \psi) \frac{j^2_{\ell}(x)}{x^2} \,\,e^{-  \frac{k^2}{k_{\mathrm{t}}^2}}.
\label{GEN6}
\end{eqnarray}
It is practical to adopt the following general parametrization for the three relevant power spectra 
of the problem
\begin{equation}
{\mathcal P}_{\mathcal R}(k) = {\mathcal A}_{{\mathcal R}} \biggl(\frac{k}{k_{\mathrm{p}}}\biggr)^{n_{\mathrm{s}}-1},\qquad {\mathcal P}_{\Omega}(k) = {\mathcal E}_{\mathrm{B}} 
\biggl(\frac{k}{k_{\mathrm{L}}}\biggr)^{2(n_{\mathrm{B}}-1)},\qquad {\mathcal P}_{\sigma}(k) = 
r_{\mathrm{B}}{\mathcal P}_{\Omega}(k),
\label{GEN8}
\end{equation}
where ${\mathcal A}_{{\mathcal R}}$ denotes  the amplitude of the curvature perturbations at the pivot scale 
$k_{\mathrm{p}}$;  
 ${\mathcal E}_{\mathrm{B}}$ denotes the amplitude of the power spectrum of $\Omega_{\mathrm{B}}$ (see also 
 \cite{max3,max4,faraday1}; $r_{\mathrm{B}}$ denotes the ratio\footnote{It is often practical to assign ratios of power spectra at the same pivot scale; this is what happens also when assigning tensor power spectra in standard CMB studies.}  between the power spectrum of $\sigma_{\mathrm{B}}$ and the 
 power spectrum of $\Omega_{\mathrm{B}}$ at the same magnetic pivot scale $k_{\mathrm{L}}$.
To leading order ${\mathcal E}_{\mathrm{B}}$ and $r_{\mathrm{B}}$ are independent 
upon the wave-number. There are however corrections which imply that ${\mathcal E}_{\mathrm{B}}$ and $r_{\mathrm{B}}$ do depend upon the wave-number. If ${\mathcal E}(k)$ and $r_{\mathrm{B}}(k)$ the form of the integrals listed below as well as the 
related discussion slightly changes but the explicit results are more 
cumbersome and will not be reported here.  

Using Eqs.  (\ref{TP7a}) and (\ref{TP7b}) into Eqs. (\ref{GEN4}), 
(\ref{GEN5}) and (\ref{GEN6}) the explicit form of the temperature and polarization 
observables can be derived.  Since some of the subsequent expressions are rather lengthy, the following rescaled amplitudes 
will be defined:
\begin{eqnarray}
&&{\mathcal Q}_{{\mathcal R}{\mathcal R}}=  {\mathcal A}_{{\mathcal R}} \biggl(\frac{k_{0}}{k_{\mathrm{p}}}\biggr)^{n_{\mathrm{s}}-1}\, e^{- 2\epsilon_{\mathrm{re}}}, \qquad 
{\mathcal Q}_{\mathrm{B}\mathrm{B}} = {\mathcal E}_{{\mathrm{B}}} \biggl(\frac{k_{0}}{k_{\mathrm{L}}}\biggr)^{2(n_{\mathrm{B}}-1)}
\,e^{- 2\epsilon_{\mathrm{re}}},
\nonumber\\
&& {\mathcal Q}_{{\mathcal R}{\mathrm{B}}}= \sqrt{{\mathcal A}_{\mathcal R}} \sqrt{{\mathcal E}_{\mathrm{B}}} \biggl(\frac{k_{0}}{k_{\mathrm{p}}}\biggr)^{\frac{n_{\mathrm{s}}-1}{2}} \biggl(\frac{k_{0}}{k_{\mathrm{L}}}\biggr)^{(n_{\mathrm{B}}-1)}\,e^{- 2\epsilon_{\mathrm{re}}}, 
\label{GEN8a}
\end{eqnarray}
where $k_{\mathrm{p}} =0.002 \, \mathrm{Mpc}^{-1}$ is the pivot scale of curvature perturbations and, as already mentioned in section \ref{sec2},  $k_{\mathrm{L}} = 1\, \mathrm{Mpc}^{-1}$ is the magnetic pivot scale. 

\subsection{Temperature autocorrelations}
The temperature autocorrelations are hereby written in terms of four basic 
integrals, i.e.  
\begin{equation}
 G_{\ell}^{(\mathrm{TT})} = 
{\mathcal I}_{(1)}^{(\mathrm{TT})}(\ell, \ell_{\mathrm{t}})   + {\mathcal I}_{(2)}^{(\mathrm{TT})}(\ell, \ell_{\mathrm{S}}) + {\mathcal I}_{(3)}^{(\mathrm{TT})}(\ell, \ell_{\mathrm{S}}) + {\mathcal I}_{(4)}^{(\mathrm{TT})}(\ell, \ell_{\mathrm{t}}, \ell_{\mathrm{S}}).
\label{GEN27}
\end{equation}
Each of the terms appearing in Eq. (\ref{GEN27}) contains 
three contributions proportional, respectively, to ${\mathcal Q}_{{\mathcal R}{\mathcal R}}$, ${\mathcal Q}_{\mathrm{B}\mathrm{B}}$ and to 
${\mathcal Q}_{{\mathcal R}{\mathrm{B}}}$ whose explicit form can be written as:
\begin{eqnarray}
&& {\mathcal I}_{(1)}^{(\mathrm{TT})}(\ell,\ell_{\mathrm{t}}) 
={\mathcal V}^{(1)}_{{\mathcal R}{\mathcal R}}(\ell,\ell_{\mathrm{t}}) 
+ {\mathcal V}_{\mathrm{BB}}^{(1)}(\ell,\ell_{\mathrm{t}})  +2 \cos{\beta} {\mathcal V}^{(1)}_{{\mathcal R}\mathrm{B}}(\ell,\ell_{\mathrm{t}}),
\label{PR1}\\
&&  {\mathcal I}_{(2)}^{(\mathrm{TT})}(\ell,\ell_{\mathrm{S}}) 
={\mathcal V}^{(2)}_{{\mathcal R}{\mathcal R}}(\ell,\ell_{\mathrm{t}}) 
+ {\mathcal V}_{\mathrm{BB}}^{(2)}(\ell,\ell_{\mathrm{t}})  + 2\cos{\beta} {\mathcal V}^{(2)}_{{\mathcal R}\mathrm{B}}(\ell,\ell_{\mathrm{t}}),
\label{PR2}\\
&&  {\mathcal I}_{(3)}^{(\mathrm{TT})}(\ell,\ell_{\mathrm{S}}) 
={\mathcal V}^{(3)}_{{\mathcal R}{\mathcal R}}(\ell,\ell_{\mathrm{t}}) 
+ {\mathcal V}_{\mathrm{BB}}^{(3)}(\ell,\ell_{\mathrm{t}})  + 2\cos{\beta} {\mathcal V}^{(3)}_{{\mathcal R}\mathrm{B}}(\ell,\ell_{\mathrm{t}}),
\label{PR3}\\
&&  {\mathcal I}_{(4)}^{(\mathrm{TT})}(\ell,\ell_{\mathrm{S}},\ell_{\mathrm{t}}) 
={\mathcal V}^{(4)}_{{\mathcal R}{\mathcal R}}(\ell,\ell_{\mathrm{S}},\ell_{\mathrm{t}}) 
+ {\mathcal V}_{\mathrm{BB}}^{(4)}(\ell,\ell_{\mathrm{S}},\ell_{\mathrm{t}})  
+ \cos{\beta} ({\mathcal V}^{(4)}_{{\mathcal R}\mathrm{B}} + {\mathcal V}^{(4)}_{\mathrm{B}{\mathcal R}})(\ell,\ell_{\mathrm{S}},\ell_{\mathrm{t}}),
\label{PR4}
\end{eqnarray}
where  $\cos{\beta}$ parametrizes the correlation between the purely adiabatic and the purely magnetized components\footnote{This correlation  arises, in explicit models, because magnetic fields are produced 
during some stages of inflationary expansion \cite{spectator} (see also \cite{bamba1,bamba2,camp1,camp2}). 
In a model-independent  perspective the correlation between different components should 
also be considered in full analogy with what happens for entropic initial conditions \cite{h1,h2,h3,h4,h5}.}.
The terms appearing in Eqs.  (\ref{PR1}) are expressible as 
\begin{eqnarray} 
&& {\mathcal V}^{(1)}_{{\mathcal R}{\mathcal R}}(\ell,\ell_{\mathrm{t}})= {\mathcal Q}_{{\mathcal R}{\mathcal R}}\, \ell^{n_{\mathrm{s}}-1} I^{(1)}_{{\mathcal R}{\mathcal R}}(\ell,\ell_{\mathrm{t}}, n_{\mathrm{s}}),
\label{PR5}\\
&&  {\mathcal V}^{(1)}_{\mathrm{BB}}(\ell,\ell_{\mathrm{t}})= 
{\mathcal Q}_{\mathrm{B}\mathrm{B}} \, \ell^{2(n_{\mathrm{B}} -1)}
I^{(1)}_{{\mathrm{BB}}}(\ell,\ell_{\mathrm{t}}, 2 n_{\mathrm{B}}-1),
\label{PR6}\\
&& {\mathcal V}^{(1)}_{{\mathcal R}\mathrm{B}}(\ell,\ell_{\mathrm{t}})=  {\mathcal Q}_{{\mathcal R}{\mathrm{B}}}\,\ell^{\frac{n_{\mathrm{s}}+ 2 n_{\mathrm{B}} -3}{2}} 
I_{{\mathcal R}\mathrm{B}}^{(1)}\biggl(\ell,\ell_{\mathrm{t}}, \frac{n_{\mathrm{s}} + 2n_{\mathrm{B}} -1}{2}\biggr).
\label{PR7}
\end{eqnarray}
The basic integral appearing in Eqs. (\ref{PR5})--(\ref{PR7}) is given by \footnote{It is 
relevant to point out that the arguments of the integrals contain the multipole, the 
diffusion scales and the relevant spectral index. These are the basic 
quantities which define the eight basic integrals which will now be listed.}:
\begin{equation}
I_{XY}^{(1)}(\ell,\ell_{\mathrm{t}},n) = \int_{1}^{\infty} \frac{w^{n -3}}{\sqrt{w^2 -1}} L_{X}(w,\ell) L_{Y}(w,\ell)
e^{- 2(\frac{\ell^2}{\ell_{\mathrm{t}}^2}) w^2}\, dw.
\label{PR8}
\end{equation}
In Eqs. (\ref{PR8}) the functions $L_{X}(w,\ell)$ and $L_{Y}(w,\ell)$ account for the contribution of large-scale magnetic fields to the tight coupling solutions and also depend upon the resolution of the calculation, i.e. upon $\ell_{\mathrm{max}}$ (which denotes the maximal multipole at which the calculation is trustable). The four functions which enter Eq. (\ref{PR8}) as well as the other seven integrals which will be discussed below are: 
 \begin{eqnarray}
 && L_{\mathcal R}(w,\ell)= \alpha_{\mathcal R} - \beta_{\mathcal R} \ln{(w\,q_{\ell})},\qquad L_{\mathrm{B}}(w,\ell)=  \alpha_{\mathrm{B}} - \beta_{\mathrm{B}} \ln{(w\,q_{\ell})},
 \label{GEN13}\\
 && M_{\mathcal R}(w,\ell)= \overline{\alpha}_{\mathcal R} + \overline{\beta}_{\mathcal R} \ln{(w\,q_{\ell})},\qquad  M_{\mathrm{B}}(w,\ell)= \overline{\alpha}_{\mathrm{B}} + \overline{\beta}_{\mathrm{B}} \ln{(w\,q_{\ell})},
 \label{GEN15}
 \end{eqnarray}
 where 
 \begin{eqnarray}
 && \alpha_{\mathcal R} = \frac{R_{\mathrm{b}}}{6} \ln{\biggl(\frac{7}{100}\biggr)},\qquad \beta_{\mathcal R} = 
 \frac{R_{\mathrm{b}}}{6},
\label{GEN16}\\ 
&&  \overline{\alpha}_{\mathcal R} = - \frac{6}{25} \ln{(96)},\qquad  \overline{\beta}_{\mathcal R} = - \frac{6}{25},
 \label{GEN17}\\
 && \alpha_{\mathrm{B}} = r_{\mathrm{B}} - \frac{3 \,R_{\gamma} r_{\mathrm{B}} + 5}{20},\qquad \beta_{\mathrm{B}} =0,
 \label{GEN18}\\
&&\overline{\alpha}_{\mathrm{B}} = [3 (R_{\mathrm{b}}+1)]^{1/4} \, \biggl[ \frac{R_{\gamma} +5}{20} - r_{\mathrm{B}}\biggr]
 ,\qquad \overline{\beta}_{\mathrm{B}} =0,
 \label{GEN19}
 \end{eqnarray}
 where $R_{\mathrm{b}}$ is the baryon to photon ratio at the recombination
 and $q_{\ell}$ is given by:
 \begin{equation}
 q_{\ell} = \biggl(\frac{\ell}{200\, d_{\mathrm{A}}(z_{*})}\biggr) \sqrt{\frac{r_{\mathrm{R}*}}{z_{*} +1}}.
\label{GEN20}
\end{equation}
Concerning the notations of Eqs. (\ref{GEN16})--(\ref{GEN19}) we remind, as defined 
after Eq. (\ref{GEN8}) that $r_{\mathrm{B}} = {\mathcal P}_{\Omega}/{\mathcal P}_{\sigma}$: $r_{\mathrm{B}}$ is, therefore, the ratio between the power spectrum
associated with $\Omega_{\mathrm{B}}$ and the power spectrum associated 
with $\sigma_{\mathrm{B}}$. 

Since the aim of the present analysis 
is to have analytic estimates of the modifications induced by large-scale magnetic fields 
especially at small angular scales (i.e. in the limit $\ell \gg 1$).  In the latter 
limit the spherical Bessel functions $j_{\ell}(x)$ can be approximated 
in their large-order limit  and the acoustic multipole fixes the oscillatory 
structure of the angular power spectra:
\begin{equation}
\ell (\ell + 1) j_{\ell}^2(x) \simeq  \ell (\ell +1) \frac{\cos^2{[\beta(x,\ell)]}}{x \sqrt{x^2 - \ell^2}} \simeq \frac{1}{2} \frac{1}{w \sqrt{w^2 -1}}, \qquad x = w\, \ell,
\label{GEN23}
\end{equation}
where $\beta(x,\ell)= \sqrt{x^2 -\ell^2} - \ell\arccos{(\ell/x )} - \frac{\pi}{4}$ \cite{abr1,abr2}.
Recall that, often, changes of variables are required to evaluate the integrals. In particular, a practical choice is:
\begin{equation}
w\to y^2 +1, \qquad dw \to 2 y d y, \qquad \frac{d w}{\sqrt{w^2 -1}} \to \frac{ 2\, d y}{\sqrt{y^2 +2}}.
\label{GEN21}
\end{equation}
The change of variable $ w^2 = y^2 + 1$ is also possible in some cases 
and it leads to a simpler structure of the integrands, in some cases. In spite of the change of variables, the numerical values of the various integrals do not change.
At the same time, since the integrals will be evaluated numerically, the 
time of evaluation can also change as a function of the algebraic 
form of the various integrands.

The contribution labeled by 
${\mathcal I}_{(2)}^{(\mathrm{TT})}(\ell,\ell_{\mathrm{S}})$ in Eqs. (\ref{GEN27}) and (\ref{PR2}) leads to the following explicit results 
\begin{eqnarray}
&& {\mathcal V}^{(2)}_{{\mathcal R}{\mathcal R}}(\ell,\ell_{\mathrm{S}})= {\mathcal Q}_{{\mathcal R}{\mathcal R}}\, \ell^{n_{\mathrm{s}}-1} I^{(2)}_{{\mathcal R}{\mathcal R}}(\ell,\ell_{\mathrm{S}}, n_{\mathrm{s}}),
\label{PR9}\\
&&  {\mathcal V}^{(2)}_{\mathrm{BB}}(\ell,\ell_{\mathrm{S}})= 
{\mathcal Q}_{\mathrm{B}\mathrm{B}}\,
\, \ell^{2(n_{\mathrm{B}}-1)} I^{(2)}_{{\mathrm{BB}}}(\ell,\ell_{\mathrm{S}}, 2 n_{\mathrm{B}}-1),
\label{PR10}\\
&& {\mathcal V}^{(2)}_{{\mathcal R}\mathrm{B}}(\ell,\ell_{\mathrm{t}})= 
 {\mathcal Q}_{{\mathcal R}{\mathrm{B}}}\, \ell^{\frac{n_{\mathrm{s}}+ 2 n_{\mathrm{B}} -3}{2}} I_{{\mathcal R}\mathrm{B}}^{(2)}\biggl(\ell,\ell_{\mathrm{S}}, \frac{n_{\mathrm{s}} + 2n_{\mathrm{B}} -1}{2}\biggr),
\label{PR11}
\end{eqnarray}
where the second basic integral appearing in Eqs. (\ref{PR9})--(\ref{PR11}) 
can be written as:
\begin{equation}
I_{XY}^{(2)}(\ell,\ell_{\mathrm{t}},n) = \frac{1}{2}
\int_{1}^{\infty} {\mathcal W}_{+}(w, c_{\mathrm{sb}}) \, w^{n -5}\,M_{X}(w,\ell) M_{Y}(w,\ell)
e^{- 2(\frac{\ell^2}{\ell_{\mathrm{S}}^2}) w^2}\, dw.
\label{PR12}
\end{equation}
For practical convenience, the two functions ${\mathcal W}_{\pm}(w, c_{\mathrm{sb}})$ are introduced, respectively, in Eq. (\ref{PR12}) and in Eq. (\ref{PR16}):
\begin{equation}
{\mathcal W}_{\pm}(w, c_{\mathrm{sb}}) = \frac{c_{\mathrm{sb}} ( 1 \pm 9 c_{\mathrm{sb}}^2) w^2 \mp 9 c_{\mathrm{sb}}^3}{\sqrt{w^2 -1}},
\label{PR12a}
\end{equation}
where $c_{\mathrm{sb}}$ is the photon-baryon sound speed already introduced, for instance, in Eq. (\ref{Eq3}).
The third contribution appearing in Eq. (\ref{GEN27}), i.e.  
${\mathcal I}_{3}^{(\mathrm{TT})}(\ell,\ell_{\mathrm{S}})$ is determined 
by the terms appearing in Eq. (\ref{PR3}) whose explicit expressions are:
\begin{eqnarray}
&& {\mathcal V}^{(3)}_{{\mathcal R}{\mathcal R}}(\ell,\ell_{\mathrm{S}})= 
{\mathcal Q}_{{\mathcal R}{\mathcal R}}\, \ell^{n_{\mathrm{s}}-1} I^{(3)}_{{\mathcal R}{\mathcal R}}(\ell,\ell_{\mathrm{S}}, n_{\mathrm{s}}),
\label{PR13}\\
&&  {\mathcal V}^{(3)}_{\mathrm{BB}}(\ell,\ell_{\mathrm{S}})= 
{\mathcal Q}_{\mathrm{B}\mathrm{B}}
\, \ell^{2(n_{\mathrm{B}}-1)} I^{(3)}_{{\mathrm{BB}}}(\ell,\ell_{\mathrm{S}}, 2 n_{\mathrm{B}}-1),
\label{PR14}\\
&& {\mathcal V}^{(3)}_{{\mathcal R}\mathrm{B}}(\ell,\ell_{\mathrm{t}})= 
 {\mathcal Q}_{{\mathcal R}{\mathrm{B}}}\,
 \ell^{\frac{n_{\mathrm{s}}+ 2 n_{\mathrm{B}} -3}{2}} 
I_{{\mathcal R}\mathrm{B}}^{(3)}\biggl(\ell,\ell_{\mathrm{S}}, \frac{n_{\mathrm{s}} + 2n_{\mathrm{B}} -1}{2}\biggr).
\label{PR15}
\end{eqnarray}
Recalling Eq. (\ref{PR12a}) the basic integral appearing in Eqs. (\ref{PR13})--(\ref{PR15}) is given by:
\begin{equation}
I_{XY}^{(3)}(\ell,\ell_{\mathrm{S}},n) = \frac{1}{2}
\int_{1}^{\infty}  {\mathcal W}_{-}(w, c_{\mathrm{sb}})\,w^{n -5}\,\cos{(2 \gamma_{\mathrm{A}}\ell w)}\,M_{X}(w,\ell) M_{Y}(w,\ell)
e^{- 2(\frac{\ell^2}{\ell_{\mathrm{S}}^2}) w^2}\, dw,
\label{PR16}
\end{equation}
where $\gamma_{\mathrm{A}}$ has been introduced in Eq. (\ref{B3}) and ${\mathcal W}_{-}(w, c_{\mathrm{sb}})$
is defined in Eq. (\ref{PR12a}).
Finally, the fourth basic term of Eqs. (\ref{GEN27}) and (\ref{PR4})
is completely specified by the four expressions:
\begin{eqnarray}
&& {\mathcal V}^{(4)}_{{\mathcal R}{\mathcal R}}(\ell,\ell_{\mathrm{S}})= {\mathcal Q}_{{\mathcal R}{\mathcal R}}\,\ell^{n_{\mathrm{s}}-1} I^{(4)}_{{\mathcal R}{\mathcal R}}(\ell,\ell_{\mathrm{S}}, \ell_{\mathrm{t}}, n_{\mathrm{s}}),
\label{PR17}\\
&&  {\mathcal V}^{(4)}_{\mathrm{BB}}(\ell,\ell_{\mathrm{S}})= 
{\mathcal Q}_{\mathrm{B}\mathrm{B}} \,
\ell^{2(n_{\mathrm{B}}-1)} I^{(4)}_{{\mathrm{BB}}}(\ell,\ell_{\mathrm{S}}, \ell_{\mathrm{t}}, 2 n_{\mathrm{B}}-1),
\label{PR18}\\
&& {\mathcal V}^{(4)}_{{\mathcal R}\mathrm{B}}(\ell,\ell_{\mathrm{t}})= 
{\mathcal Q}_{{\mathcal R}{\mathrm{B}}} \ell^{\frac{n_{\mathrm{s}}+ 2 n_{\mathrm{B}} -3}{2}} 
I_{{\mathcal R}\mathrm{B}}^{(4)}\biggl(\ell,\ell_{\mathrm{S}}, \frac{n_{\mathrm{s}} + 2n_{\mathrm{B}} -1}{2}\biggr).
\label{PR19}
\end{eqnarray}
The basic integral appearing 
in Eqs. (\ref{PR13})--(\ref{PR15}) is given by:
\begin{equation}
I_{XY}^{(4)}(\ell,\ell_{\mathrm{S}},\ell_{\mathrm{t}},n) = 2
\int_{1}^{\infty} \frac{\sqrt{c_{\mathrm{sb}}} \, w^{n -3}}{\sqrt{w^2 -1}}\,\cos{(\gamma_{\mathrm{A}}\ell w)}\,L_{X}(w,\ell) M_{Y}(w,\ell)
e^{- [(\frac{\ell^2}{\ell_{\mathrm{S}}^2})+ (\frac{\ell^2}{\ell_{\mathrm{t}}^2})]w^2}\, dw.
\label{PR20}
\end{equation}
Equations (\ref{PR8}), (\ref{PR12}), (\ref{PR16}) and (\ref{PR20}) define 
the primary form of the integrals determining the temperature autocorrelations. 
In what follows the EE and TE correlations will be more specifically studied.
\subsection{E-mode autocorrelations}
Within the same logical scheme already employed in the case of 
the TT correlations, the EE angular power spectra of Eq. (\ref{GEN2}) can be written in terms of two (further) basic integrals, i.e. 
\begin{equation}
 G_{\ell}^{(\mathrm{EE})} =  {\mathcal I}^{(\mathrm{EE})}_{(5)}(\ell, \ell_{\mathrm{D}}) - {\mathcal I}^{(\mathrm{EE})}_{(6)}(\ell, \ell_{\mathrm{D}}).
 \label{EE14}
 \end{equation}
Both the EE and the TE angular power spectra are suppressed with 
respect to the TT correlations. It is therefore useful to define the quantity
\begin{equation}
{\mathcal N}^{(\mathrm{EE})}(\ell, \sigma_{*}) = 0.132\,  (k_{0} \sigma_{*})^2\, (\ell+1)^2 (\ell -1) (\ell+2) \, \ell^{-4},
\label{EE14a}
\end{equation}
which is independent of $\ell$ in the range 
of multipoles where the calculation can be trusted 
(i.e., in practice, $\ell >20$).  
The explicit form of the integrals appearing in Eq. (\ref{EE14} can be 
written, in full analogy with Eqs. (\ref{PR1})--(\ref{PR4}), as 
\begin{eqnarray}
 && {\mathcal I}_{(5)}^{(\mathrm{EE})}(\ell,\ell_{\mathrm{D}}) 
={\mathcal V}^{(5)}_{{\mathcal R}{\mathcal R}}(\ell,\ell_{\mathrm{D}}) 
+ {\mathcal V}_{\mathrm{BB}}^{(5)}(\ell,\ell_{\mathrm{D}})  +2 \cos{\beta} {\mathcal V}^{(5)}_{{\mathcal R}\mathrm{B}}(\ell,\ell_{\mathrm{D}}),
\label{PREE1}\\
&& {\mathcal I}_{(6)}^{(\mathrm{EE})}(\ell,\ell_{\mathrm{D}}) 
={\mathcal V}^{(6)}_{{\mathcal R}{\mathcal R}}(\ell,\ell_{\mathrm{D}}) 
+ {\mathcal V}_{\mathrm{BB}}^{(6)}(\ell,\ell_{\mathrm{D}})  +2 
\cos{\beta} {\mathcal V}^{(6)}_{{\mathcal R}\mathrm{B}}(\ell,\ell_{\mathrm{D}}).
\label{PREE2}
\end{eqnarray}
where 
\begin{eqnarray}
&& {\mathcal V}^{(5)}_{{\mathcal R}{\mathcal R}}(\ell,\ell_{\mathrm{D}}) = 
{\mathcal Q}_{{\mathcal R}{\mathcal R}} {\mathcal N}^{(\mathrm{EE})}(\ell, \sigma_{*})
 \ell^{n_{\mathrm{s}}+1} \, c_{\mathrm{sb}}^3  I^{(5)}_{{\mathcal R}{\mathcal R}} (\ell,\ell_{\mathrm{D}}, n_{\mathrm{s}}),
\label{PREE3}\\
&& {\mathcal V}^{(5)}_{\mathrm{BB}}(\ell,\ell_{\mathrm{D}}) = 
{\mathcal Q}_{\mathrm{B}\mathrm{B}}  {\mathcal N}^{(\mathrm{EE})}
(\ell, \sigma_{*})  
\ell^{2 n_{\mathrm{B}}} \, c_{\mathrm{sb}}^3 
I^{(5)}_{\mathrm{B}{\mathcal R}}(\ell,\ell_{\mathrm{D}}, 2 n_{\mathrm{B}}-1)
\label{PREE4}\\
&& {\mathcal V}^{(5)}_{{\mathcal R}\mathrm{B}}(\ell,\ell_{\mathrm{D}}) = 
 {\mathcal Q}_{{\mathcal R}{\mathrm{B}}}
 {\mathcal N}^{(\mathrm{EE})}(\ell, \sigma_{*})\, 
\ell^{\frac{n_{\mathrm{s}} + 2 n_{\mathrm{B}}+1}{2}} \, c_{\mathrm{sb}}^3 
 I^{(5)}_{{\mathcal R}\mathrm{B}}\biggl(\ell,\ell_{\mathrm{D}}, 
\frac{n_{\mathrm{s}} +2 n_{\mathrm{B}}-1}{2}\biggr),
\label{PREE5}
\end{eqnarray}
where 
\begin{equation}
I^{(5)}_{XY}(\ell,\ell_{\mathrm{D}}, n) = \int_{1}^{\infty} \frac{w^{n - 5}}{\sqrt{w^2 -1}}  \, M_{X}(w,\ell) M_{Y}(w,\ell) e^{- 2 (\frac{\ell^2}{\ell_{\mathrm{D}}^2})w^2}\, dw.
\label{PREE6}
\end{equation}
The three terms defining ${\mathcal I}_{(6)}^{(\mathrm{EE})}(\ell,\ell_{\mathrm{D}})$ are:
\begin{eqnarray}
&& {\mathcal V}^{(6)}_{{\mathcal R}{\mathcal R}}(\ell,\ell_{\mathrm{D}}) = 
{\mathcal Q}_{{\mathcal R}{\mathcal R}} \,
{\mathcal N}^{(\mathrm{EE})}(\ell,\sigma_{*})
\ell^{n_{\mathrm{s}}-3} \, c_{\mathrm{sb}}^3  I^{(6)}_{{\mathcal R}{\mathcal R}} (\ell,\ell_{\mathrm{D}}, n_{\mathrm{s}}),
\label{PREE7}\\
&& {\mathcal V}^{(6)}_{\mathrm{BB}}(\ell,\ell_{\mathrm{D}}) = 
{\mathcal Q}_{\mathrm{B}\mathrm{B}}\,
{\mathcal N}^{(\mathrm{EE})}(\ell, \sigma_{*})
\ell^{2 n_{\mathrm{B}}-4} \, c_{\mathrm{sb}}^3 \, I^{(6)}_{\mathrm{B}{\mathcal R}}(\ell,\ell_{\mathrm{D}}, 2 n_{\mathrm{B}}-1),
\label{PREE8}\\
&& {\mathcal V}^{(6)}_{{\mathcal R}\mathrm{B}}(\ell,\ell_{\mathrm{D}}) = 
 {\mathcal Q}_{{\mathcal R}{\mathrm{B}}} {\mathcal N}^{(\mathrm{EE})}(\ell, \sigma_{*})\,
\ell^{\frac{n_{\mathrm{s}} + 2 n_{\mathrm{B}}-7}{2}} \, c_{\mathrm{sb}}^3 
 I^{(6)}_{{\mathcal R}\mathrm{B}}\biggl(\ell,\ell_{\mathrm{D}}, 
\frac{n_{\mathrm{s}} +2 n_{\mathrm{B}}-1}{2}\biggr).
\label{PREE9}
\end{eqnarray}
The sixth basic integral appearing in Eqs. (\ref{PREE7})--(\ref{PREE9}) is 
\begin{equation}
I^{(6)}_{XY}(\ell,\ell_{\mathrm{D}}, n) = \int_{1}^{\infty} \frac{\cos{(2\gamma_{\mathrm{A}} \ell w)}\, w^{n - 5}}{\sqrt{w^2 -1}} \, dw \, M_{X}(w,\ell) M_{Y}(w,\ell) e^{- 2 (\frac{\ell^2}{\ell_{\mathrm{D}}^2})w^2}.
\label{PREE10}
\end{equation}
Equations (\ref{PREE6}) and (\ref{PREE10}) represent the primary 
form of the two basic integrals determining the polarization 
autocorrelations.  From the purely algebraic point of view the EE angular power 
spectra have a single periodicity is insofar as Eq. (\ref{PREE6})
has an integrand which is not oscillating while the integrand of 
Eq. (\ref{PREE10}) depends on a single oscillating term. Conversely, 
the TT angular power  spectra are given by the weighted 
superposition of the integrals appearing in Eqs. (\ref{PR8}), (\ref{PR12}), (\ref{PR16}) and (\ref{PR20}) whose corresponding integrals do not depend upon the same 
oscillating term. The single periodicity of the EE angular power spectra will, have, in the present context, interesting consequences. 
\subsection{Temperature-polarization cross-correlations}
The last angular power spectrum considered here is the one 
arising from the cross-correlations between temperature and polarization, i.e. 
the TE power spectrum leading to the following integrals 
\begin{equation}
G_{\ell}^{(\mathrm{TE})} = 
{\mathcal I}_{(7)}^{(\mathrm{TE})}(\ell,\ell_{\mathrm{S}}) + 
{\mathcal I}_{(8)}^{(\mathrm{TE})}(\ell,\ell_{\mathrm{S}}, \ell_{\mathrm{D}}).
\label{TE4}
\end{equation}
In full analogy with what has been done in the case of the EE correlations 
(see Eq. (\ref{EE14a})) it is practical to define 
\begin{equation}
{\mathcal N}^{(\mathrm{TE})}(\ell,\sigma_{*}) = 0.515\, k_{0} \sigma_{*}\,(\ell+1) \sqrt{(\ell +1)(\ell -1) (\ell+2)}\, \ell^{- 5/2}.
\label{TE4a}
\end{equation}
Consequently, 
the explicit form of the integrals appearing in Eq. (\ref{TE4}) is 
\begin{eqnarray}
&& {\mathcal I}_{7}^{(\mathrm{TE})}(\ell,\ell_{\mathrm{S}}) = 
{\mathcal V}^{(7)}_{{\mathcal R}{\mathcal R}}(\ell,\ell_{\mathrm{S}}) 
+ {\mathcal V}_{\mathrm{BB}}^{(7)}(\ell,\ell_{\mathrm{S}})  + \cos{\beta} [{\mathcal V}^{(7)}_{{\mathcal R}\mathrm{B}}(\ell,\ell_{\mathrm{S}}) + {\mathcal V}^{(7)}_{\mathrm{B}{\mathcal R}}(\ell,\ell_{\mathrm{S}})],
\label{TE7}\\
&& {\mathcal I}_{8}^{(\mathrm{TE})}(\ell,\ell_{\mathrm{S}},\ell_{\mathrm{D}}) ={\mathcal V}^{(8)}_{{\mathcal R}{\mathcal R}}(\ell,\ell_{\mathrm{S}},\ell_{\mathrm{D}}) 
+ {\mathcal V}_{\mathrm{BB}}^{(8)}(\ell,\ell_{\mathrm{S}}, \ell_{\mathrm{D}})  + 2 \cos{\beta} {\mathcal V}^{(8)}_{{\mathcal R}\mathrm{B}}(\ell,\ell_{\mathrm{S}},\ell_{\mathrm{D}}).
\label{TE8}
\end{eqnarray}
where
\begin{eqnarray}
&& {\mathcal V}^{(7)}_{{\mathcal R}{\mathcal R}}(\ell,\ell_{\mathrm{S}}) = 
{\mathcal Q}_{{\mathcal R}{\mathcal R}} {\mathcal N}^{(\mathrm{TE})}(\ell,\sigma_{*})  \ell^{n_{\mathrm{s}}} \, c_{\mathrm{sb}}^{3/2}  I^{(7)}_{{\mathcal R}{\mathcal R}} (\ell,\ell_{\mathrm{S}}, n_{\mathrm{s}}),
\label{TE9}\\
&& {\mathcal V}^{(7)}_{\mathrm{BB}}(\ell,\ell_{\mathrm{S}}) = 
 {\mathcal Q}_{\mathrm{B}\mathrm{B}} {\mathcal N}^{(\mathrm{TE})}(\ell,\sigma_{*})   \ell^{2 n_{\mathrm{B}}-1} \, c_{\mathrm{sb}}^{3/2} I^{(7)}_{\mathrm{B}{\mathcal R}}(\ell,\ell_{\mathrm{S}}, 2 n_{\mathrm{B}}-1)
\label{TE10}\\
&& {\mathcal V}^{(7)}_{{\mathcal R}\mathrm{B}}(\ell,\ell_{\mathrm{S}}) = 
{\mathcal Q}_{{\mathcal R}{\mathrm{B}}}
{\mathcal N}^{(\mathrm{TE})}(\ell,\sigma_{*}) \, c_{\mathrm{sb}}^{3/2}
 \ell^{\frac{n_{\mathrm{s}} + 2 n_{\mathrm{B}}-1}{2}} \, I^{(7)}_{{\mathcal R}\mathrm{B}}\biggl(\ell,\ell_{\mathrm{S}}, 
\frac{n_{\mathrm{s}} +2 n_{\mathrm{B}}-1}{2}\biggr).
\label{TE11}
\end{eqnarray}
The integral appearing in Eqs. (\ref{TE9})--(\ref{TE11})
\begin{equation}
I^{(7)}_{XY}(\ell,\ell_{\mathrm{S}}, n) = \int_{1}^{\infty} \frac{\sin{(\gamma_{\mathrm{A}} \ell w)}\, w^{n - 4}}{\sqrt{w^2 -1}} \, dw \, L_{X}(w,\ell) M_{Y}(w,\ell) e^{-  (\frac{\ell^2}{\ell_{\mathrm{S}}^2})w^2}.
\label{TE12}
\end{equation}
The last bunch of terms contributing to ${\mathcal I}_{8}^{(\mathrm{TE})}(\ell,\ell_{\mathrm{S}}, \ell_{\mathrm{D}})$ is given by 
\begin{eqnarray}
&& {\mathcal V}^{(8)}_{{\mathcal R}{\mathcal R}}(\ell,\ell_{\mathrm{S}},\ell_{\mathrm{D}}) = {\mathcal Q}_{{\mathcal R}{\mathcal R}}
{\mathcal N}^{(\mathrm{TE})}(\ell, \sigma_{*})\ell^{n_{\mathrm{s}}} \,c_{\mathrm{sb}}^{2}  I^{(8)}_{{\mathcal R}{\mathcal R}} (\ell,\ell_{\mathrm{S}}, \ell_{\mathrm{D}}, n_{\mathrm{s}}),
\label{TE13}\\
&& {\mathcal V}^{(8)}_{\mathrm{BB}}(\ell,\ell_{\mathrm{S}}) = {\mathcal Q}_{\mathrm{B}\mathrm{B}} 
{\mathcal N}^{(\mathrm{TE})}(\ell,\sigma_{*}) \ell^{2 n_{\mathrm{B}}-1} \, c_{\mathrm{sb}}^{2} I^{(8)}_{\mathrm{B}{\mathcal R}}(\ell,\ell_{\mathrm{S}}, \ell_{\mathrm{D}}, 2 n_{\mathrm{B}}-1),
\label{TE14}\\
&& {\mathcal V}^{(8)}_{{\mathcal R}\mathrm{B}}(\ell,\ell_{\mathrm{S}},\ell_{\mathrm{D}}) = {\mathcal Q}_{{\mathcal R}{\mathrm{B}}} \, {\mathcal N}^{(\mathrm{TE})}(\ell,\sigma_{*}) \ell^{\frac{n_{\mathrm{s}} + 2 n_{\mathrm{B}}-7}{2}} \, 
c_{\mathrm{sb}}^{2}\, I^{(8)}_{{\mathcal R}\mathrm{B}}\biggl(\ell,\ell_{\mathrm{S}}, \ell_{\mathrm{D}}
\frac{n_{\mathrm{s}} +2 n_{\mathrm{B}}-1}{2}\biggr),
\label{TE15}
\end{eqnarray}
where 
\begin{equation}
I^{(8)}_{XY}(\ell,\ell_{\mathrm{S}},\ell_{\mathrm{D}}, n) = \frac{1}{2}\int_{1}^{\infty} \frac{\sin{(2\gamma_{\mathrm{A}} \ell w)}\, w^{n - 4}}{\sqrt{w^2 -1}} \, dw \, M_{X}(w,\ell) M_{Y}(w,\ell) e^{-  (\frac{\ell^2}{\ell_{\mathrm{S}}^2}+ \frac{\ell^2}{\ell_{\mathrm{D}}^2})w^2}.
\label{TE16}
\end{equation}
The integrals of Eqs. (\ref{TE12}) and (\ref{TE16}) give the last 
pair of primary integrals. The results obtained in the 
present section allow for an explicit evaluation 
of the TT, EE and TE angular power spectra. The following 
section is devoted to the derivation of a number of scaling relations 
which are the magnetized analog of the  standard scaling relations 
which constitute the basis of any sound strategy of parameter estimation. 

\renewcommand{\theequation}{5.\arabic{equation}}
\setcounter{equation}{0}
\section{Scaling properties and form factors}
\label{sec5}
The 8 basic integrals derived in section \ref{sec4} can be 
exploited to study the deviations induced by the ambient magnetic field
on the CMB observables. In the present section the semi-analytic 
results will be confronted with the numerical estimates. The purpose 
will not be to touch upon all the possible themes of the analysis 
but rather to mention only some of the most notable aspects which emerged 
from an exhaustive study of these matters. 
\begin{figure}[!ht]
\centering
\includegraphics[height=6.5cm]{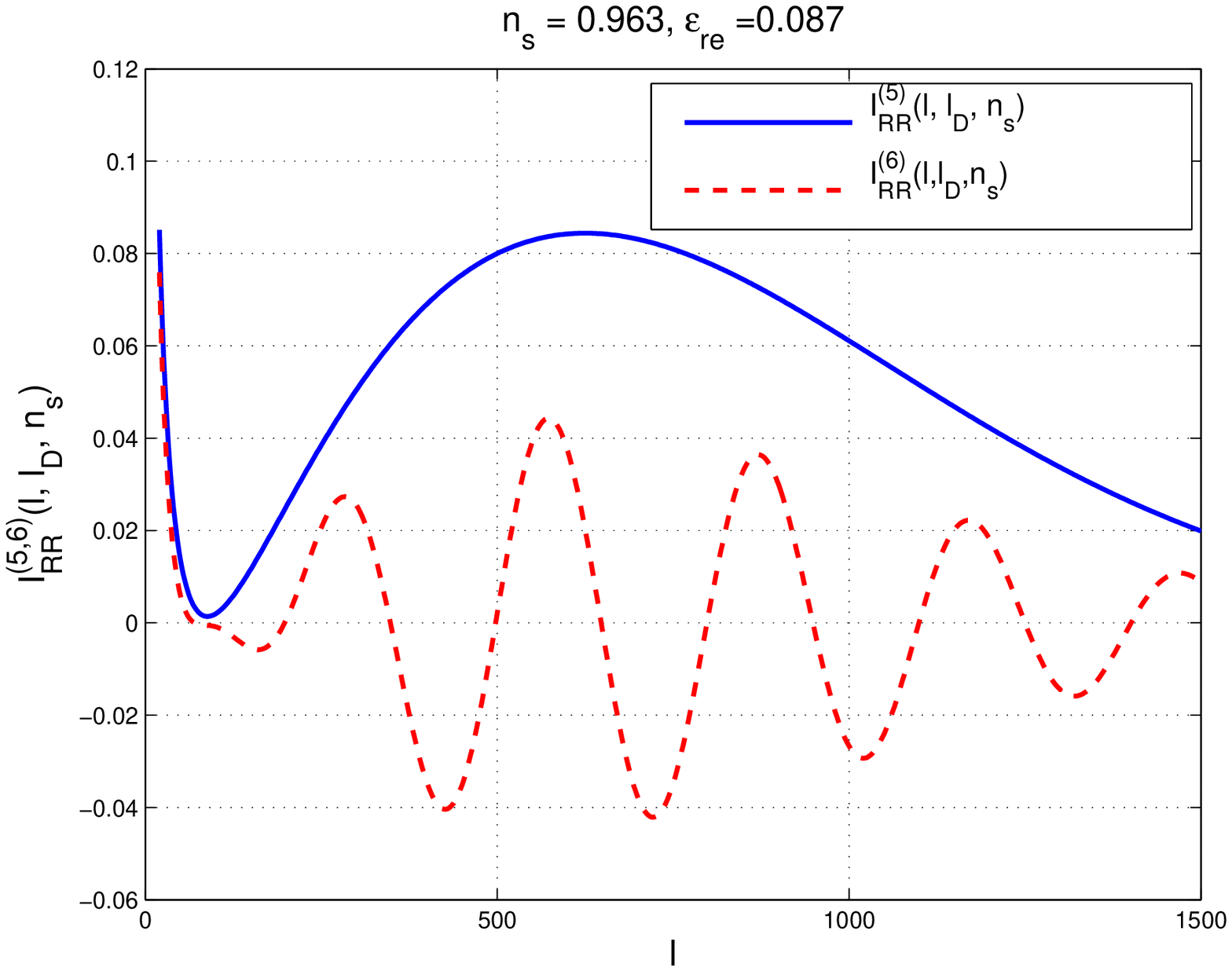}\includegraphics[height=6.5cm]{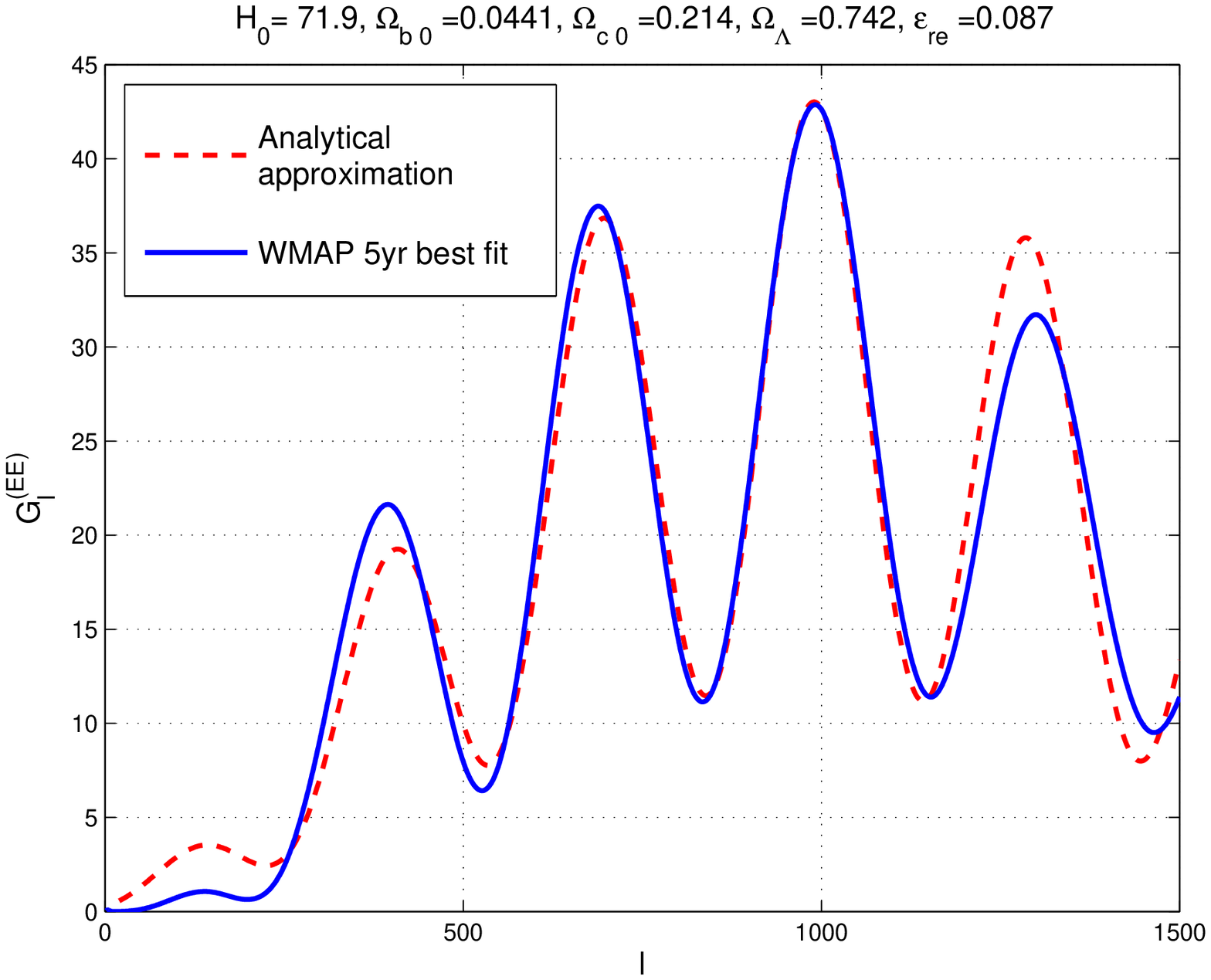}
\caption[a]{The semi-analytic results for the polarization autocorrelations are illustrated 
in the absence of magnetic fields. In the plot at the left, the explicit result for the integrals 
of Eq. (\ref{EE21}) is reported. In the plot at the right the explicit result of Eqs. 
(\ref{EE22a})--(\ref{EE22}) is confronted to the WMAP 5yr best fit. }
\label{figure3}      
\end{figure}
The polarization autocorrelations  are sensitive  to 2 out of 8 basic integrals and, as 
previously discussed (see, e.g. discussion after Eq. (\ref{PREE10})), 
they have the simpler periodicity.
They also depend upon $\ell_{\mathrm{D}}$ since the integral over the optical depth allows for an explicit integration of the source term (see Eqs. (\ref{TP8})--(\ref{TP12})).
The present section is organized as follows: in subsection \ref{SUBS51} the EE 
angular power spectra will be discussed and the semi-analytical results will be compared with the numerical evaluation. In subsection \ref{SUBS52} the semi-analytical 
results for the TT and TE correlations will be illustrated. Finally, subsection 
\ref{SUBS53} will be focussed on the scaling properties 
of the temperature and polarization autocorrelations.  
\subsection{EE angular power spectra}
\label{SUBS51}
In the absence of any ambient magnetic field, Eqs. (\ref{EE14})  and (\ref{PREE1})--(\ref{PREE2}) lead to the complete expression of the EE correlation which can be written as
\begin{eqnarray}
G_{\ell}^{(\mathrm{EE})} &=&  {\mathcal I}^{(\mathrm{EE})}_{(5)}(\ell, \ell_{\mathrm{D}}) - {\mathcal I}^{(\mathrm{EE})}_{(6)}(\ell, \ell_{\mathrm{D}}), 
\nonumber\\
 {\mathcal I}_{(5)}^{(\mathrm{EE})}(\ell,\ell_{\mathrm{D}}) &=& {\mathcal V}^{(5)}_{{\mathcal R}{\mathcal R}}(\ell,\ell_{\mathrm{D}}),\qquad 
{\mathcal I}_{(6)}^{(\mathrm{EE})}(\ell,\ell_{\mathrm{D}})={\mathcal V}^{(6)}_{{\mathcal R}{\mathcal R}}(\ell,\ell_{\mathrm{D}}).
\label{EE18a}
\end{eqnarray}
Bearing in mind the explicit form of the different contributions, Eq. (\ref{EE18a}) becomes 
\begin{eqnarray}
&&G_{\ell}^{(\mathrm{EE})} = \overline{{\mathcal A}}^{(\mathrm{EE})}\,(\ell -1) (\ell +1)^2 (\ell +2) \ell^{n_{s} -3} [ I^{(5)}_{{\mathcal R}{\mathcal R}}
 (\ell, \ell_{\mathrm{D}},n_{\mathrm{s}}) - I^{(6)}_{{\mathcal R}{\mathcal R}}(\ell, \ell_{\mathrm{D}}, n_{\mathrm{s}})],
 \label{EE18}\\
 && \overline{{\mathcal A}}^{(\mathrm{EE})} = (0.132) \, 
 (k_{0} \sigma_{*})^{2} \,  \biggl(\frac{k_{0}}{k_{\mathrm{p}}}\biggr)^{n_{\mathrm{s}} -1}  c_{\mathrm{sb}}^3\, e^{- 2\epsilon_{\mathrm{re}}} \, {\mathcal A}_{{\mathcal R}}\, T_{\gamma 0}^2
 \label{EE18b}
 \end{eqnarray}
where $ \overline{{\mathcal A}}^{(\mathrm{EE})}$ is the rescaled amplitude grouping 
all the factors which are independent on the multipole $\ell$. If
 $n_{\mathrm{s}}=0.963$ and $\epsilon_{\mathrm{re}}=0.087$ (as in the 5yr best fit to the WMAP data alone), Eq. (\ref{EE18b}) implies that \footnote{It should be noticed that 
 the expression for $\overline{{\mathcal A}}^{(\mathrm{EE})}$ 
is dimension-full since the result has been multiplied, as customary, by 
$T_{\gamma 0}^2 = ( 2.725\times 10^{6})^2 \, (\mu \mathrm{K})^2$ where $T_{\gamma 0}$ is the inferred 
value of the CMB black-body  spectrum in units of $\mu \mathrm{K}$.}
 $\overline{{\mathcal A}}^{(\mathrm{EE})} = 4.25 \times 10^{-4}\, (\mu \mathrm{K})^2$. 
The factor $(k_{0} \sigma_{*})$ can be 
estimated within the WMAP data and it is given by $1.43 \times 10^{-3}$. The latter figure arises, as  discussed after Eq. (\ref{LS5}), by computing 
$\Delta \tau_{*}/\tau_{0}$ where $\Delta \tau_{*}$ is the 
thickness of the last scattering  surface in conformal time.
\begin{figure}[!ht]
\centering
\includegraphics[height=6cm]{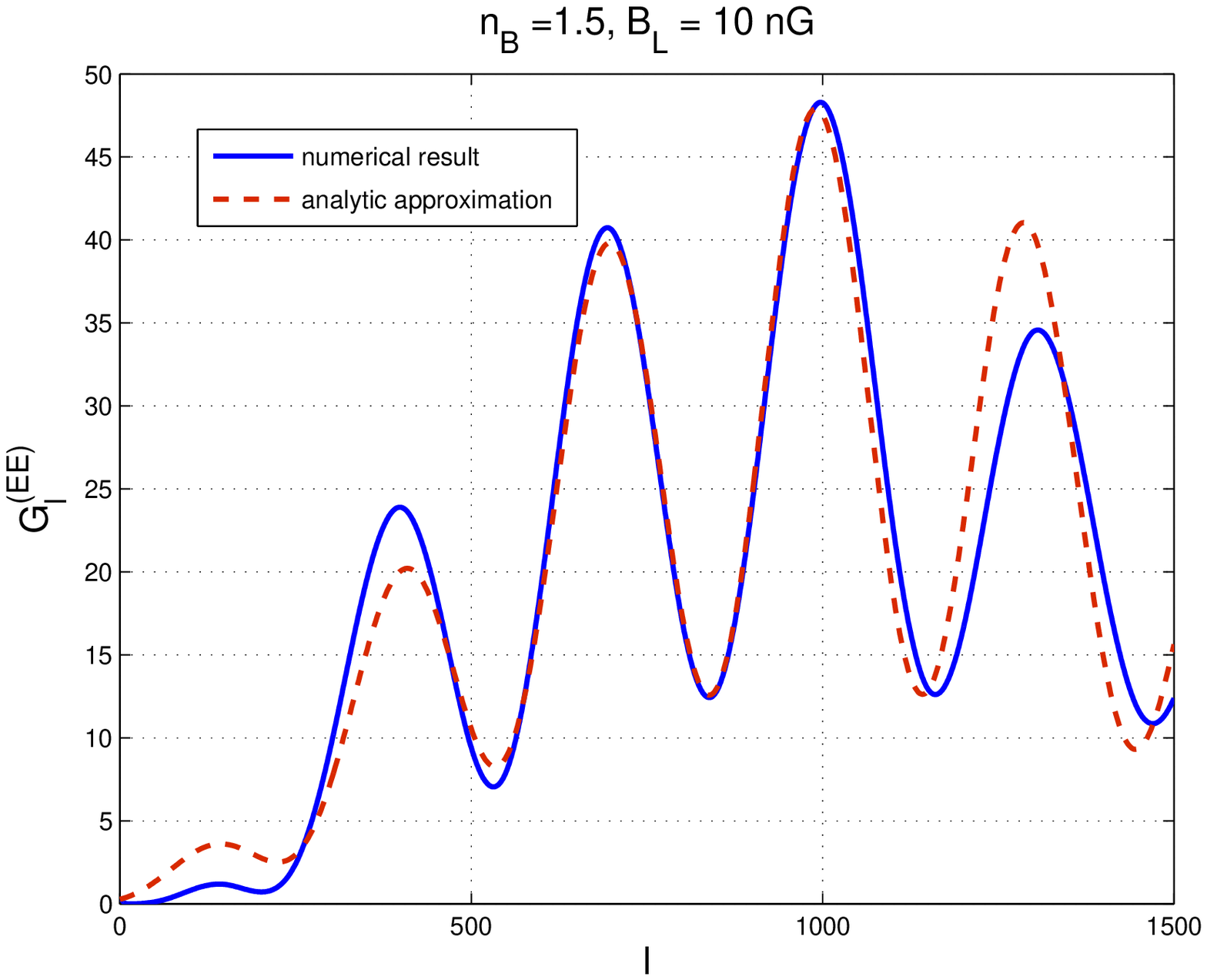}\includegraphics[height=6cm]{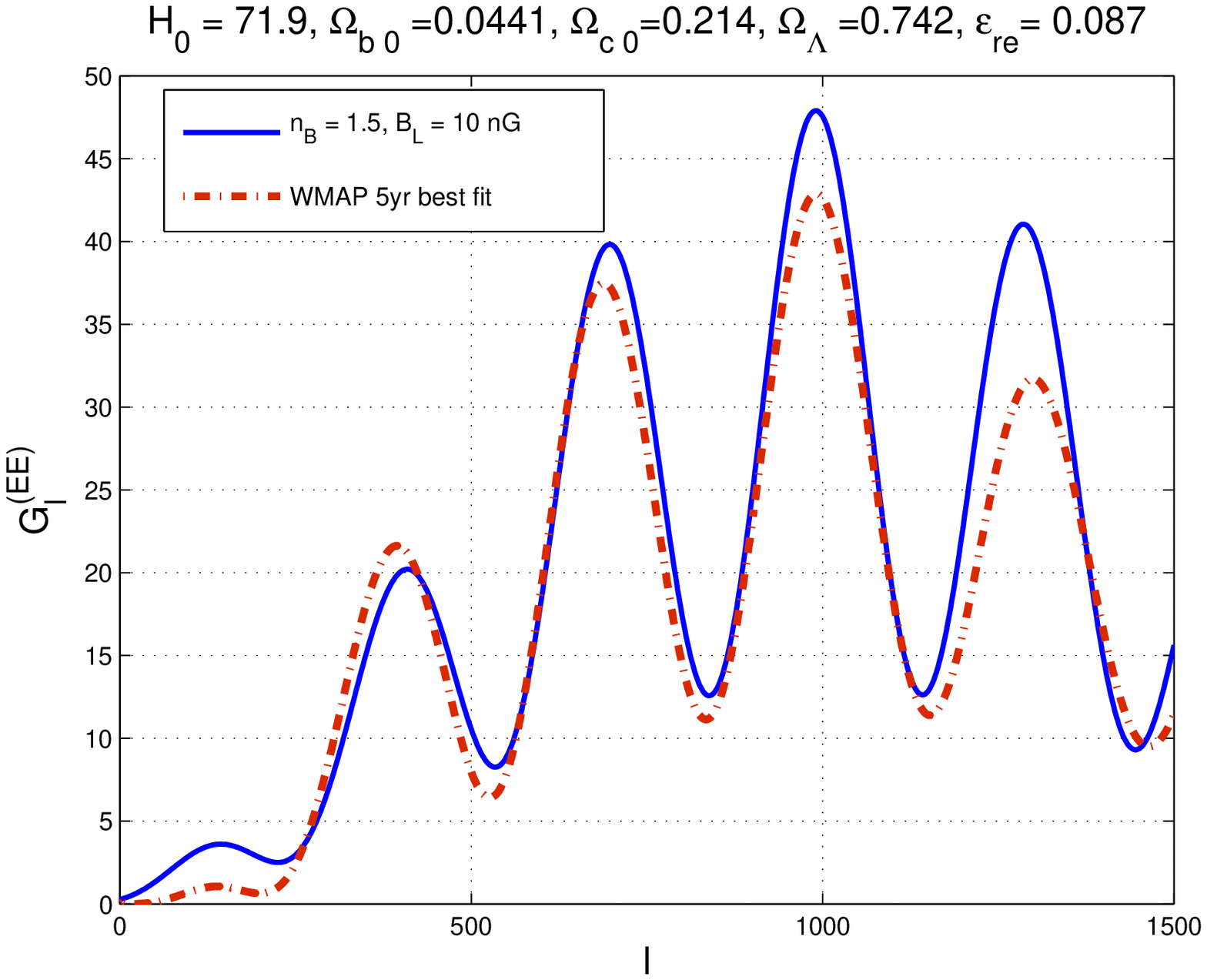}
\caption[a]{The semi-analytic results for the polarization autocorrelations are illustrated 
in the presence of magnetic fields. In the left plot the full line denotes the numerical result while the 
dashed line denotes the analytic approximation. In the right plot the magnetized 
result is compared with the WMAP best fit. In this and in the following plots $\beta=0$.}
\label{figure4}      
\end{figure}
The integrals of  Eq. (\ref{EE18}) appeared already in Eqs. (\ref{PREE6}) and (\ref{PREE10}) and their explicit expressions, for the case at hand, is:
\begin{eqnarray}
I^{(5)}_{{\mathcal R}{\mathcal R}}(\ell, \ell_{\mathrm{D}}, n_{\mathrm{s}}) &=&  \int_{1}^{\infty} d w \frac{ w^{n_{\mathrm{s}}-5} }{ \, \sqrt{w^2 -1}} \, M^2_{\mathcal R}(w,\ell)\,e^{-2 (\ell/\ell_{\mathrm{D}})^2 w^2},
 \nonumber\\
I^{(6)}_{{\mathcal R}{\mathcal R}}(\ell, \ell_{\mathrm{D}}, n_{\mathrm{s}})&=&  \int_{1}^{\infty}d w \frac{ w^{n_{\mathrm{s}}-5} }{ \, \sqrt{w^2 -1}}  \cos{[ 2 \gamma_{\mathrm{A}} \ell w]}\, M^2_{\mathcal R}(w,\ell)\,e^{-2 (\ell/\ell_{\mathrm{D}})^2 w^2}.
 \label{EE20}
\end{eqnarray}
It is possible to change integration variable in Eqs. (\ref{EE20}). By positing $w= y^2 +1$ we do get\footnote{The change of variable $ w^2 = y^2 + 1$ is also possible and, 
in this particular case, will lead, of course, to the same results. In the case 
of other integrals, however, mathematically equivalent change of variables 
might lead to different evaluation times of the corresponding numerical integrals.}:
 \begin{eqnarray}
 {\mathcal I}^{(5)}_{{\mathcal R}{\mathcal R}}(\ell, \ell_{\mathrm{D}},n_{\mathrm{s}}) &=&  2 \int_{0}^{\infty} d y \frac{ (y^2+1)^{n_{\mathrm{s}}-5} }{ \, \sqrt{y^2 +2}} M^2_{\mathcal R}(y,\ell)\,e^{-2 (\frac{\ell^2}{\ell_{\mathrm{D}}^2}) (y^2 +1)^2},
 \nonumber\\
 {\mathcal I}^{(6)}_{{\mathcal R}{\mathcal R}}(\ell, \ell_{\mathrm{D}},n_{\mathrm{s}}) &=&  2 \int_{0}^{\infty}d y \frac{ (y^2 +1)^{n_{\mathrm{s}}-5} }{ \, \sqrt{y^2 +2}}  \cos{[ 2 \gamma_{\mathrm{A}} \ell (y^2 +1)]}\, M^2_{\mathcal R}(y,\ell)\, e^{-2 (\frac{\ell^2}{\ell_{\mathrm{D}}^2})(y^2 +1)^2}.
 \label{EE21}
\end{eqnarray}
The integrals of Eq. (\ref{EE21}) converge  rapidly and can be estimated, for instance, with numerical techniques; the final result can be expressed in a closed form  for $\ell > \ell_{1}$ as 
\begin{eqnarray}
&&G_{\ell}^{(\mathrm{EE})}= \overline{{\mathcal A}}^{(\mathrm{EE})} (\ell + \ell_{1})^{n_{\mathrm{s}}+1}\,\biggl\{ a_{\mathrm{E}} - b_{\mathrm{E}} \cos{[2 \gamma_{\mathrm{A}} (\ell + \ell_{1})]}\biggr\}\, e^{- 2 (\ell/\ell_{\mathrm{D}})^2},
\label{EE22a}\\
&& \overline{{\mathcal A}}^{(\mathrm{EE})} = 4.476 \times 10^{-4} \, (0.0354)^{n_{\mathrm{s}} -1} \, \biggl(\frac{{\mathcal A}_{{\mathcal R}}}{2.41\times 10^{-9}}\biggr) e^{-2 \epsilon_{\mathrm{re}}} \, (\mu\mathrm{K})^2,
\label{EE22}\\
&& a_{\mathrm{E}}= 1.67,\qquad b_{\mathrm{E}}=3.38, \qquad \ell_{1} = 65,
\label{EE22b}
\end{eqnarray}
 where  $\ell_{1}$ appears because  the analytic derivations of the previous sections assume a large-order expression for the spherical Bessel functions. 
Concerning Eq. (\ref{EE22}) few comments are in order:
\begin{itemize}
\item{} Eq. (\ref{EE22}) assumes the simplified treatment of reionization which has been spelled out in Eqs. (\ref{TP18})--(\ref{TP19}) and which is less accurate for low multipoles (i.e. in the region of the reionization peaks) than for large multipoles;
\item{} $a_{\mathrm{E}}$ and $b_{\mathrm{E}}$ are, respectively,  the form factors coming from the integral  ${\mathcal I}^{(5)}_{{\mathcal R}{\mathcal R}}(\ell, \ell_{\mathrm{D}},n_{\mathrm{s}})$ and from ${\mathcal I}^{(6)}_{{\mathcal R}{\mathcal R}}(\ell, \ell_{\mathrm{D}},n_{\mathrm{s}})$;
\item{} $\ell_{\mathrm{D}}$ (i.e. the diffusion damping scale) is given 
by Eq. (\ref{TP3}) and has been also discussed prior to Eq. (\ref{B1}) in connection with 
the estimate of Silk damping;
\item{} the numerical value of ${\mathcal A}^{(\mathrm{EE})}$  follows from the pivotal
value  of $k_{\mathrm{p}}$ (i.e. $0.002\, \mathrm{Mpc}^{-1}$) and by 
computing $k_{0}$ from the (comoving) angular diameter distance of Eq. (\ref{TP4});
\end{itemize}
In terms of the values of the cosmological parameters obtainable from the 
WMAP 5yr best fit \cite{WMAP5a,WMAP5b,WMAP5c}
\begin{equation}
(\omega_{\mathrm{M}},\, \omega_{\mathrm{c}}, \, \omega_{\mathrm{b}}, \omega_{\Lambda},\, h_{0},\,n_{\mathrm{s}},\, \epsilon_{\mathrm{re}}) \equiv 
(0.1326,\, 0.1099,\, 0.02273,\,0.385,\,0.719,\, 0.963,\,0.087),
\label{Par1}
\end{equation}
the values of the derived parameters of Eqs. (\ref{TP3})--(\ref{TP6}) are\footnote{Different best-fit parameters, obtained 
by combining CMB data with other data sets (e.g. \cite{SN,LSS}) lead to different values of the derived parameters which can be 
however computed always using the general formulae of the previous sections. }
\begin{eqnarray}
&& [z_{*}, \, c_{\mathrm{sb}}(z_{*}), \, D_{\mathrm{A}}(z_{*}), \, \ell_{\mathrm{A}}, \, 
\ell_{\mathrm{D}},\ \ell_{\mathrm{t}}, \, \ell_{\mathrm{S}}] = 
\nonumber\\
&&[1099.5,\,0.451,\, 14110.8 \, \mathrm{Mpc},\,301.578,\, 1422,\, 1211,\, 922].
\label{Par2}
\end{eqnarray}
The results for the polarization autocorrelations are illustrated in Fig. \ref{figure3}. In the 
plot at the left   ${\mathcal I}^{(5)}_{{\mathcal R}{\mathcal R}}(\ell, \ell_{\mathrm{D}},n_{\mathrm{s}})$ and $ {\mathcal I}^{(6)}_{{\mathcal R}{\mathcal R}}(\ell, \ell_{\mathrm{D}},n_{\mathrm{s}})$ are reported, respectively, with the full and with the dashed lines. In the plot at the right of Fig. \ref{figure3}
the analytic result for $G_{\ell}^{(\mathrm{EE})}$ (dashed line) 
is compared with the WMAP 5yr best fit (full line) holding for exactly the same 
set of parameters (i.e. Eq. (\ref{Par1})). 
In Fig. \ref{figure4}  the analytic results for the magnetized polarization 
autocorrelations are compared with the numerical results.
In both plots of Fig. \ref{figure4} the numerical result is reported with the 
full line. The dashed line denotes the analytical approximation (plot at the 
left). The dot-dashed line denotes the WMAP 5yr best fit (for the same value of 
cosmological parameters).  In both plots the correlation angle has been chosen 
as $\beta =0$.  In summary we can therefore say that Figs. \ref{figure3} and 
\ref{figure4} show that, in spite of the different approximations, the analytic 
result is in fair agreement with the numerical one.
Finally, the numerical results illustrated in Figs. \ref{figure4} and \ref{figure5} 
follow from an improved version of the approach already mentioned 
in the introduction \cite{max4,max5} which is based on a modification of CMBFAST \cite{cmbf1,cmbf2} (which 
is, in turn, a modified version of Cosmics \cite{cosm1,cosm2}). 
\subsection{TT and TE angular power spectra}
\label{SUBS52}
\begin{figure}[!ht]
\centering
\includegraphics[height=6cm]{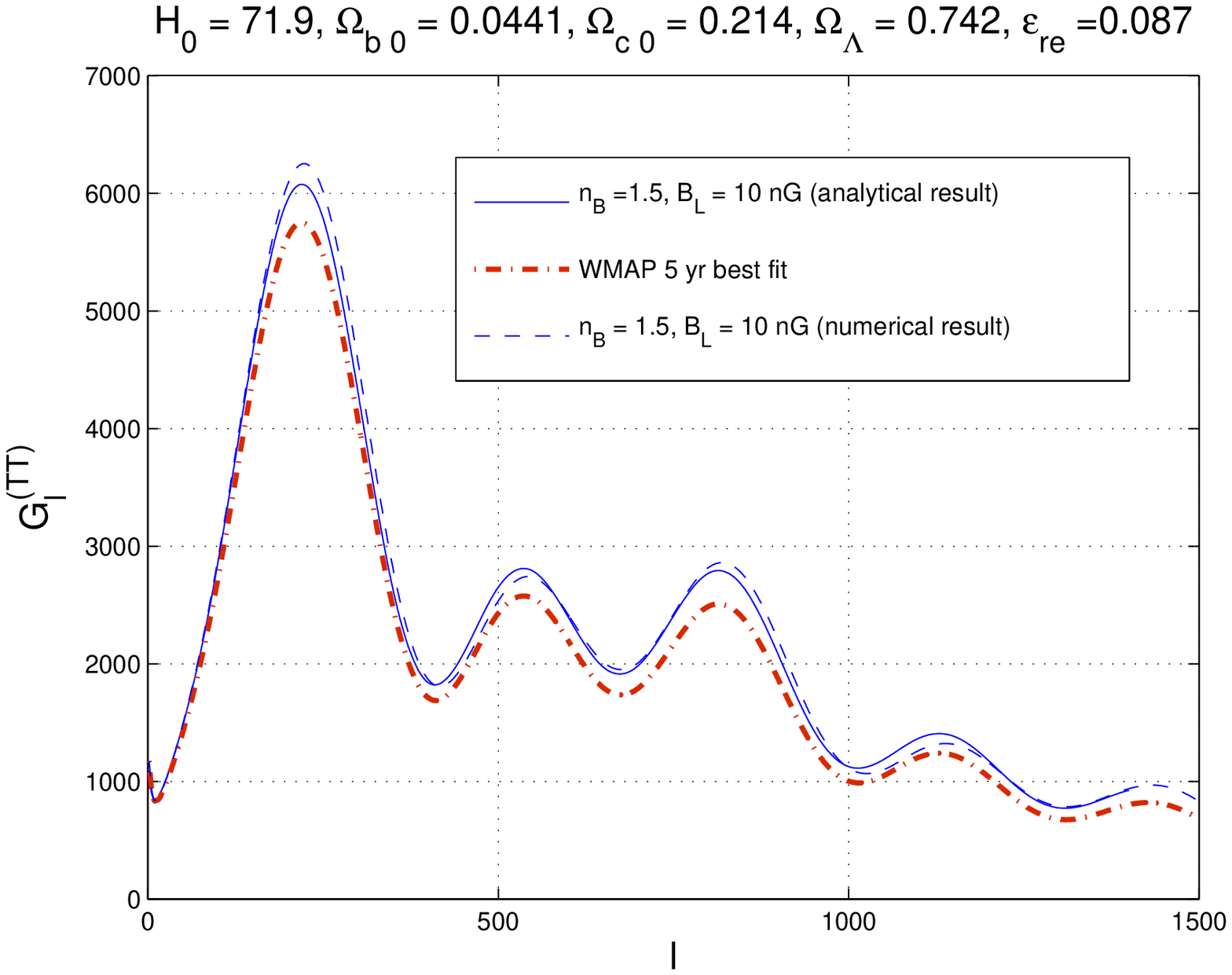}\includegraphics[height=6cm]{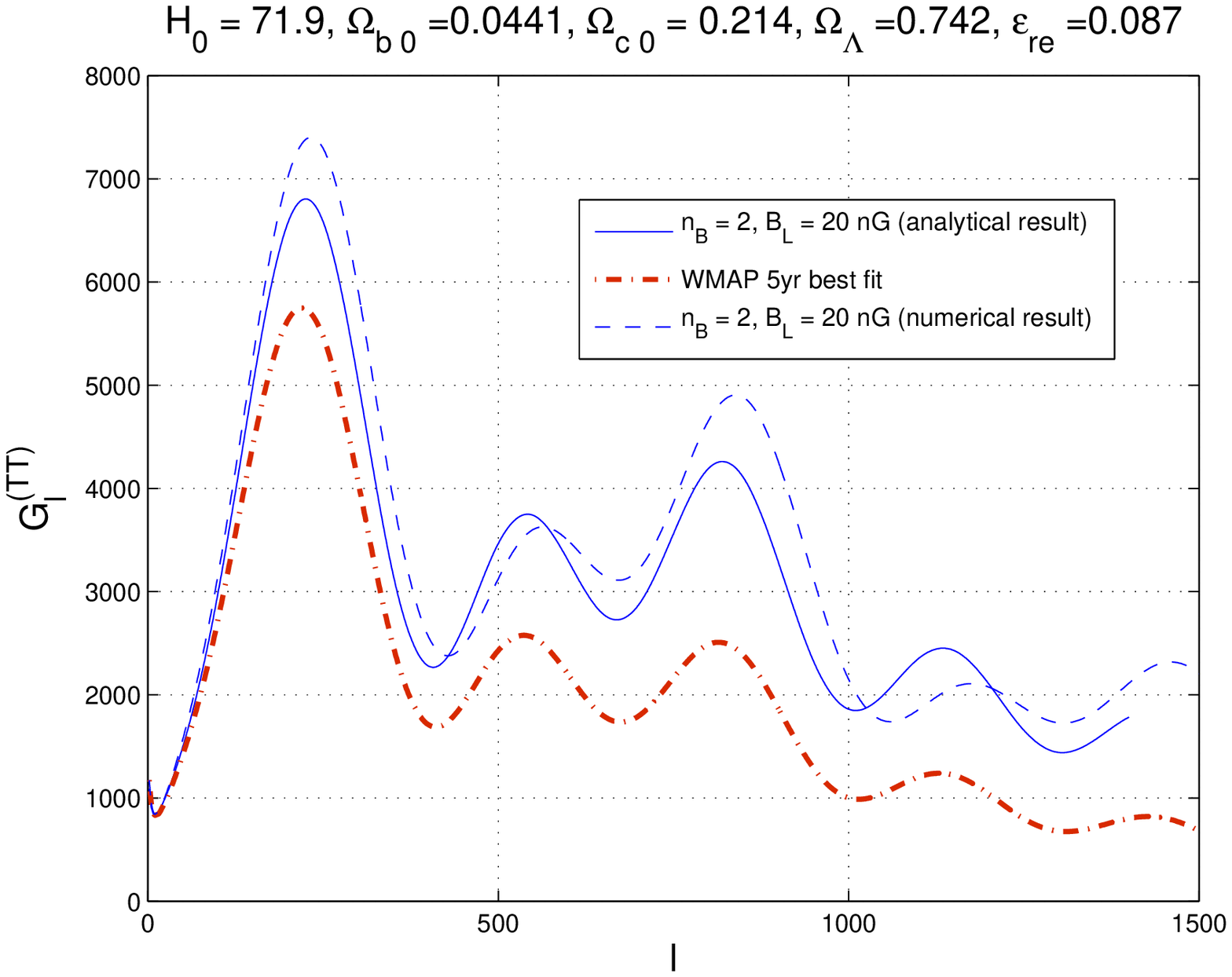}
\caption[a]{The semi-analytic results for the temperature autocorrelations are illustrated in the case $\beta=0$.}
\label{figure5}      
\end{figure}

The TT and TE correlations share similar features from the point of view of the 
analytic results discussed here. The periodicities of the TT and TE angular power 
spectra arise as the weighted interference of the periodicities of the monopole and of the dipole of the radiation field.  The TT correlations, have been partially discussed with a similar semi-analytic method in \cite{max4} (first paper) and corroborated by subsequent numerical estimates (second paper of \cite{max4}). The improved 
analytical understanding developed in the present paper allows for a better assessment of the accuracy of the results. In Fig. \ref{figure5} the results for the TT angular power spectra 
are illustrated. In both plots the full lines denote the analytical estimate while the dashed lines represent the numerical result.  In Fig. \ref{figure5}, from left to right, the magnetic field intensity 
and the spectra index increase. In both plots of Fig. \ref{figure5} the dot-dashed lines denote the WMAP 5yr best fit.
The results illustrated in Fig. \ref{figure5} are representative of a general trend which has been observed also in other cases and can be summarized as follows:
\begin{itemize}
\item{} the analytic result for the TT correlations stemming from the basic integrals 
studied in this paper always underestimates the numerical result;
\item{} the analytic result becomes progressively inaccurate 
as the field strength increases above $10$ nG. 
\end{itemize}
Various other examples show that the polarization autocorrelations (i.e. the EE angular power spectra) are better captured by the analytical results, as already shown in Figs. 
\ref{figure3} and \ref{figure4}. 
The temperature-polarization cross-correlations (i.e. the TE angular power spectra) share the same levels of accuracy 
of the TT correlations and they are illustrated in Fig. \ref{figure6}. 
As in Fig. \ref{figure5} the dot-dashed lines denote the best fit to the WMAP 5yr data alone.
For $B_{\mathrm{L}} \leq 10 $ n G 
the analytic results are rather accurate for the TT, TE and EE 
angular power spectra. In the case $B_{\mathrm{L}} > 10 $ nG 
the results become progressively less accurate in the case of the 
TT and TE correlations but  remain reasonable for the EE 
autocorrelations.  

\subsection{Envelopes and wiggles}
\label{SUBS53}
Having tested the accuracy of the analytical results, the handiness of the approach developed in the present investigation resides in the determination of the scaling properties of the various correlation functions.  
\begin{figure}[!ht]
\centering
\includegraphics[height=6cm]{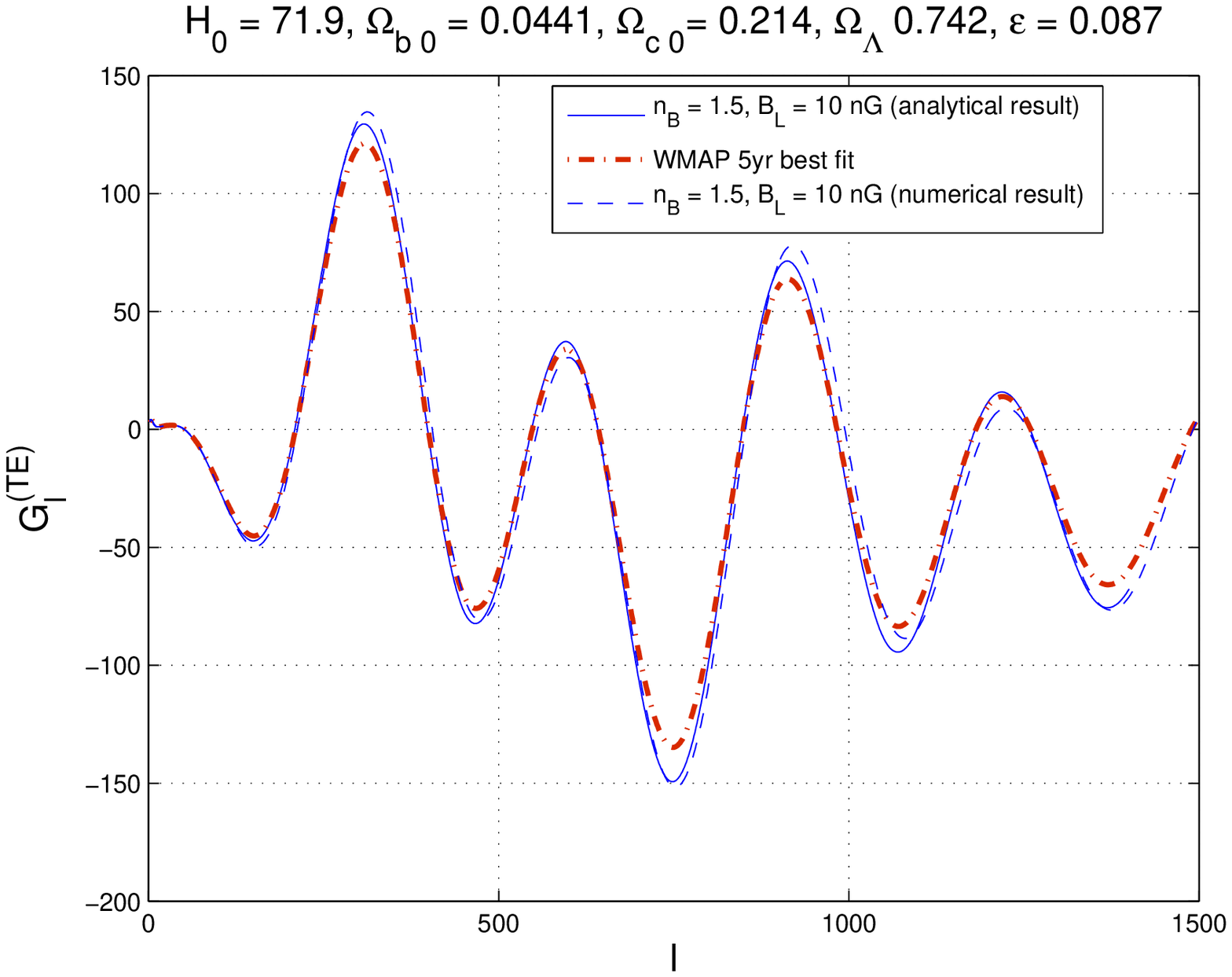}\includegraphics[height=6cm]{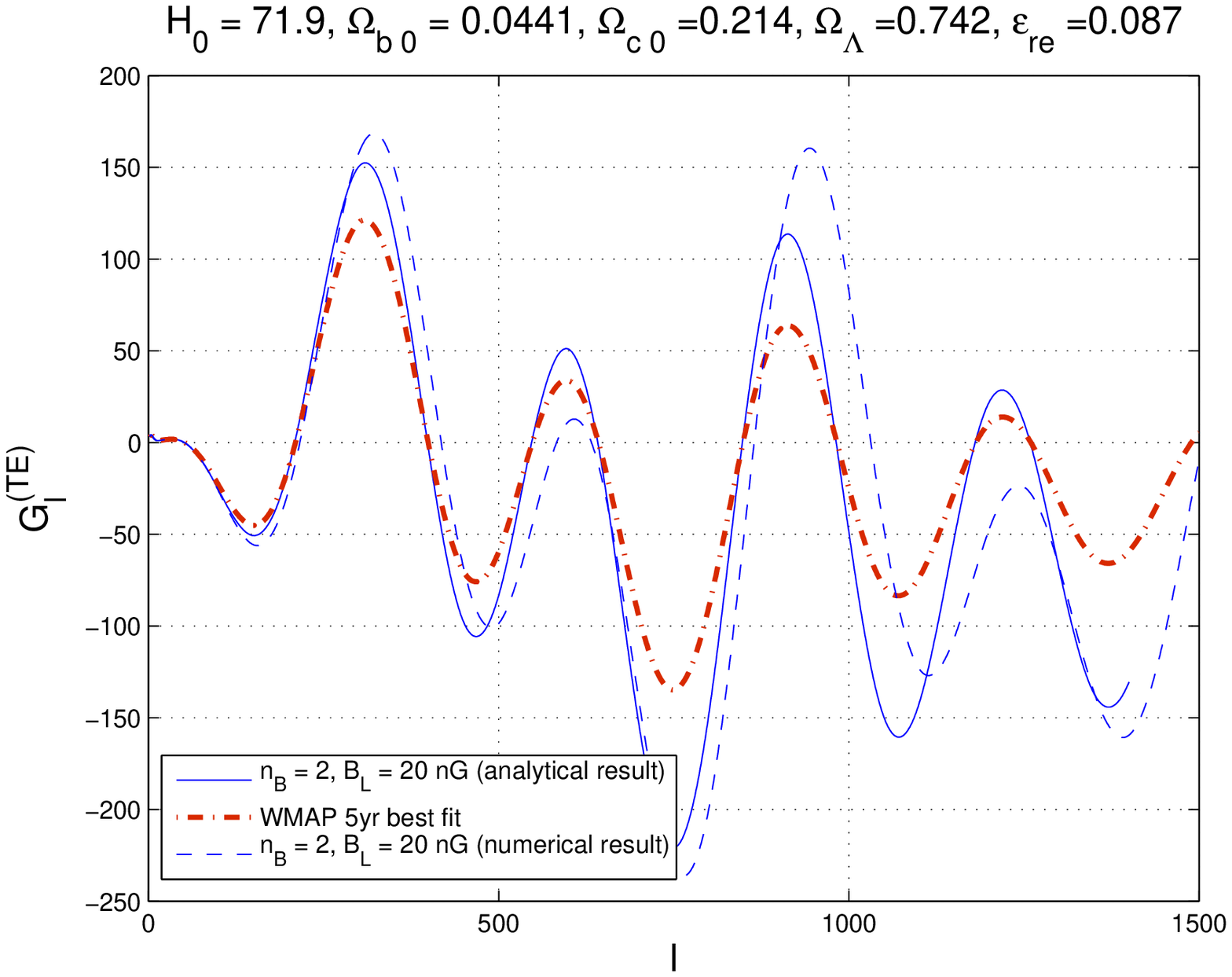}
\caption[a]{The semi-analytic results for the temperature-polarization cross-correlations 
are illustrated always in the case $\beta=0$.}
\label{figure6}      
\end{figure}
\begin{figure}[!ht]
\centering
\includegraphics[height=6cm]{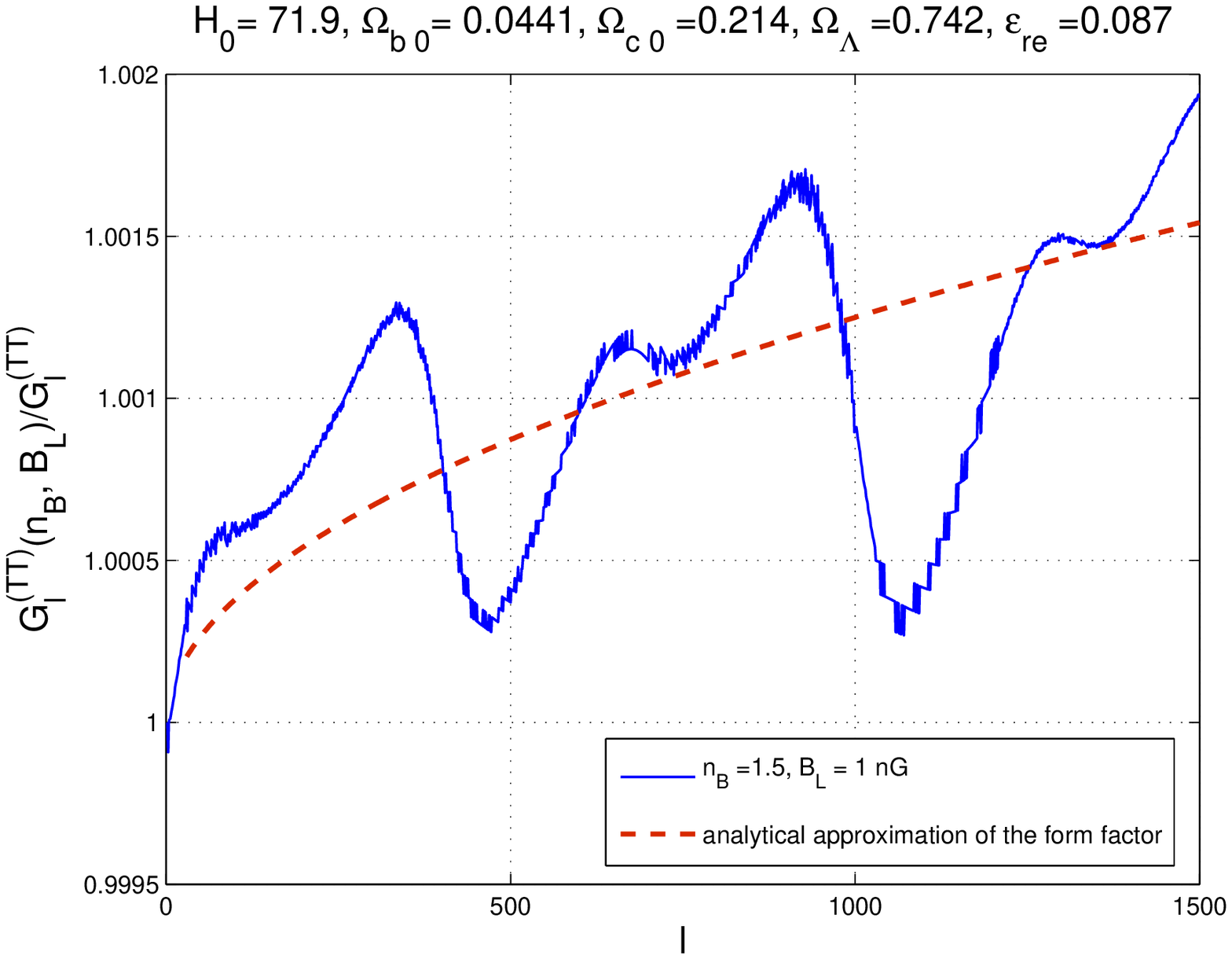}\includegraphics[height=6cm]{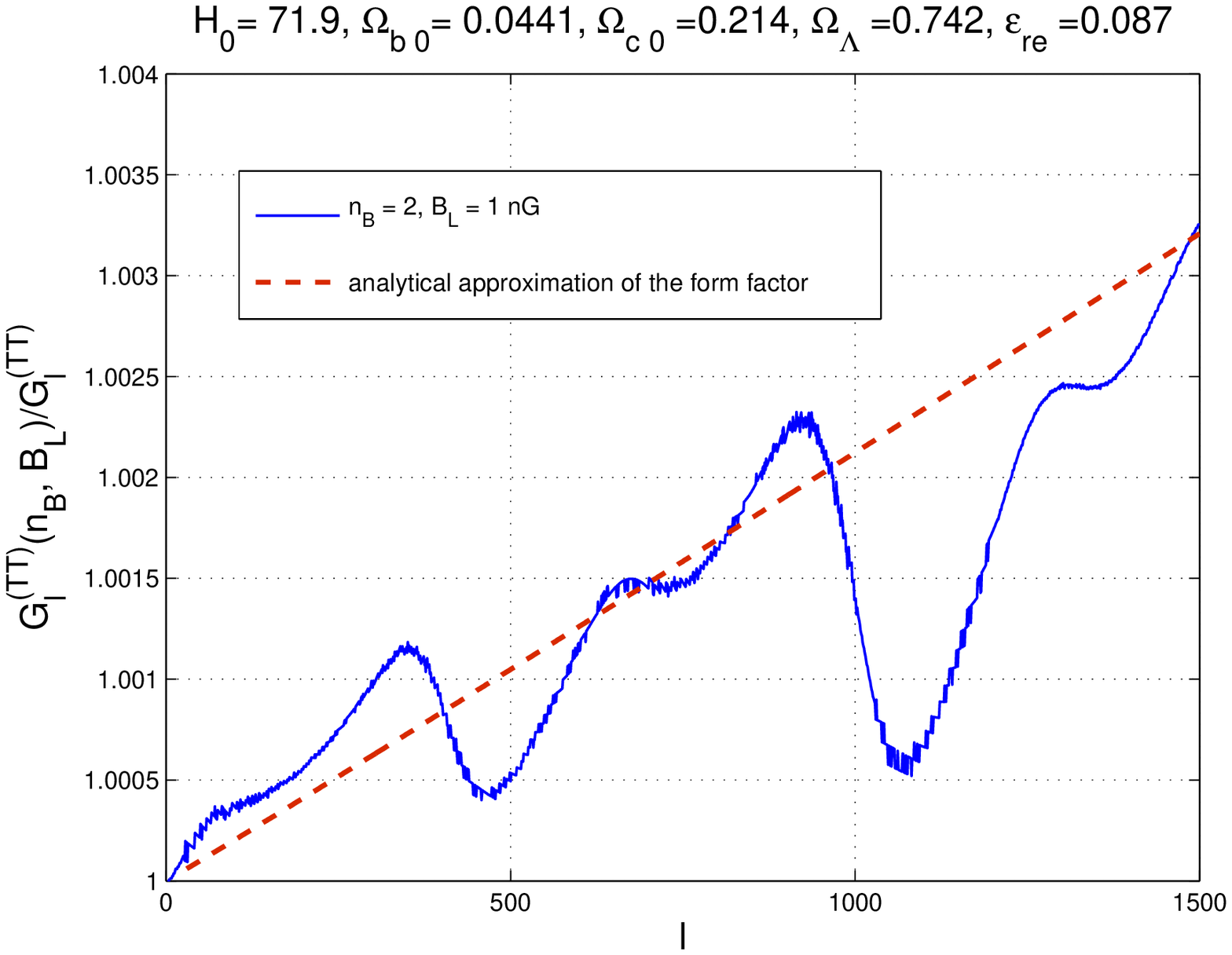}
\includegraphics[height=6cm]{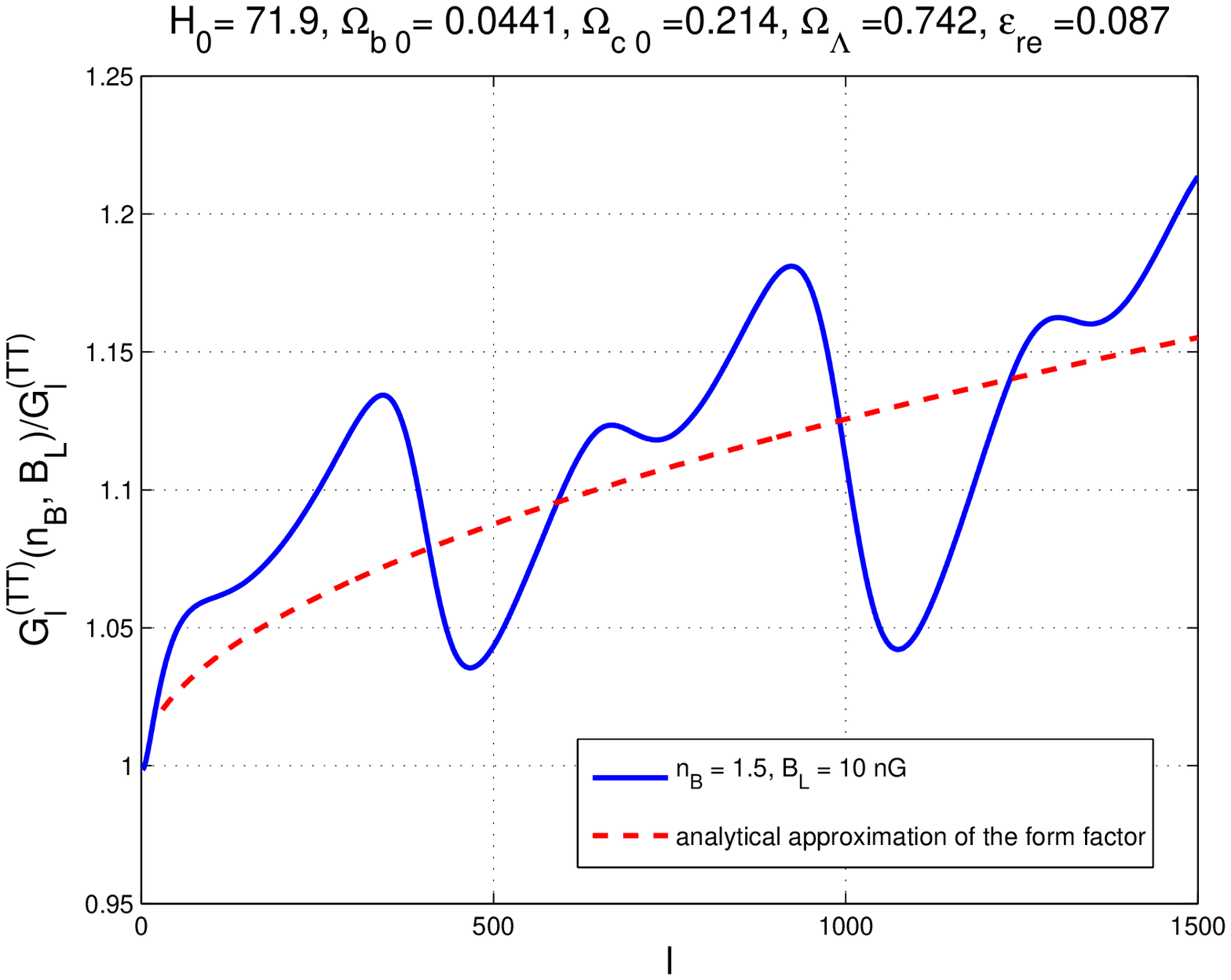}\includegraphics[height=6cm]{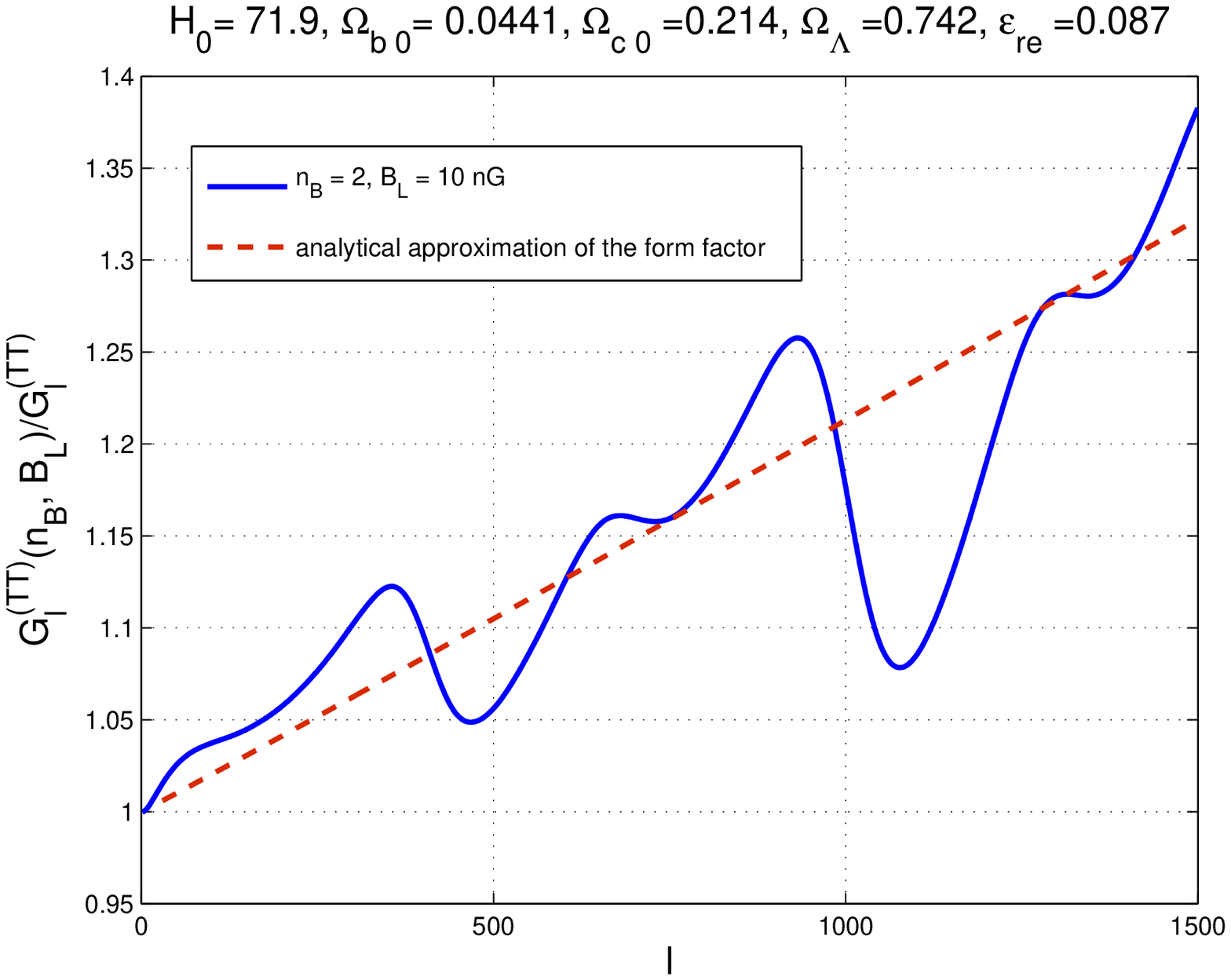}
\includegraphics[height=6cm]{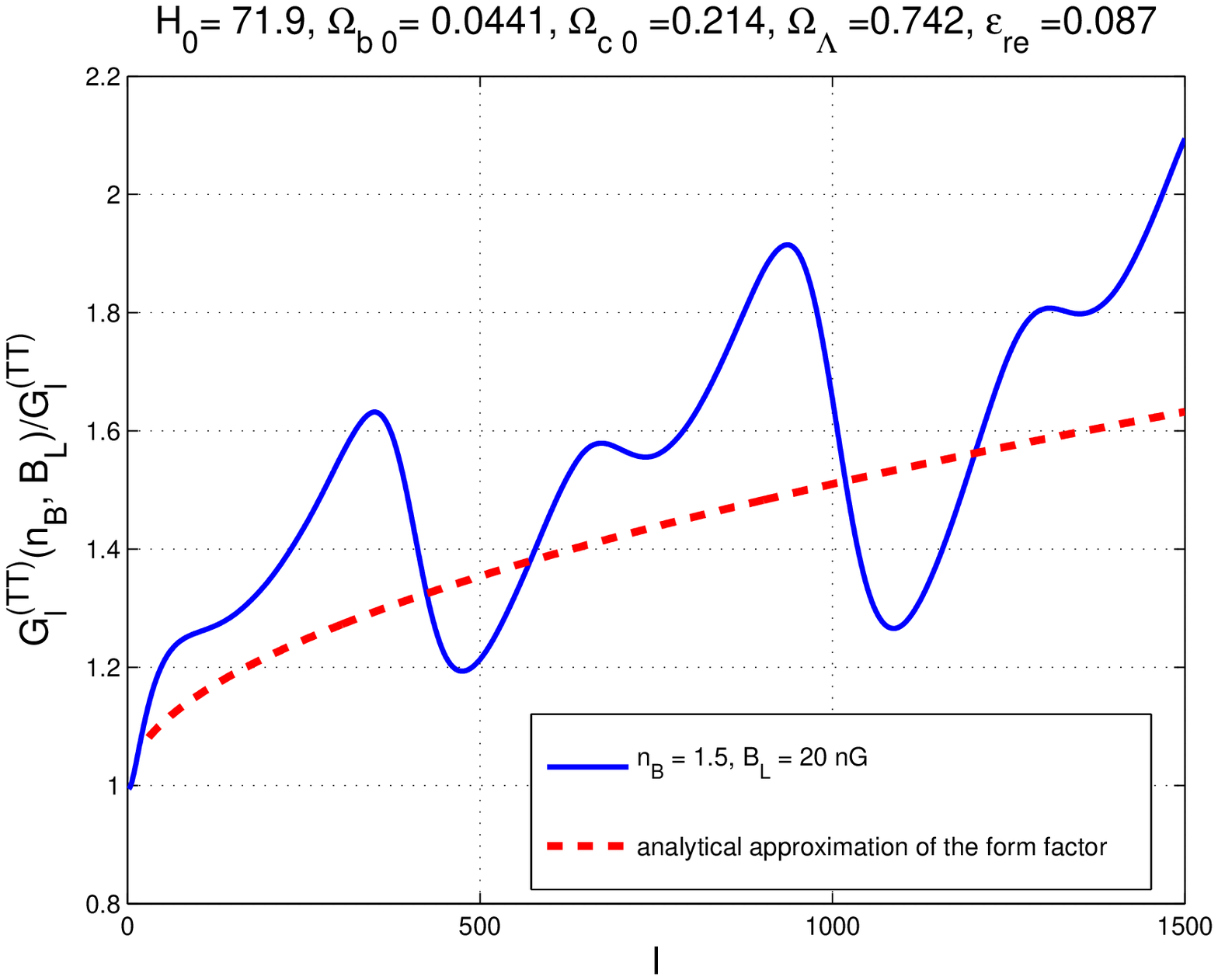}\includegraphics[height=6cm]{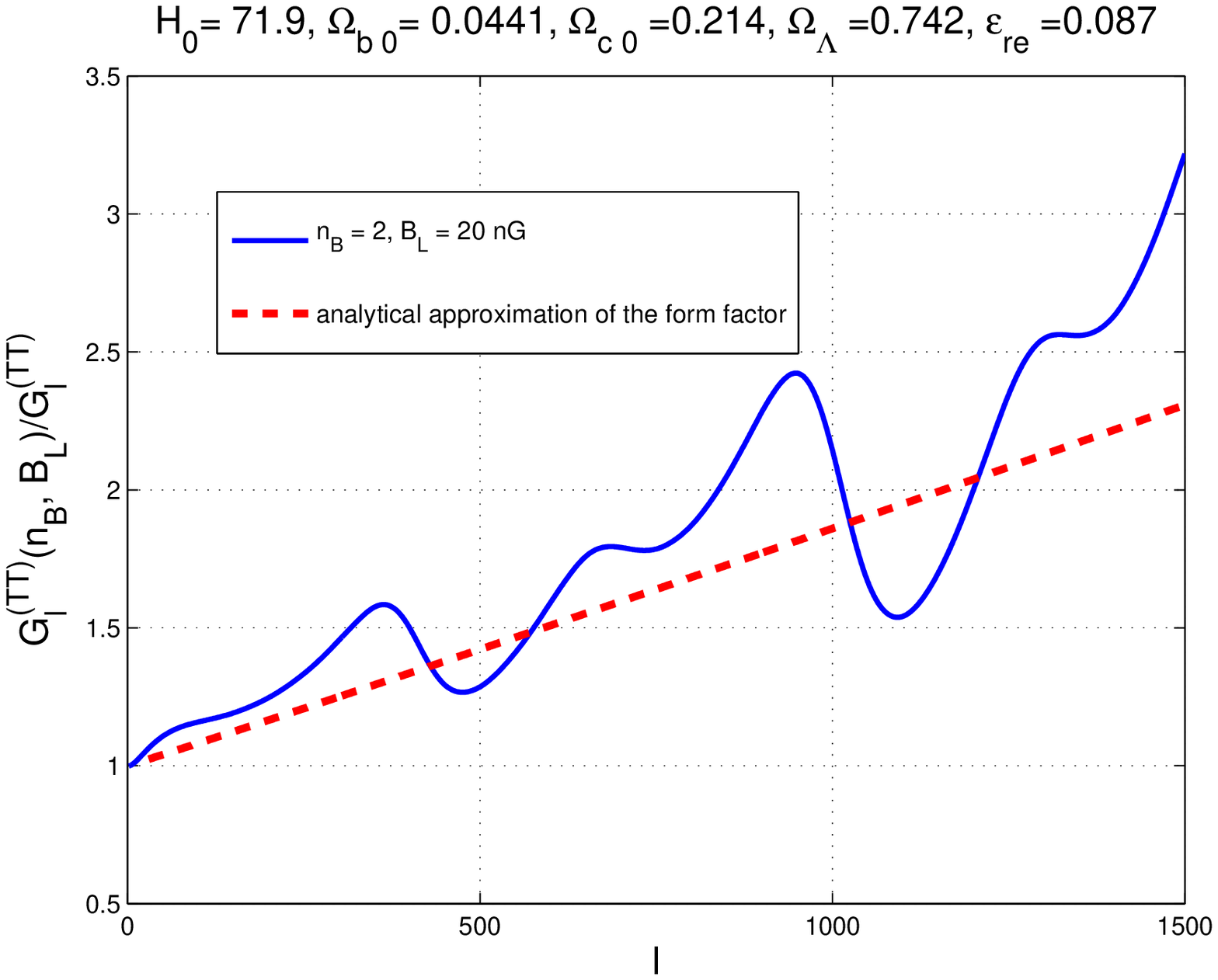}
\caption[a]{The magnetized form factor from the TT correlations for different values of the magnetic field background. The correlation angle is fixed to $\beta=0$.}
\label{figure7}      
\end{figure}
To proceed in this direction, the idea is to compute numerically the ratios 
\begin{equation}
R_{\ell}^{(\mathrm{TT})}= \frac{G_{\ell}^{(\mathrm{TT})}(n_{\mathrm{B}}, B_{\mathrm{L}})}{G_{\ell}^{(\mathrm{TT})}},\qquad  R_{\ell}^{(\mathrm{EE})} = \frac{G_{\ell}^{(\mathrm{EE})}(n_{\mathrm{B}}, B_{\mathrm{L}})}{G_{\ell}^{(\mathrm{EE})}},
\label{ratios}
\end{equation}
where, by definition, $G_{\ell}^{(\mathrm{TT})}$ and $G_{\ell}^{(\mathrm{EE})}$ 
denote the angular power spectra in the absence of ambient magnetic field.
The same procedure can be carried on also for the TE correlations. 
However, since the TE corrrelations are not positive definite and pass 
through zero, the resulting plots are not as revealing as the ones 
obtainable from the TT and EE angular power spectra.
Figures \ref{figure7}  and \ref{figure8}  illustrate, respectively, $R_{\ell}^{(\mathrm{TT})}$ and $R_{\ell}^{(EE)}$  for different values of the magnetic field intensities and for 
different values of the spectral indices. 
\begin{figure}[!ht]
\centering
\includegraphics[height=6cm]{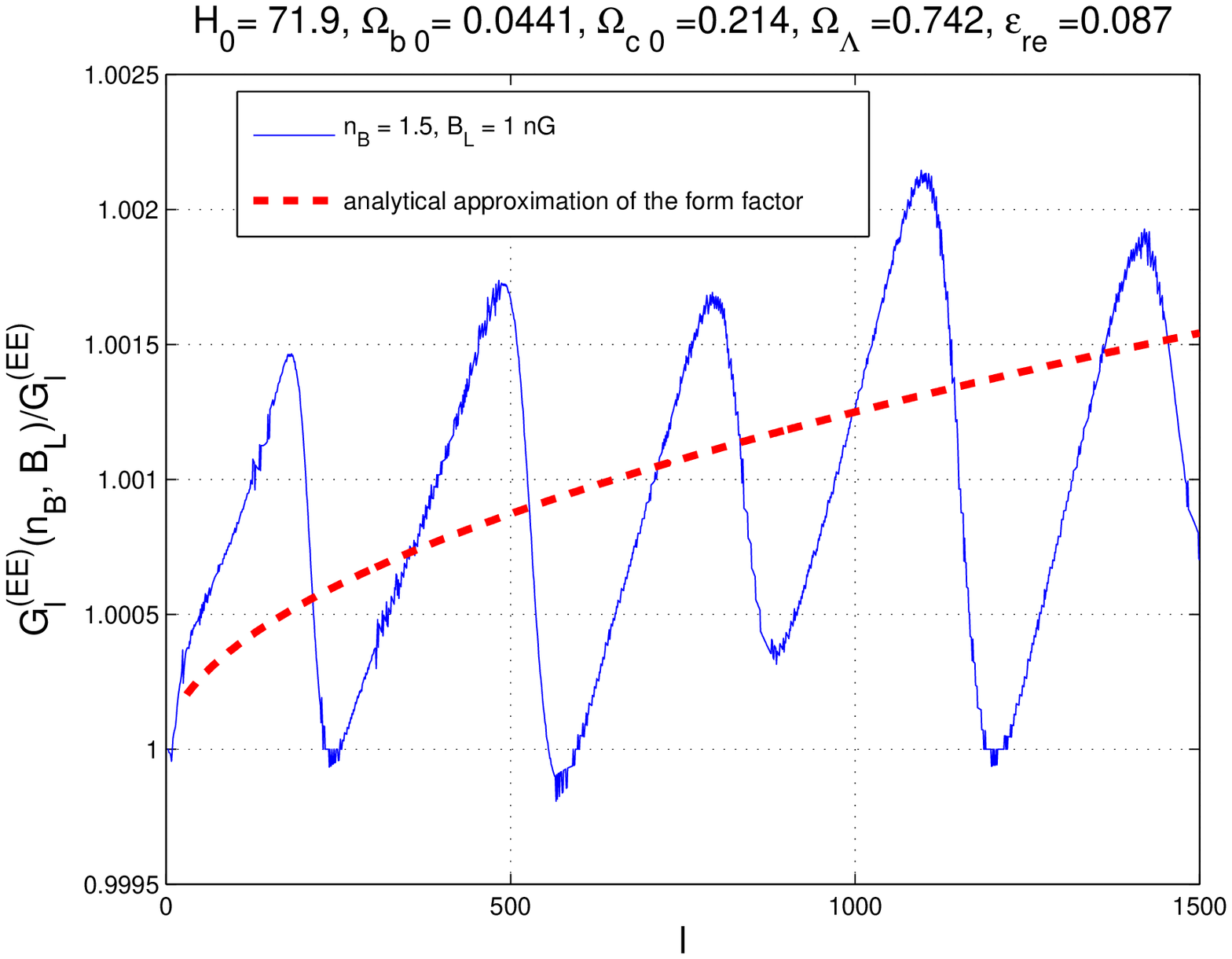}\includegraphics[height=6cm]{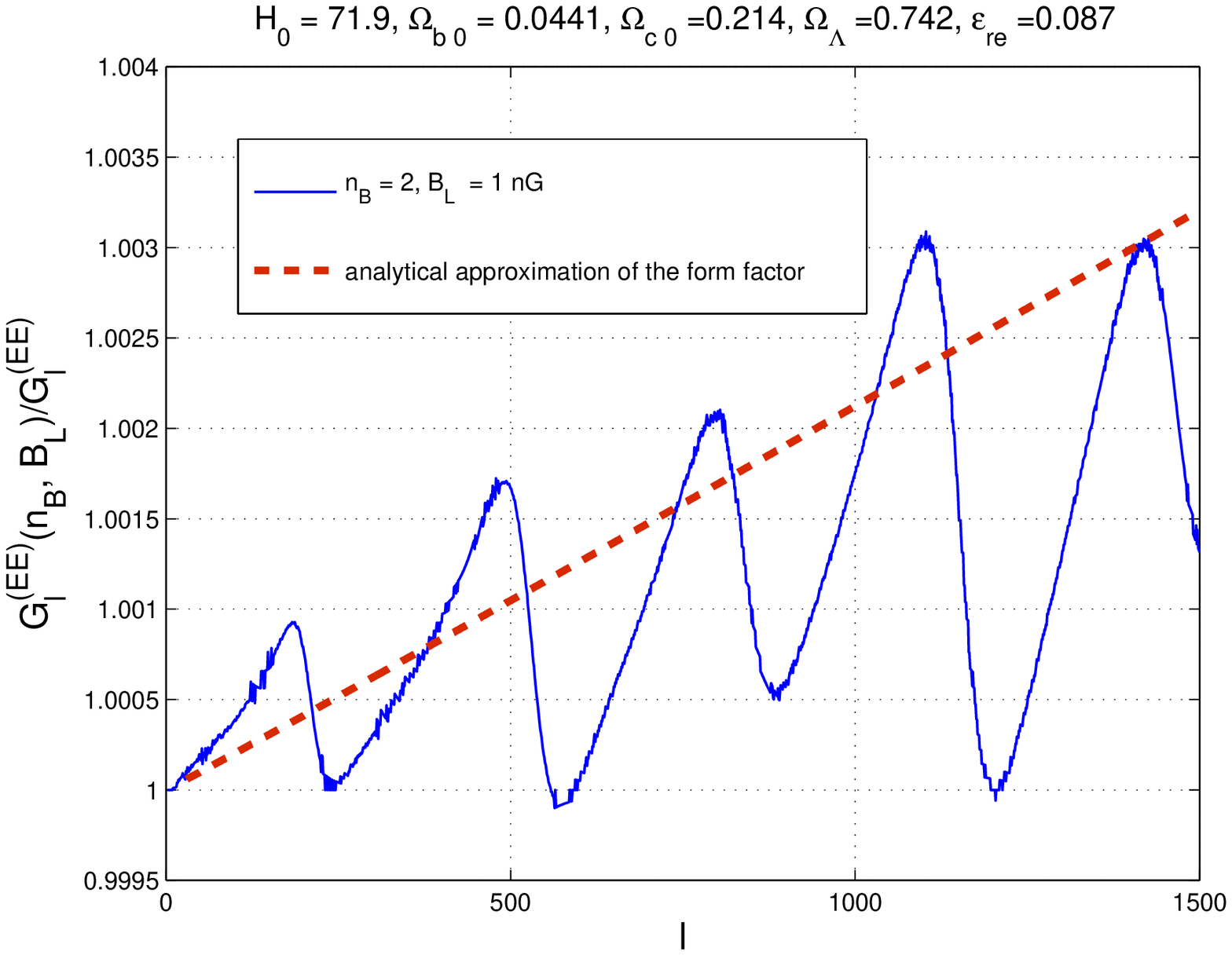}
\includegraphics[height=6cm]{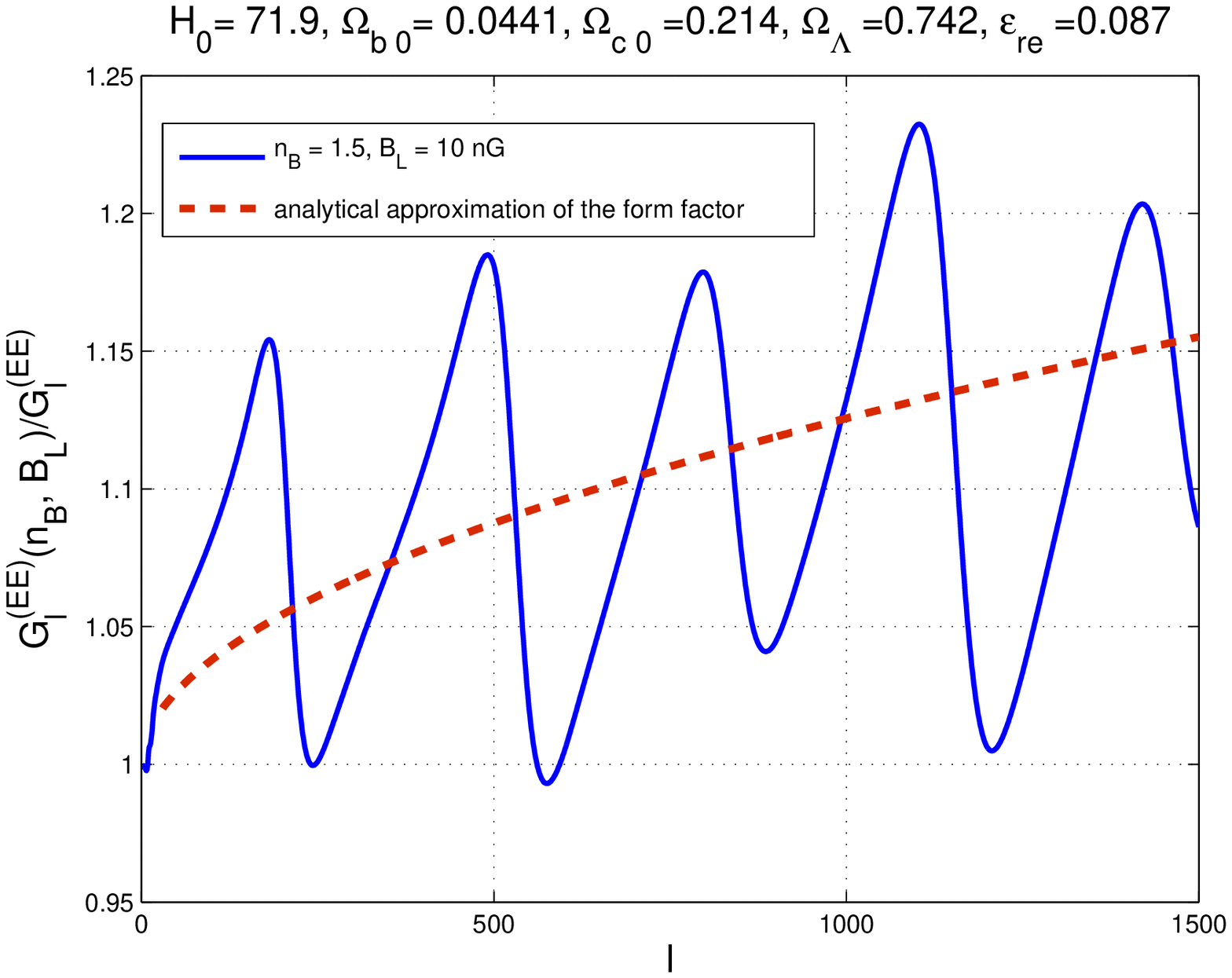}\includegraphics[height=6cm]{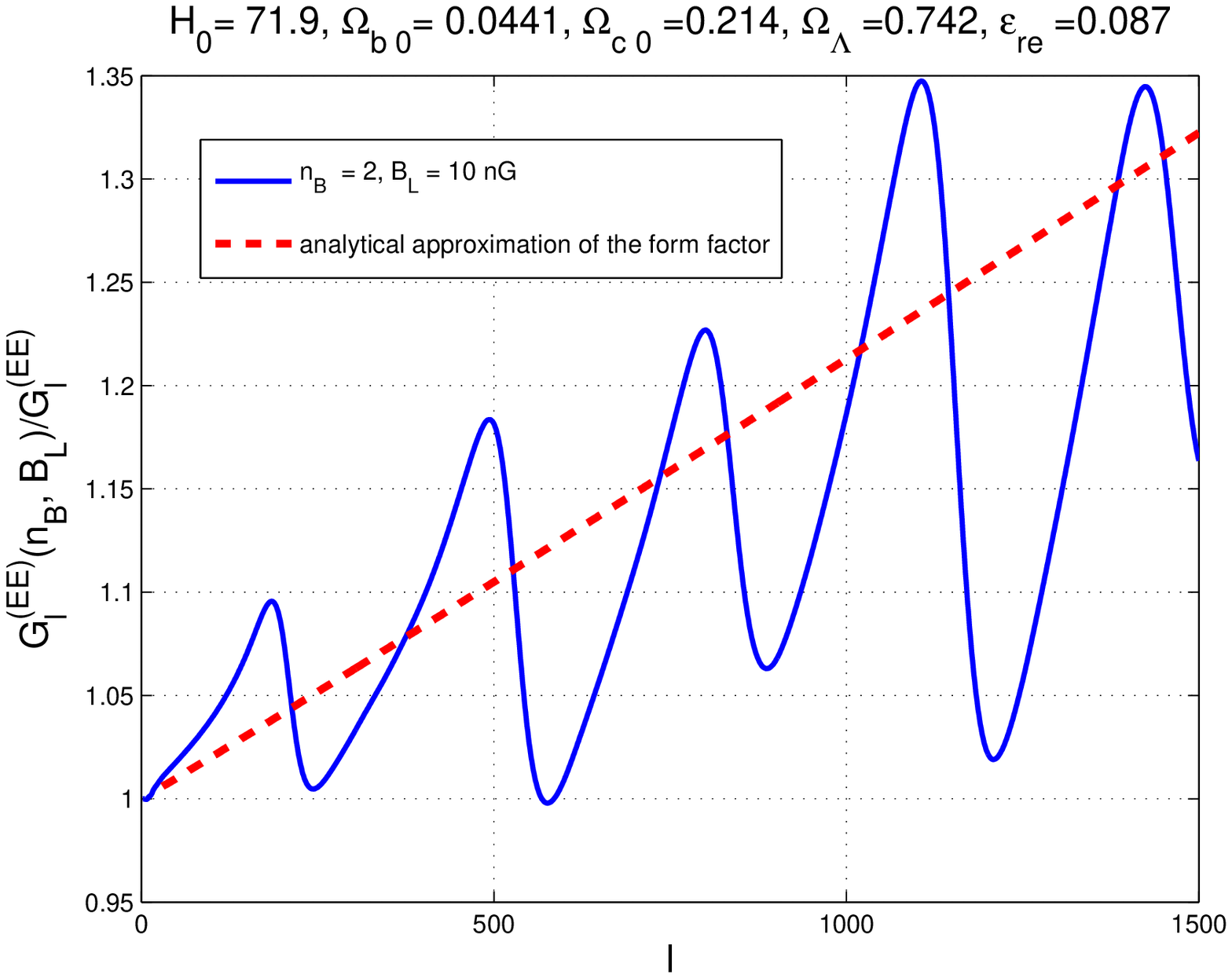}
\includegraphics[height=6cm]{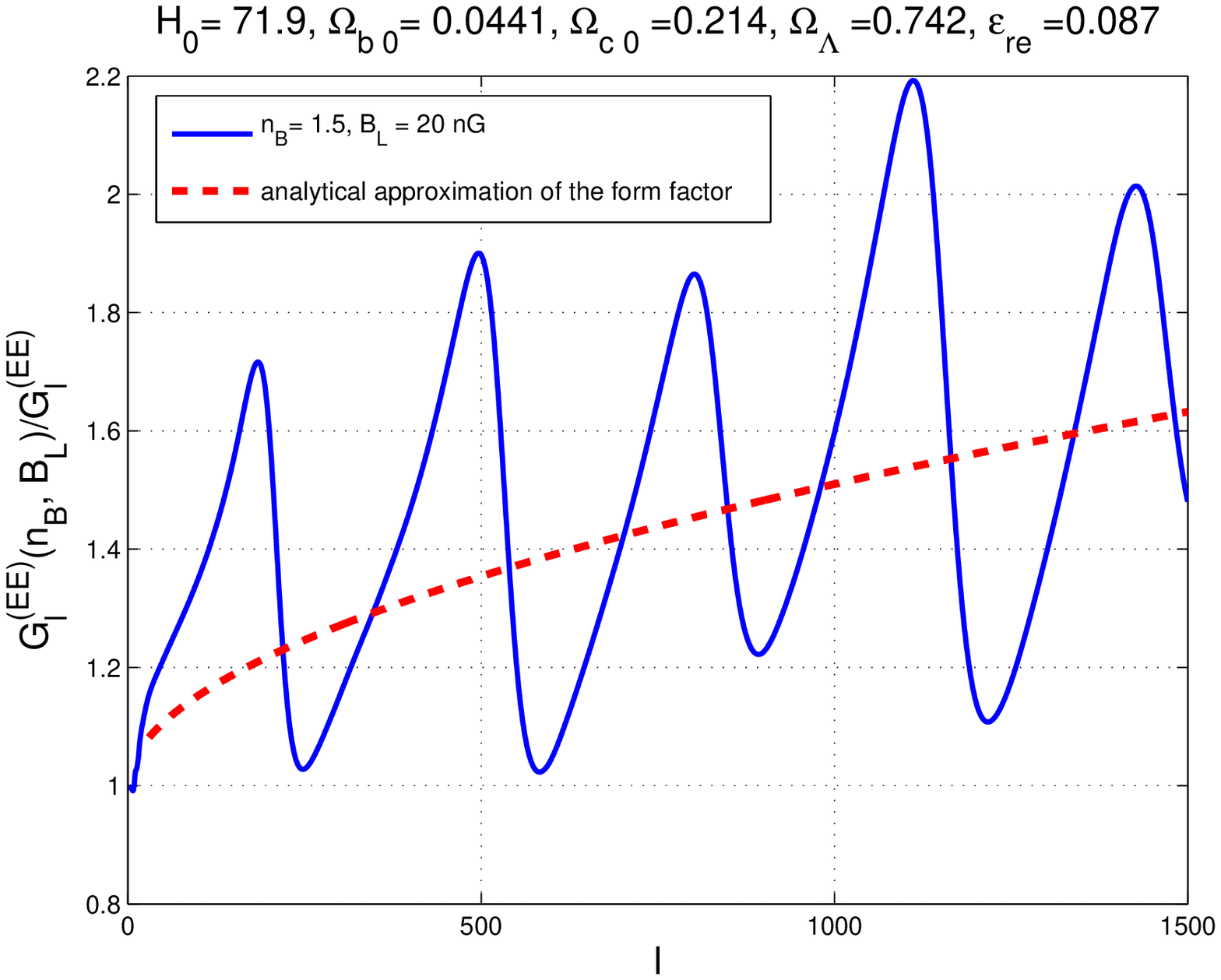}\includegraphics[height=6cm]{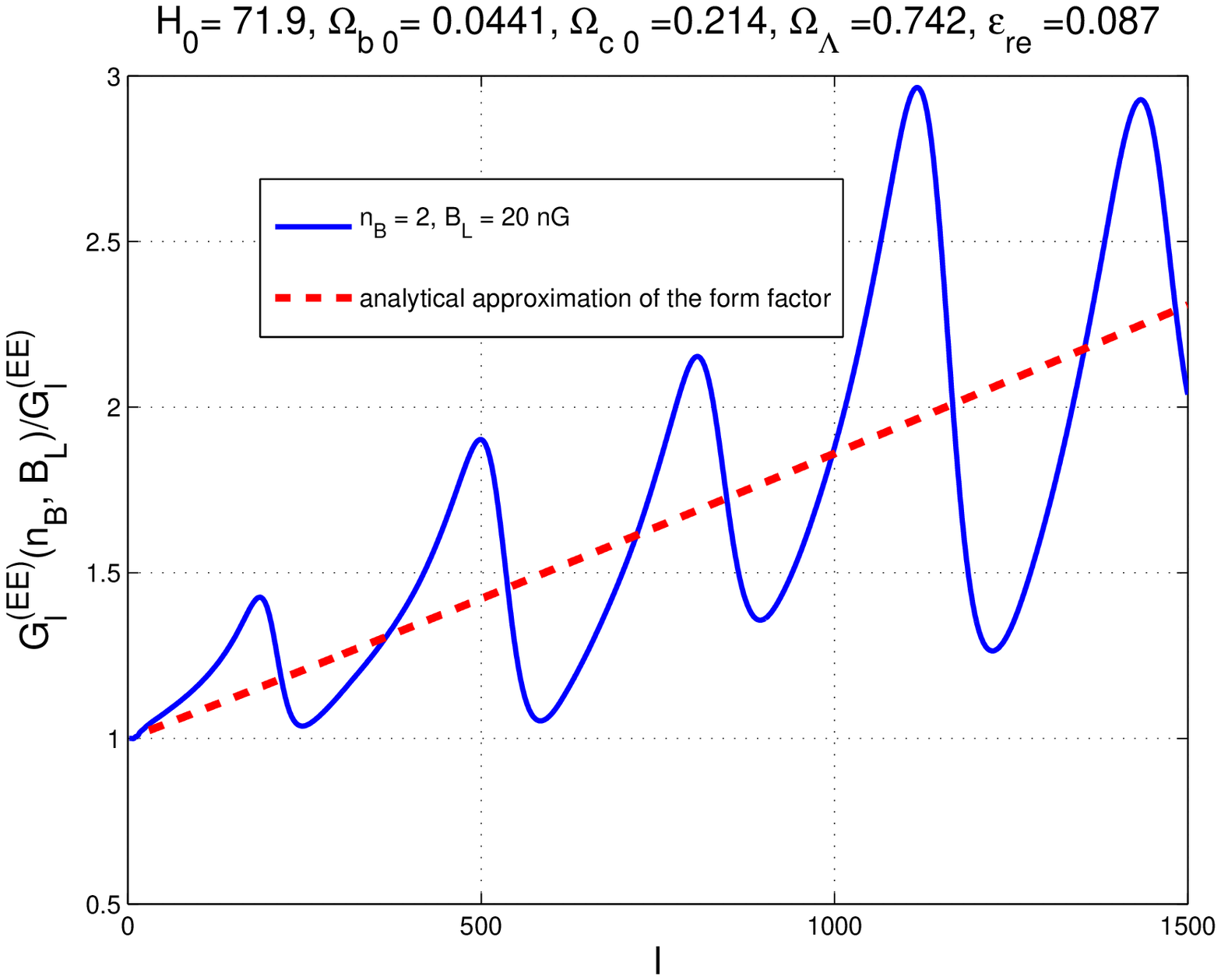}
\caption[a]{The magnetized form factor from the EE correlations for different values of the magnetic field background. The correlation angle is fixed to $\beta=0$.}
\label{figure8}      
\end{figure}
The ratios $R^{(\mathrm{TT})}$ and $R^{(\mathrm{EE})}$ represent an
effective numerical diagnostic of the possible influence of a putative magnetic field.
Indeed the diffusive scales, the thickness of the last scattering surface 
the optical depth at reionization are only mildly sensitive to the presence 
of an ambient magnetic field. The analytic structure of the angular power 
spectra computed in section \ref{sec4} suggests the possibility of factorizing 
the  effects of a putative magnetic field into an appropriate 
form factor which is, incidentally, illustrated in Figs. \ref{figure7} and \ref{figure8}
with a dashed line. On a qualitative ground the form factor 
represents, in some sense, the average of $R^{(XY)}_{\ell}$. 
On a more quantitative ground the results of section \ref{sec4} suggest 
that $R_{\ell}^{(XY)}$ can be factorized into the product of a non-oscillating 
factor (i.e.  $\overline{F}_{\ell}(n_{\mathrm{B}},  B_{\mathrm{L}})$) 
 and of an oscillating contribution (i.e.  ${\mathcal O}_{\ell}^{(XY)}(n_{\mathrm{s}},\,n_{\mathrm{B}},\, {\mathcal A}_{{\mathcal R}}, B_{\mathrm{L}})$). Therefore, within the notations 
followed in the present paper, we will have that $R_{\ell}^{(XY)}$ can be written as
\begin{equation}
R_{\ell}^{(XY)}(n_{\mathrm{B}},  B_{\mathrm{L}}, n_{\mathrm{s}}, {\mathcal A}_{{\mathcal R}}) 
 = \overline{F}_{\ell}(n_{\mathrm{B}}, B_{\mathrm{L}})\, {\mathcal O}_{\ell}^{(XY)}(n_{\mathrm{s}},\,n_{\mathrm{B}},\, {\mathcal A}_{{\mathcal R}}, \, B_{\mathrm{L}}).
\label{FF2a}
\end{equation}
where $X,Y = \mathrm{T}, \mathrm{E}$. The result for the magnetic form factor is
\begin{equation}
\overline{F}_{\ell}(n_{\mathrm{B}}, B_{\mathrm{L}}) = 1 +a_{1}  
\biggl(\frac{B_{\mathrm{L}}}{\mathrm{nG}}\biggr)^4 \, {\mathcal J}_{1}(n_{\mathrm{s}}, n_{\mathrm{B}}, \ell)  + a_{2} \biggl(\frac{B_{\mathrm{L}}}{\mathrm{nG}}\biggr)^2 
{\mathcal J}_{2}(n_{\mathrm{s}}, n_{\mathrm{B}}, \beta, \ell).
\label{FF7}
\end{equation}
Within the set of parameters given by Eq. (\ref{Par1}) the constants 
$a_{1}$ and $a_{2}$ are given by  
\begin{equation}
a_{1} = 1.393 \times 10^{-7},\qquad a_{2} = 1.952 \times 10^{-3},
\label{FF7a}
\end{equation}
while the functions ${\mathcal J}_{1}(n_{\mathrm{s}}, n_{\mathrm{B}}, \ell)$
and ${\mathcal J}_{2}(n_{\mathrm{s}}, n_{\mathrm{B}}, \beta, \ell)$ can be expressed as 
\begin{eqnarray}
&& {\mathcal J}_{1}(n_{\mathrm{s}}, n_{\mathrm{B}}, \ell)= 
\biggl(\frac{k_0}{k_{\rm L}}\biggr)^{2(n_{\mathrm{B}}-1)} \biggl(\frac{k_0}{k_{\rm p}}\biggl)^{(1-n_{\mathrm{s}})} \biggl(\frac{2 \ell}{\ell_{\mathrm{B}}}\biggr)^{2 n_{\mathrm{B}} -n_{\mathrm{s}} -1} \Sigma_{1}(n_{\mathrm{s}}, n_{\mathrm{B}}),
\label{FF8}\\
&& {\mathcal J}_{2}(n_{\mathrm{s}}, n_{\mathrm{B}}, \beta, \ell)=\cos{\beta} \biggl(\frac{k_0}{k_{\rm L}}\biggr)^{(n_{\mathrm{B}}-1)} \biggl(\frac{k_0}{k_{\rm p}}\biggl)^{(1-n_{\mathrm{s}})/2}\, \biggl(\frac{2\ell}{\ell_{\mathrm{B}}}\biggr)^{\frac{2 n_{\mathrm{B}} -n_{\mathrm{s}} -1}{2}} \Sigma_{2}(n_{\mathrm{s}}, n_{\mathrm{B}}).
\label{FF9}
\end{eqnarray}
In Eqs. (\ref{FF8}) and (\ref{FF9}) the functions $\Sigma_{1}(n_{\mathrm{s}}, n_{\mathrm{B}})$ and $\Sigma_{2}(n_{\mathrm{s}}, n_{\mathrm{B}})$ 
encode a milder dependence upon the spectral indices and they can be 
determined by matching the Sachs-Wolfe expression (valid for $\ell < \ell_{1}$)
with the results of the explicit numerical integration of the basic integrals 
(valid for $\ell> \ell_{1}$). The form of  $\Sigma_{1}(n_{\mathrm{s}}, n_{\mathrm{B}})$ and $\Sigma_{2}(n_{\mathrm{s}}, n_{\mathrm{B}})$ also depend upon the 
regularization scheme of the magnetic energy density and here the 
explicit expressions will be given in the case of blue magnetic spectral indices:
\begin{eqnarray}
&& \Sigma_{1}(n_{\mathrm{s}}, n_{\mathrm{B}}) = \frac{  \Gamma^2( 2 - n_{\mathrm{s}}/2)}{\Gamma(3 - n_{\mathrm{s}}) 
\Gamma^2(5/2- n_{\mathrm{B}})}  f(n_{\mathrm{B}}-1).
\label{FF8a}\\
&& \Sigma_{2}(n_{\mathrm{s}}, n_{\mathrm{B}})= \frac{ \Gamma\biggl(\frac{7}{2} - \frac{n_{\mathrm{s}}}{2} -n_{\mathrm{B}}\biggr)\, \Gamma^2( 2 - n_{\mathrm{s}}/2)}{\Gamma^2\biggl(\frac{9}{4} - \frac{n_{\mathrm{s}}}{4} - \frac{n_{\mathrm{B}}}{2}\biggr) \Gamma(3 - n_{\mathrm{s}})}\sqrt{f(n_{\mathrm{B}}-1)},
\label{FF9a}\\
&& f(x) = \frac{4 ( 6 - x) \, (2\pi)^{2 x}}{3 x ( 3 - 2x) \Gamma^2(x/2)}.
\label{FF10}\\
\end{eqnarray}
The functions appearing in Eqs. (\ref{FF8a})--(\ref{FF9a}) can also
be estimated (just in the limit of large $\ell$) from the analytic expressions 
of the different integrals, as illustrated above in this section. In the latter case the resulting expression will still have the correct scaling properties but the overall normalization will have to be adjusted. Conversely, the advantage 
of Eqs. (\ref{FF8})--(\ref{FF9}) and (\ref{FF8a})--(\ref{FF9a}) is that 
they are immediately comparable to the numerical calculation also 
for small $\ell$.  In Eqs. (\ref{FF8}) and (\ref{FF9}) we have that  $\ell_{\mathrm{B}} =1$. If  the integrals would just be estimated from their small-scale approximation the putative value of $\ell_{\mathrm{B}}$ would be larger and will fix the limits of applicability of the formula. 

As already mentioned, in Figs. \ref{figure6} and \ref{figure7} the dashed curves illustrate
the magnetic form factor of Eq. (\ref{FF7}) for the different values 
of the parameters appearing in each plot. The structure of Figs. \ref{figure7} and 
 \ref{figure8} is as follows:
\begin{itemize}
\item{}  in the three plots at the left the magnetic spectral index 
is fixed to $n_{\mathrm{B}}=1.5$ while in the three plots at the right the magnetic spectral index is fixed to $n_{\mathrm{B}} =2$;
 \item{} from top to bottom, as indicated in each plot, the values 
of the magnetic field intensity augments from $1$ nG to $20$ nG.
\end{itemize}
According to Figs. \ref{figure7} and \ref{figure8}, the magnetic form factor of 
Eq. \ref{FF7} reproduces quite faithfully the average of the numerical 
data and this is what scaling relations can provide, in this 
context. As explicitly shown by Figs. \ref{figure7} and \ref{figure8}
 the very same form factor works both for the TT and for the EE angular power spectra. The latter observation demonstrates  that the factorization of Eq. (\ref{FF2a}) is not only analytically plausible but it is also numerically justified.
\begin{figure}[!ht]
\centering
\includegraphics[height=6cm]{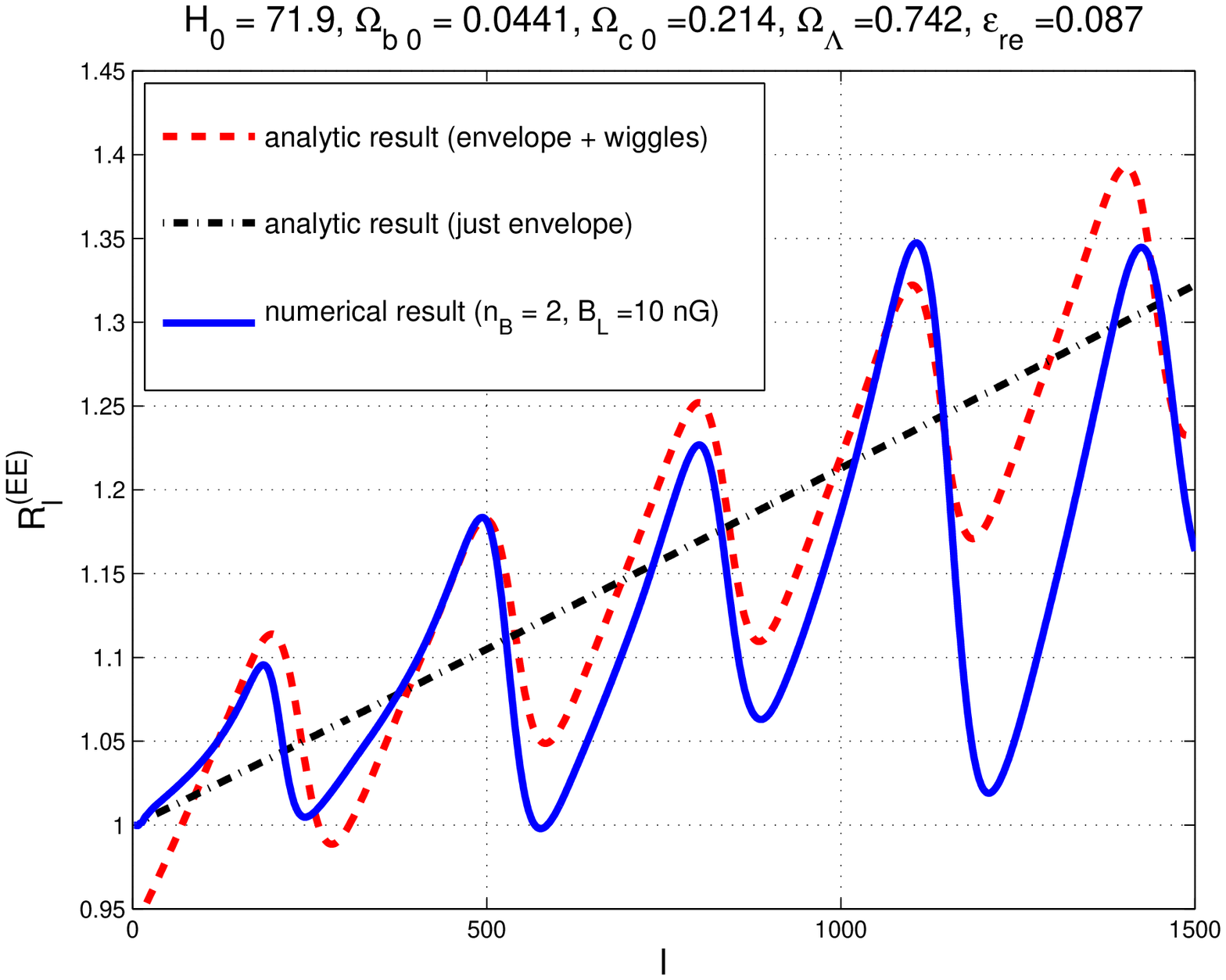}\includegraphics[height=6cm]{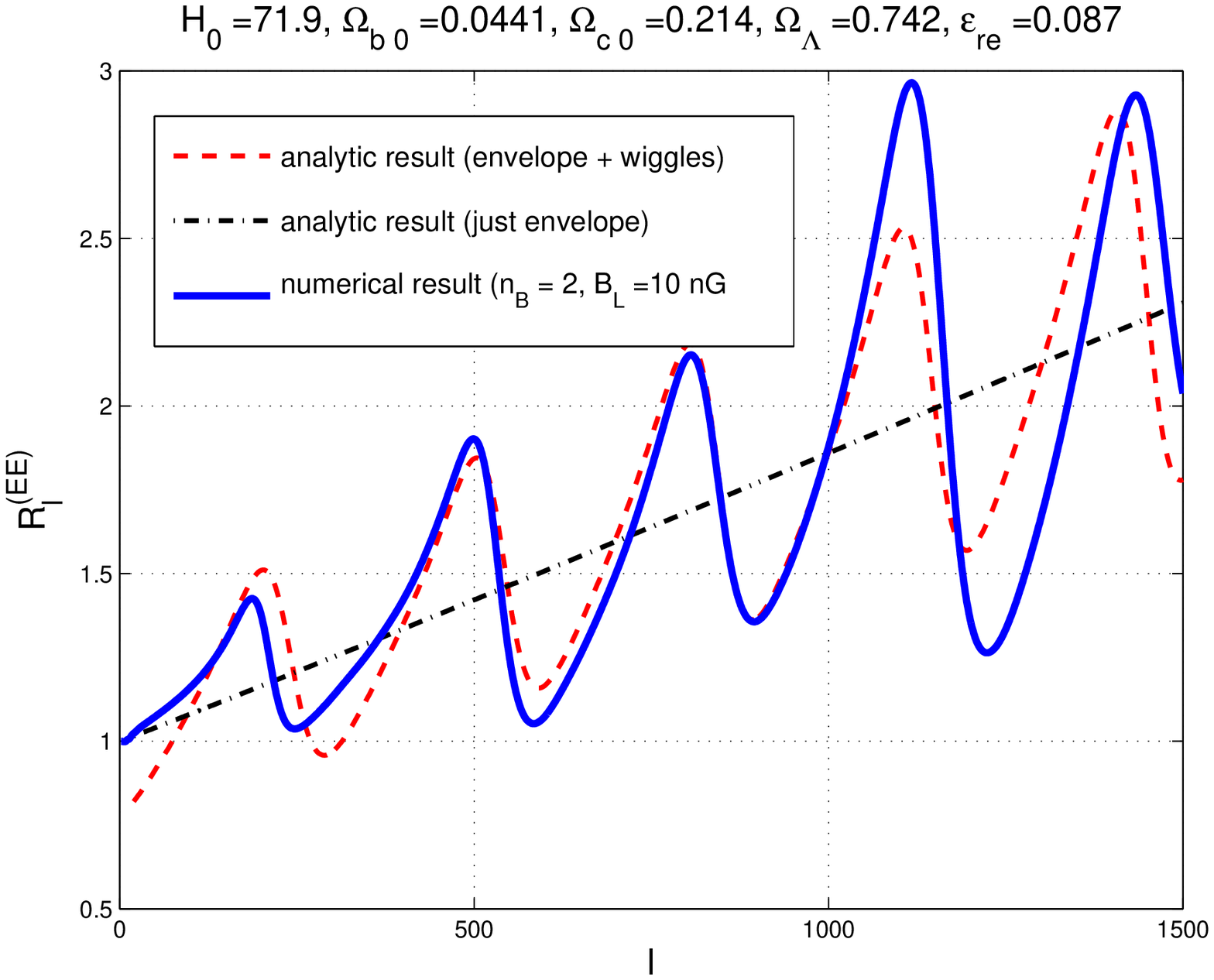}
\caption[a]{The analytic form of the wiggles is compared with the numerical results.}
\label{figure9}      
\end{figure}
The last observation brings up a further question: can we also understand semi-analytically the structure of the wiggles of Figs. \ref{figure7} and \ref{figure8}? 
While in  Fig. \ref{figure9} the wiggles exhibit a double periodicity (i.e. a hump is followed by a peak), 
in Fig. \ref{figure8} there is a single periodicity (i.e. a single peak is followed by a single peak). The difference in the two structure is understandable on the basis of the considerations of sections \ref{sec3} and \ref{sec4}:  while the TT correlations arise as the interference of the monopole and of the dipole ther EE correlations mainly feel the dipole and are, therefore, a cleaner probe where the analytic considerations 
can be more easily confronted with the numeric results. Going back to the parametrization of Eq. (\ref{FF2a}) the analytic structure of the wiggles can be written as
\begin{eqnarray}
&& O^{(EE)}(n_{\mathrm{B}}, B_{\mathrm{L}}) = \frac{a_{\mathrm{E}}  - b_{\mathrm{E}} \cos{[ 2 \gamma_{\mathrm{A}}
(\ell +\ell_{1}) -\delta_{\mathrm{B}}]}}{a_{\mathrm{E}}  - b_{\mathrm{E}} \cos{[ 2 \gamma_{\mathrm{A}}(\ell +\ell_{1}) ]}}.
\label{FF11}\\
&& \delta_{\mathrm{B}}\equiv \delta(n_{\mathrm{B}}, B_{\mathrm{L}}) = 9.2 \times 10^{-3}\biggl[
 \biggl(\frac{B_{\mathrm{L}}}{\mathrm{nG}}\biggr)^{2} + \frac{2 n_{\mathrm{B}} - n_{\mathrm{s}} -1}{2} 
 \biggl(\frac{B_{\mathrm{L}}}{\mathrm{nG}}\biggr)\biggr],
 \label{FF12}
\end{eqnarray}
where $\ell_{1}$, $a_{\mathrm{E}}$ and $b_{\mathrm{E}}$  have been already introduced in 
Eqs. (\ref{EE22a})--(\ref{EE22}). The rationale behind Eqs. (\ref{FF11})--(\ref{FF12}) 
is rather simple. In the denominator of Eq. (\ref{FF11}) there is the analytic form of the best fit, while 
in the numerator the ambient magnetic field introduces a phase difference. 
In Fig. \ref{figure9} the analytic expressions for the wiggles are compared with the numerical results.  The remaining offsets are within the accuracy of the analytic approach and improve on the pure scaling estimate which lead to the derivation of the envelope. 
The results derived here are relevant for the dedicated strategies of parameter extraction which have been 
suggested in \cite{ESTIMATE}.

\renewcommand{\theequation}{6.\arabic{equation}}
\setcounter{equation}{0}
\section{Parameter space of magnetized CMB observables}
\label{sec6}
In the present study the values of the magnetic field intensities and of the corresponding 
spectral indices have been taken to be, in some cases, rather large in the sense 
that the selected values lead to CMB observables which are 
incompatible with the observed ones. As already mentioned the largeness 
of some of the selected values is evident from the comparison of the computed CMB 
observables with the best fit to the WMAP 5yr data alone (see, e.g. Fig. \ref{figure2}) 

The choice of dealing with some of these extreme values is, in some sense, 
dictated by the logic followed in the present study: for large values 
of the magnetic fields, the scaling properties of the angular power spectra are more 
transparent and the distortions enhanced. We hope it is clear, from the results of the previous section, that 
indeed, the distortion patterns scale with the amplitude but their morphology 
remains unchanged. This is, after all, closely related to the intuitive notion of scaling (see, e.g. Figs. \ref{figure7} 
and \ref{figure8}).

At the beginning of Section \ref{sec2} it has been mentioned that, for instance, 
the values $(n_{\mathrm{B}},\, B_{\mathrm{L}}) = (2,\, 10\, \mathrm{nG})$ are 
excluded, in a frequentist perspective, to 95 \% confidence level. In what follows 
the latter statement will be made more quantitative by deriving and by discussing 
the relevant exclusion plots in terms of the spectral index $n_{\mathrm{B}}$ and of the 
magnetic field intensity $B_{\mathrm{L}}$. 
\begin{figure}[!ht]
\centering
\includegraphics[height=6cm]{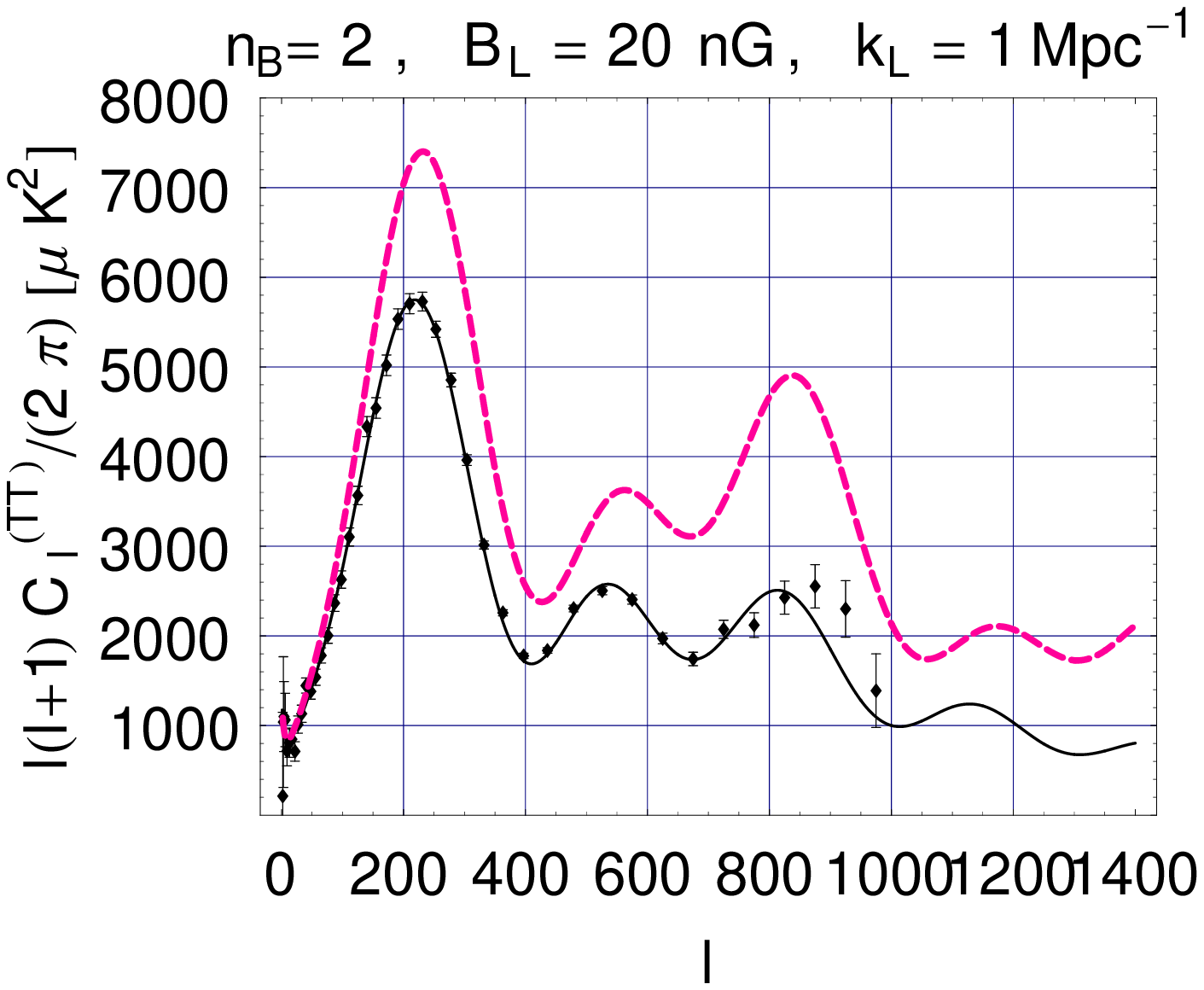}
\includegraphics[height=6cm]{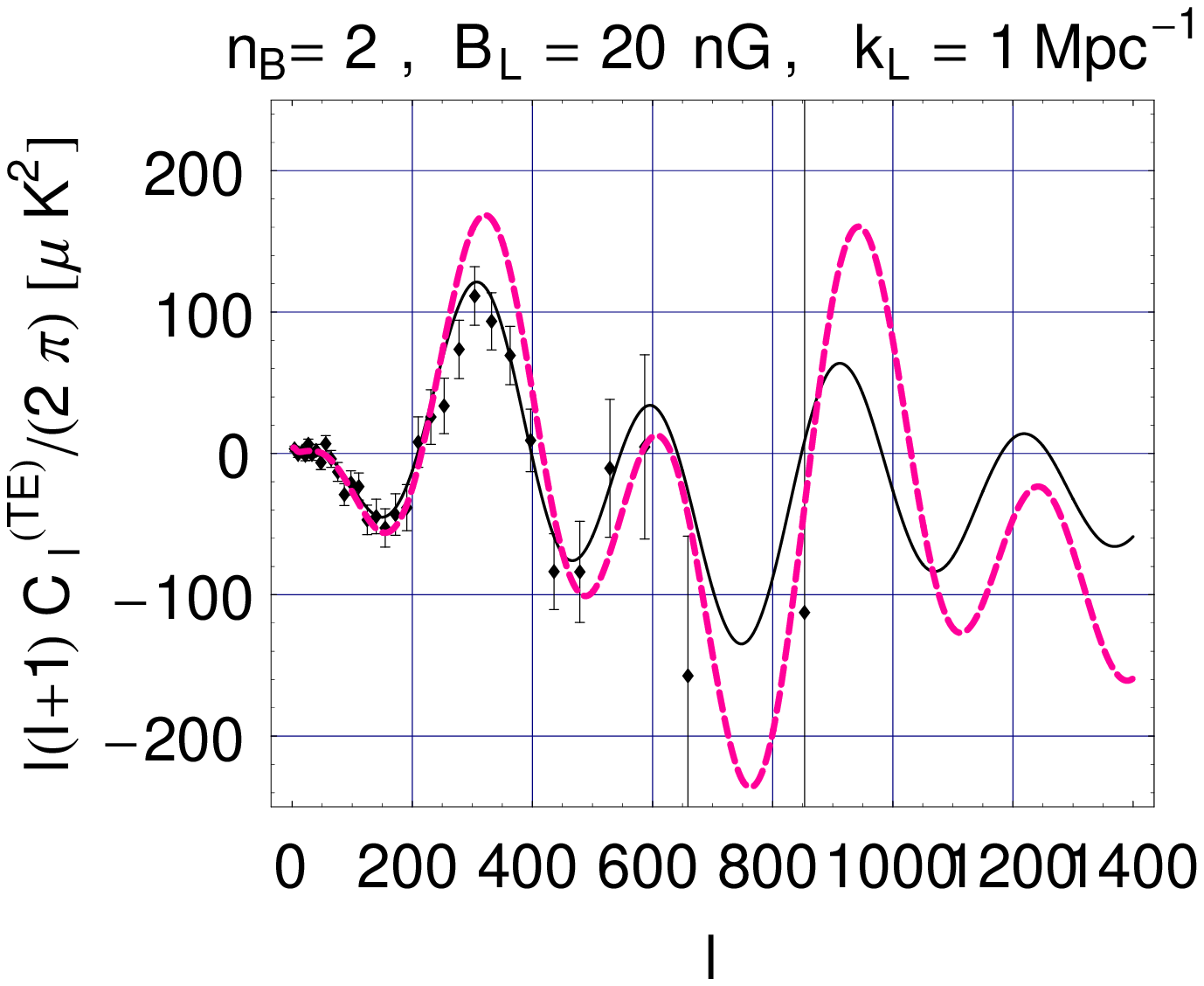}
\caption[a]{The TT and TE angular power spectra are illustrated in the case  $(n_{\mathrm{B}},\, B_{\mathrm{L}}) = (2,\, 10\, \mathrm{nG})$ (dashed lines) and for the best fit to the WMAP 5yr data alone (full line). In both cases $\beta =0$ 
for consistency with the analysis of the previous sections.}
\label{figure10}      
\end{figure}
Let us, first of all, convince ourselves that, indeed,  the model $(n_{\mathrm{B}},\, B_{\mathrm{L}}) = (2,\, 10\, \mathrm{nG})$
does not correctly reproduce the data. In Fig. \ref{figure10} the full line illustrates the best fit to the WMAP 5yr data 
alone  while the dashed line is the numerical result for the case $(n_{\mathrm{B}},\, B_{\mathrm{L}}) = (2,\, 10\, \mathrm{nG})$, when all the other parameters are fixed as in Eq. (\ref{Par1}).  If Fig. \ref{figure10} the data points (with the 
corresponding error bars) refer to the binned data both for the TT and TE angular power spectra. The binned 
data  contain $34$ (effective) multipoles in the TE correlation and $43$ (effective)multipoles in the TT spectrum. Following the usual habit, to make the plots more readable,  the binned data points have been included. 
Conversely, the unbinned data (which are the ones used in the following analyses) contemplate all the 
 multipoles from $\ell =2$ to $\ell = 1000$ both for the TT and for the TE (observed)
 power spectra. Finally, we included the TT and the TE angular power spectra since they are 
 the best measured spectra in the context of the WMAP 5yr data.

Having established that  the parameters $(n_{\mathrm{B}},\, B_{\mathrm{L}}) = (2,\, 10\, \mathrm{nG})$ are 
excluded let us now try to understand to what confidence level they can be excluded. 
The simplest way of exploring the parameter space of the magnetized models goes, in short, as follows:
\begin{itemize}
\item{} the joined two-dimensional marginalized contours for the various cosmological parameters identified already by the analyses of the WMAP 3yr data  are ellispses with an approximate Gaussian dependence on the confidence level;
\item{}   the confidence intervals for the $2$ supplementary  parameters of the model (i.e. $n_{\mathrm{B}},\, B_{\mathrm{L}}$) can then be determined by using an appropriate gridding  approach;
\item{} the remaining parameters of  Eq. (\ref{Par1}) are assumed to be known and follow a Gaussian probability densityfunction. 
\end{itemize}
This approach is rather standard when exploring the impact of new scenarios on the CMB observables 
(see, e.g.,  \cite{h1,h2,h5} for the case of non-adiabatic modes supplementing the standard $\Lambda$CDM scenario).
In the approach we just described, the shape of the likelihood function can be determined by evaluating the least square estimator
\begin{equation}
\chi^2(n_{\mathrm{B}}, B_{\mathrm{L}}) = 
\sum_{\ell} \biggl[\frac{ C_{\ell}^{(\mathrm{obs})} - C_{\ell}(n_{\mathrm{B}},  B_{\mathrm{L}})}{\sigma_{\ell}^{(\mathrm{obs})}} \biggr]^2, 
\label{grid1}
\end{equation}
where $\sigma_{\ell}^{\mathrm{obs}}$ are the estimated errors from the observations 
for each multipole and where the functional dependence of  $C_{\ell}(n_{\mathrm{B}}, B_{\mathrm{L}})$ is given by the underlying theory (i.e. the magnetized $\Lambda$CDM model) which we try to falsify by comparing its predictions to the actual observations.  The observed angular power spectra 
(i.e. $C_{\ell}^{\mathrm{obs}}$) 
are derived by using the (further) estimators $\hat{C}_{\ell}^{(\mathrm{TT})}$ and $\hat{C}_{\ell}^{(\mathrm{TE})}$, i.e.  
\begin{equation}
\hat{C}_{\ell}^{(\mathrm{TT})}= \frac{1}{2 \ell +1} \sum_{m = - \ell}^{\ell} |\hat{a}^{(\mathrm{T})}_{\ell \, m}|^2, \qquad 
 \hat{C}_{\ell}^{(\mathrm{TE})}= \frac{1}{2 \ell +1} \sum_{m = - \ell}^{\ell} |\hat{a}^{(\mathrm{T})}_{\ell \, m}\,\hat{a}^{(\mathrm{E})*}_{\ell \, m}|, 
\label{grid2}
\end{equation}
whose distribution becomes Gaussian, according to the central limit theorem, when $\ell \to \infty$. 
The minimization of Eq. (\ref{grid1}) is equivalent to the minimization of the lognormal likelihood function ${\mathcal L} = - 2 \, \ln{L}$ where $L$ is given by
\begin{equation}
L(\mathrm{data}|\,n_{\mathrm{B}}, \, B_{\mathrm{L}}) = L_{\mathrm{max}}
e^{- \chi^2(n_{\mathrm{B}}, B_{\mathrm{L}})/2}.
\label{grid4}
\end{equation}
Thus, the minimization of Eq. (\ref{grid1})  is equivalent to the maximization of the likelihood of Eq. (\ref{grid4}). 
\begin{figure}[!ht]
\centering
\includegraphics[height=6cm]{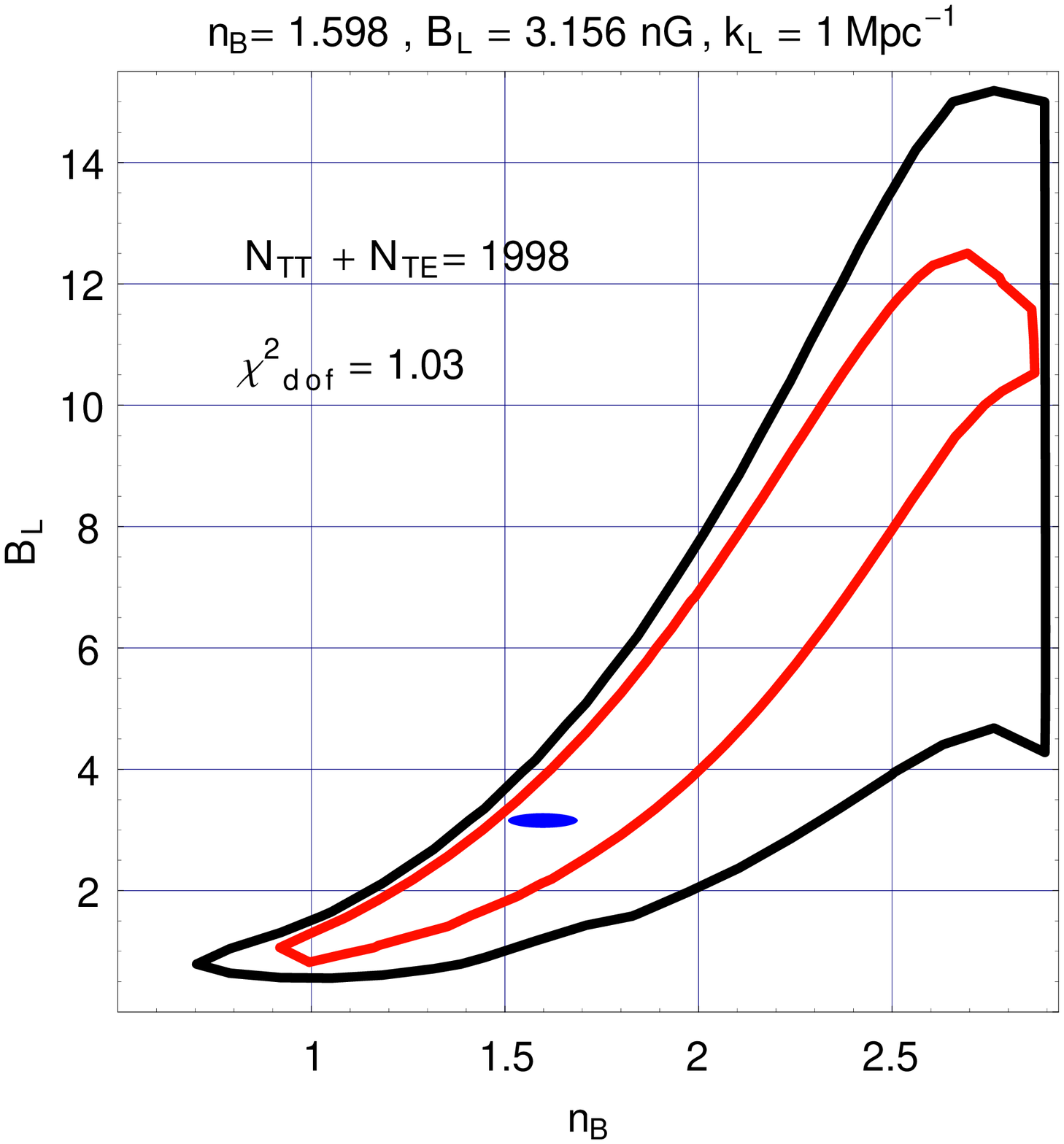}
\includegraphics[height=6cm]{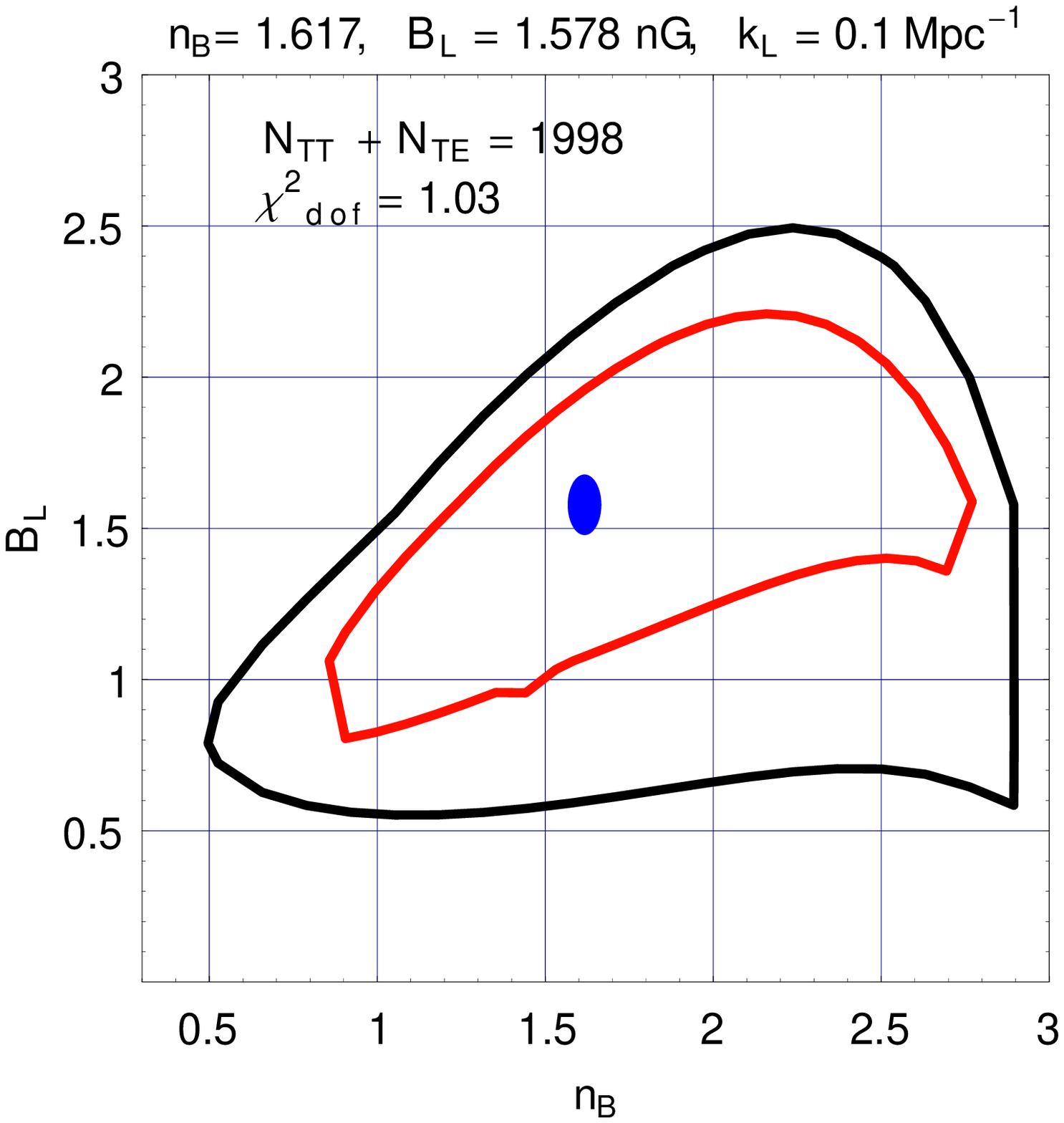}
\caption[a]{The parameter space of the magnetized CMB anisotropies for two illustrative choices of the magnetic pivot scale. In both plots $\beta =0$.}
\label{figure11}      
\end{figure}
In Fig. \ref{figure11} the parameter space of the model is illustrated, respectively, for two different choices 
of the magnetic pivot scale $k_{\mathrm{L}}$, i.e. $k_{\mathrm{L}} = \mathrm{Mpc}^{-1}$ (plot at the left) and $k_{\mathrm{L}} =0.1 \mathrm{Mpc}^{-1}$ (plot at the  right). The shaded spots, in both plots of Fig. \ref{figure11}, are meant 
to emphasize the value for which the estimator of Eq. (\ref{grid1}) is minimized; note that  $\chi^2_{\mathrm{dof}} = \chi_{\mathrm{min}}^2/N_{\mathrm{dof}}$ is the value of the (reduced) $\chi^2$ at the minimum (i.e. when $\chi^2 \equiv \chi^2_{\mathrm{min}}$).  In both plots the data points 
for the TE and TT angular power spectra have been used in their unbinned form. Overall the total number of data points 
is $N_{\mathrm{TT}} + N_{\mathrm{TE}} = 1998$ since $N_{\mathrm{TT}} =999$  and $N_{\mathrm{TE}}=999$.
In both plots of Fig. \ref{figure11} the boundaries of the two regions 
contain $68.3\,$\% and $95.4\,$\% of likelihood as the 
values for which the $\chi^2$ has increased, respectively, by an amount $\Delta \chi^2 = 2.3$ and 
$\Delta \chi^2 = 6.17$.  The latter figures stem directly from the fact that we are 
dealing with a two-dimensional parameter space. 
Figure \ref{figure11} offers a more quantitative interpretation of the plots reported in Fig. \ref{figure10}.
In Fig. \ref{figure10} the value of the magnetic pivot scale is $k_{\mathrm{L}} = \mathrm{Mpc}^{-1}$. Therefore 
the results of Fig. \ref{figure11} do apply. In Fig. \ref{figure11} the point $(n_{\mathrm{B}},\, B_{\mathrm{L}}) = (2,\, 10\, \mathrm{nG})$ is located very far from the outer contour. 
In a frequentist perspective, a  model located beyond the outer contours of Figs. \ref{figure11} is excluded, 
by the current WMAP data on the TT and TE correlations, to $95$\% confidence level.

The exclusion plots reported in Fig. \ref{figure11} have been obtained by means of a grid approach and by using 
directly the numerical results. We can now ask ourselves the following question. What happens  if we use 
the approximate form of the angular power spectra derived in the present paper and use, simultaneously, the 
same data but in their binned form? 
\begin{figure}[!ht]
\centering
\includegraphics[height=6cm]{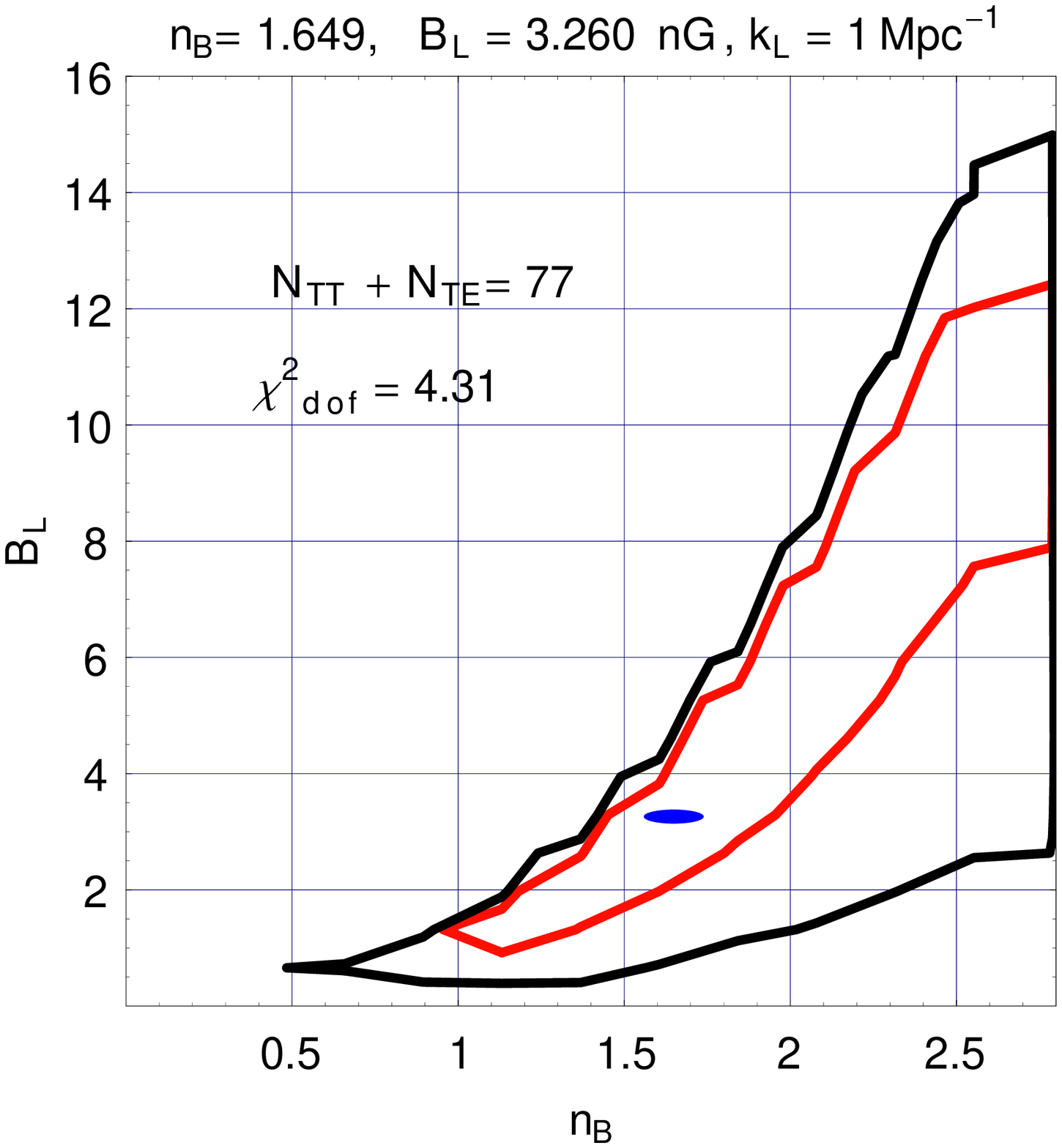}
\includegraphics[height=6cm]{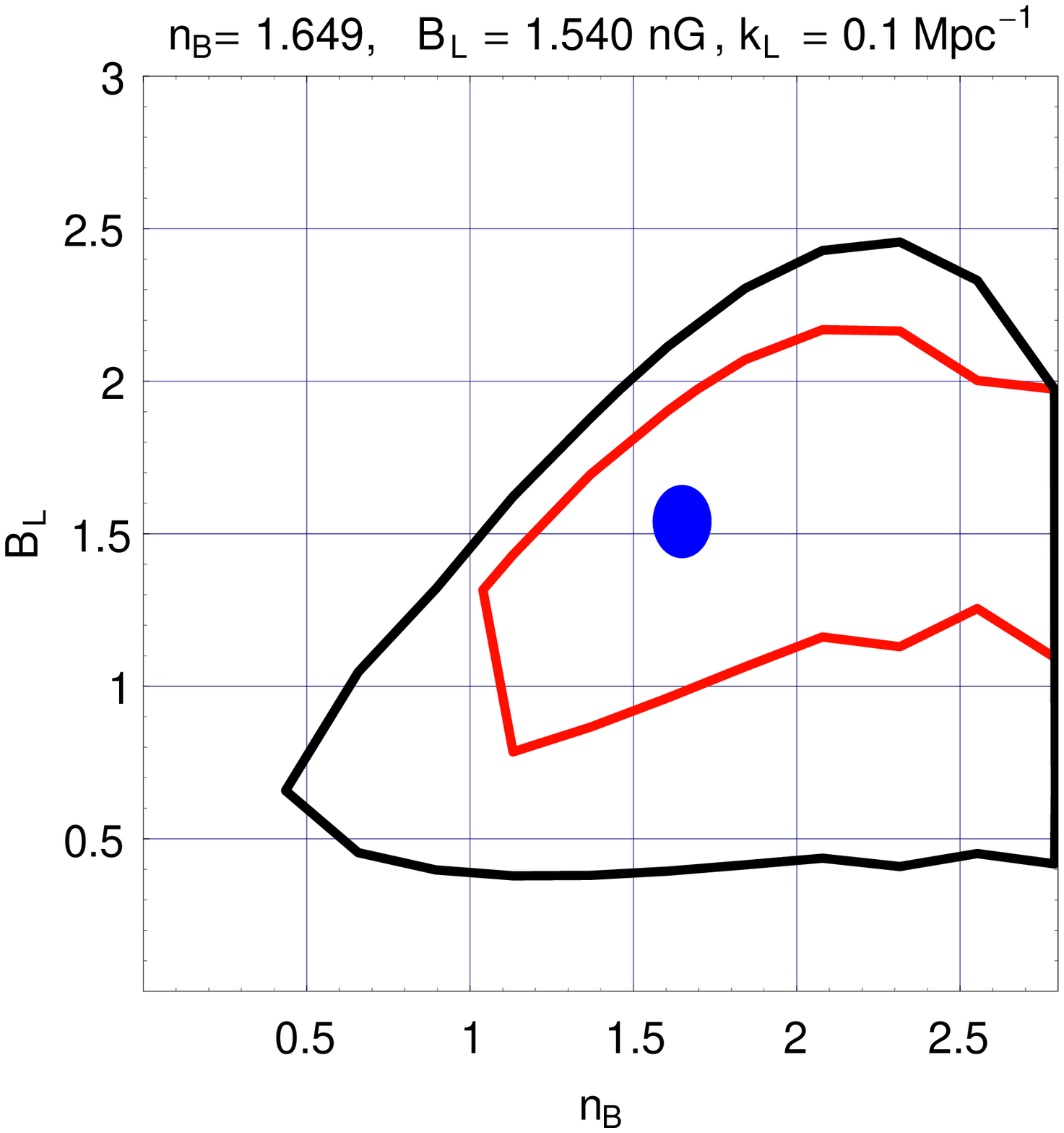}
\caption[a]{The parameter space of the magnetized CMB anisotropies as derived by using the semi-analytic expressions for the angular power spectra together with the data in their binned form. In both plots $\beta =0$.}
\label{figure12}      
\end{figure}
The results are illustrated in Fig. \ref{figure12}. It is amusing to notice that, in spite 
of numerical differences which are fully justified the shapes and the regions of the exclusion 
plots of Fig. \ref{figure12} are consistent with the ones of Fig. \ref{figure11}. The results 
of Fig. \ref{figure12} have been obtained by parametrizing the angular power spectra as   
in Section \ref{sec5}. Note, however, that in Section \ref{sec5} we only reported explicit expressions 
in the case $n_{\mathrm{B}}>1$. To obtain the result of Fig. \ref{figure12} we also need the analog formulas 
but in the case\footnote{Recall, as stressed in \cite{faraday1} (third paper in the reference), that the cases
 $n_{\mathrm{B}} =1$ and $n_{\mathrm{B}} =5/2$ 
should be separately regularized at the level of the magnetic energy density. } $n_{\mathrm{B}} < 1$.
The results of Fig. \ref{figure12} show a fair consistency of our analytical approach. At the same time 
it is clear that the value of the reduced $\chi^2$ is larger than in the case of Fig. \ref{figure11}: in Fig. \ref{figure12}, 
the data have been used in their binned form. The largeness of the reduced $\chi^2$ 
simply means, within the present approach,  that the uncertainties entering Eq. (\ref{grid1}) have been underestimated.
With the last proviso, the results derived in this paper seem to allow for  approximate evaluations of the parameter space of the 
magnetized CMB anisotropies which are compatible with the fully numerical results. 

\renewcommand{\theequation}{7.\arabic{equation}}
\setcounter{equation}{0}
\section{Concluding considerations}
\label{sec7}
The (limited) question addressed in this paper  is 
how the angular power spectra of the CMB anisotropies scale with 
the parameters of an ambient magnetic field when the remaining parameters  
are close to the ones of the standard $\Lambda$CDM scenario. 

To get a definite answer, the calculation of the magnetized temperature and polarization anisotropies had to  be reduced to the evaluation of a set of basic integrals whose explicit form simplifies in the limit 
of sufficiently small angular scales, i.e. in the limit of sufficiently large multipoles. 
It has been shown that the temperature and polarization observables obtainable by semi-analytic means are sufficiently accurate to infer a set of scaling relations which can be used to determine the effect of large-scale 
magnetic fields on the TT, TE and EE angular 
power spectra.  It has been also shown explicitly 
how the distortions patterns induced by large-scale 
magnetism can be deduced from generalized magnetic 
form factors accounting for the scaling properties 
of the CMB observables as a function of the parameters 
of the ambient magnetic field.  

The cleanest probe of large-scale magnetism is represented by the polarization autocorrelations. 
Indeed the magnetized EE angular power spectra exhibit a single periodicity, a milder 
dependence upon the dissipative scale an a rather clear shift of the peaks for intermediate and large multipoles. 
The obtained results have been confronted with the numerical calculation. The same evolution 
equations studied analytically have been integrated numerically. 
The analytical derivations are corroborated by the numerical results which are based
on the same physical description. The obtained results are relevant for dedicated 
strategies of parameter extraction such as the ones mentioned at the end of the previous section.

Large-scale magnetism is a complicated phenomenon which we choose to scrutinize in its infancy, i.e. around the time of photon decoupling. This analysis must necessarily be cautious and modest.  The rationale for such a caveat does not reside in a particular ethical conviction but in the nature of the problem: 
while in the laboratory we completely dominate the initial conditions of our experiments 
in astrophysics and cosmology this is not the case. So, unless we are vigilant 
it is well possible to mistake an effect due to the peculiar nature of initial conditions 
with a missing piece of dynamics which should have been included in a particular 
regime \footnote{By this statement I mean that it is 
important to scrutinize systematically all the relevant plasma 
effects which are applicable in the pre-decoupling regimes. This 
has been partially done in the present paper as well as the original 
studies of Refs. \cite{max1,max2,max3,max5}. Even if the present estimates suggest that other (potentially relevant) plasma effects do not play a crucial numerical role, care must betaken in sharpening these 
estimates. Indeed,  a missing piece of the dynamics could be mistaken as a peculiar feature stemming from the initial conditions.}. In this respect it is important  to understand in detail all the potentially interesting physical effects which could modify the CMB observables. This study is time-consuming both at a numerical and at the analytic level. It could even be said, by some, that such an approach is slow and pedantic. While we are ready to take this risk, it is
rewarding that, at the moment, the estimates of the effects of large-scale magnetism on the scalar modes of the CMB anisotropies seem to reach (slowly) the same standards employed in the absence of ambient magnetic fields. More theoretical effort, in this direction, is certainly needed.

\end{document}